\documentclass[namedreferences]{solarphysics}
\usepackage[optionalrh,natbib]{spr-sola-addons} 
\usepackage{graphicx}        	
\usepackage{natbib}         	
\usepackage{amssymb}        	
\usepackage{url}             	
\usepackage{rotating}					
\usepackage[breaklinks]{hyperref} 

\bibpunct{(}{)}{;}{a}{}{,}
\usepackage{txfonts}

\newcommand{\asr}{    {\it Adv. Space Res.}} 
 
\newcommand{\aap}{    {\it Astron. Astrophys.}}
\newcommand{\aaps}{   {\it Astron. Astrophys. Suppl.}}

\newcommand{\ag}{     {\it Ann. Geophys.}}
 
\newcommand{\apj}{    {\it Astrophys. J.}}

\newcommand{\grl}{    {\it Geophys. Res. Lett.}}

\newcommand{\jgr}{    {\it J. Geophys. Res.}}

\newcommand{\nat}{    {\it Nature}}

\newcommand{\pss}{		{\it Pl. Sp. Sci.}}
\newcommand{\solphys}{{\it Solar Phys.}}
 
\newcommand{\ssr}{    {\it Space Sci. Rev.}} 
\newcommand{\jrasc}{  {\it Journal of the Royal Astronomical Society of Canada}}

\begin{document}

\begin{article}

\begin{opening}

\title{Heliolatitude and time variations of solar wind structure from \textit{in~situ} measurements and interplanetary scintillation observations}

\author{J.M.~\surname{Sok\'{o}{\l}}$^{1}$\sep
        M.~\surname{Bzowski}$^{1}$\sep
        M.~\surname{Tokumaru}$^{2}$\sep
        K.~\surname{Fujiki}$^{2}$\sep
        D.J.~\surname{McComas}$^{3,4}$      
       }
\runningauthor{Sok\'{o}{\l} et al.}
\runningtitle{Heliolatitude and time variations of solar wind structure}

   \institute{$^{1}$ Space Research Centre Polish Academy of Sciences, Warsaw, Poland
                     email: \url{jsokol@cbk.waw.pl} email: \url{bzowski@cbk.waw.pl}\\ 
              $^{2}$ Solar-Terrestrial Environment Laboratory, Nagoya University, Nagoya, Japan
                     email: \url{tokumaru@stelab.nagoya-u.ac.jp} email: \url{fujiki@stelab.nagoya-u.ac.jp} \\
              $^{3}$ Southwest Research Institute, San Antonio, TX 78249, USA
                     email: \url{DMcComas@swri.edu} \\
              $^{4}$ University of Texas at San Antonio, San Antonio, TX 78249, USA\\
             }

\begin{abstract}
The 3D structure of the solar wind and its evolution in time is needed for heliospheric modeling and interpretation of energetic neutral atoms observations. We present a model to retrieve the solar wind structure in heliolatitude and time using all available and complementary data sources. 
We determine the heliolatitude structure of solar wind speed on a yearly time grid over the
past 1.5 solar cycles based on remote-sensing observations of interplanetary scintillations,
\textit{in~situ} out-of-ecliptic measurements from \textit{Ulysses}, and \textit{in~situ} in-ecliptic measurements from the OMNI-2 database. Since the \textit{in~situ} information on the solar wind density structure out of ecliptic is not available apart from the \textit{Ulysses} data, we derive correlation formulae between the solar wind speed and density and use the information on the solar wind speed from interplanetary scintillation observations to retrieve the 3D structure of solar wind density. With the variations of solar wind density and speed in time and heliolatitude available we calculate variations in solar wind flux, dynamic pressure and charge exchange rate in the approximation of stationary H atoms.

\end{abstract}
\end{opening}

\section{Introduction}
The goal of this paper is to retrieve the solar wind structure at 1~AU as a function of time and heliolatitude based on available \textit{in~situ} data sources and interplanetary scintillation observations for the time interval since 1990 until present that covers 1.5 solar activity cycles. 

The existence of the solar wind was predicted on a theoretical basis by \citet{parker:57a} and discovered experimentally by \textit{Lunnik}~II and \textit{Mariner}~2 at the very beginning of the space age \citep{gringauz_etal:60a, neugebauer_snyder:62a}. Regular measurements of its parameters began in the early 1960s and data from many spacecraft are now available, obtained using various techniques of observations and data processing \citep[see for review][]{bzowski_etal:12b}. Shortly after the discovery of the solar wind (SW hereafter), a question of whether or not it is spherically symmetric was put forward. Most spacecraft with instruments to measure the SW parameters are at orbits close to the ecliptic plane and the information about the latitudinal structure of the SW is hard to obtain. There are a few sources of data on the out of ecliptic SW parameters, but only one of them from \textit{in~situ} measurements, namely from \textit{Ulysses}. 

While direct observations of the SW in the ecliptic plane have been collected for many years, information on its latitudinal structure had been available only from indirect observations of the cometary ion tails \citep{brandt_etal:75a}, until radio-astronomy observations of interplanetary scintillation \citep{hewish_etal:64a, coles_maagoe:72a} and spaceborne measurements of the Lyman-$\alpha$ helioglow \citep{lallement_etal:85a, bertaux_etal:95} became available. To our knowledge these two techniques remain the only source of global, time-resolved information on the the solar wind structure. The launch of the \textit{Ulysses} spacecraft \citep{wenzel_etal:89a} improved our understanding of the 3D behavior of the solar wind by offering direct \textit{in~situ} observations and a very high resolution in latitude, but a poor resolution in time\footnote{The same latitudes were visited only a few times during the $\sim 20$-year mission.}. 

The solar wind structure varies in latitude with the solar activity cycle. Knowledge of its evolution is needed to construct credible models of the heliosphere and its boundary regions. With the history of the SW evolution based on a homogeneous series of data and retrieved using homogeneous analysis method, one obtains a tool to interpret both present heliospheric observations, such as the ongoing measurements of Energetic Neutral Atoms (ENAs) by the NASA \textit{Interstellar Boundary Explorer}~\citep[IBEX,][]{mccomas_etal:09a} and \textit{in~situ} measurements of the heliospheric environment by the Voyagers, and to compare them with the results from past and current long-lived experiments such as \textit{Solar Wind ANisotropy}~(SWAN) onboard \textit{SOlar and Heliospheric Observatory}~(SOHO) etc. 

We assume that the solar wind expansion is purely radial, its speed does not change with solar distance, and density drops down quadratically with distance from the Sun. These assumptions are valid for close distances from Sun $\left( r < 10~AU \right)$. Their validity at farther distances will be considered in the Discussion section. 

In the following text ``CR-averaged'' data mean Carrington rotation (CR hereafter) averaged values, where Carrington rotation period is the synodic period of solar rotation equal to 27.2753 days \citep{franz_harper:02a}. Throughout the paper, ``adjusted'' means scaled from the value $x_0$ measured at a heliocentric distance $r_\mathrm{0}$ to the distance $r_\mathrm{E} = 1$~AU, where the value specific for $r_\mathrm{E}$ is calculated as $x_\mathrm{E} = x_0 \left( r_0 / r_\mathrm{E} \right)^2$.

\section{Datasets used}   
In our studies we use 3 complementary data sources: \textit{in~situ} in-ecliptic measurements from various spacecraft combined in the OMNI-2 collection, \textit{in~situ} out-of-ecliptic measurements of solar wind plasma by \textit{Ulysses}, and remote-sensing radio-observations of interplanetary scintillations (IPS), interpreted using the Computer Assisted Tomography (CAT) technique. 

\subsection{\textit{in~situ} in-ecliptic measurements collected from various spacecraft}
Solar wind in the ecliptic plane (which differs from the solar equator plane by $7.25^\circ$) is a mixture of a ``genuine'' slow solar wind, fast solar wind from coronal holes, solar wind plasma from stream-stream interaction regions, and (intermittently) interplanetary coronal mass ejections. The features of these components change with solar activity.

The parameters of the SW vary with the phases of Solar Cycle (SC hereafter) and, as recently discovered by \citet{mccomas_etal:08a}, also secularly. The OMNI-2 data collection, available at http://omniweb.gsfc.nasa.gov/ \citep[see][]{king_papitashvili:05}, gathers SW measurements from early 1960s until present and brings them to a common calibration. The absolute calibration of the current version of the OMNI-2 collection is based on the absolute calibration of \textit{Wind} measurements \citep{kasper_etal:06a}. 

The main source of data is nowadays \textit{Advance Composition Explorer}~(ACE) and \textit{Wind}. Before the \textit{Wind} era the main datasource was measurements from \textit{Interplanetary Monitoring Platform-8}~(IMP-8), with gaps filled by miscellaneous spacecraft. The data from the epoch before IMP-8 are from various experiments and could not be reliably brought to a common calibration with the IMP-8/\textit{Wind} system because of the lack of overlap between the measurement time intervals. Our analysis starts in 1985, when IMP-8 was already in operation \citep[see also review by][]{bzowski_etal:12b}.

The~solar wind parameters show considerable variations during one solar rotation period, with quasi-periodic changes from slow to fast solar wind speed and related changes in density. The time scale of the changes of the fast/slow wind streams is comparable to the solar rotation period and thus constructing a full and accurate model of the solar wind variation as a function of time and heliolongitude is currently not feasible because of the lack of sufficient data. 

\begin{figure}
		\centering
		\includegraphics[scale=0.7]{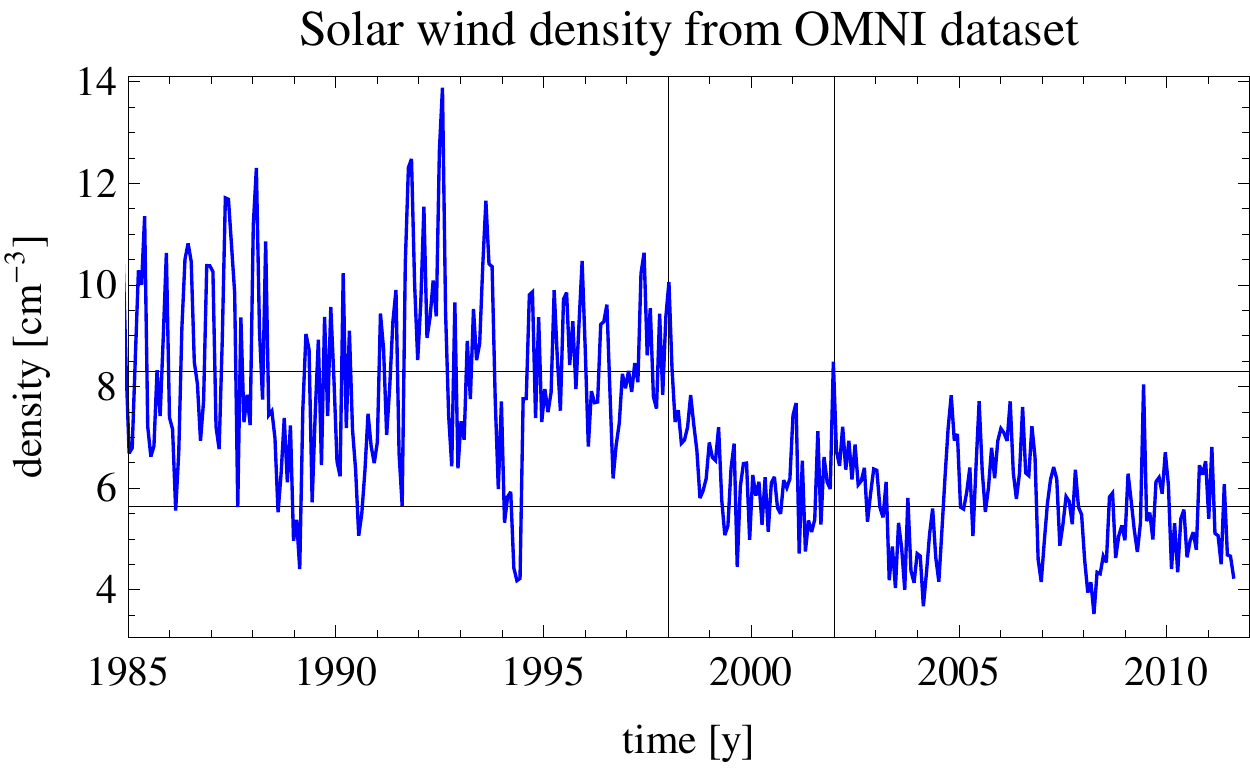}		
  	\caption{Carrington rotation averages of solar wind density in the ecliptic plane adjusted to 1~AU, calculated from the hourly averages from the OMNI-2 database, with vertical lines indicating the three phases of solar number density changes during last 20 years that are described in the text. The two horizontal lines mark the density average values calculated for the intervals 1985 - 1998 and 2002 - present.}
 		\label{figOMNIDens}
		\end{figure}

\begin{figure}
		\centering
		\includegraphics[scale=0.7]{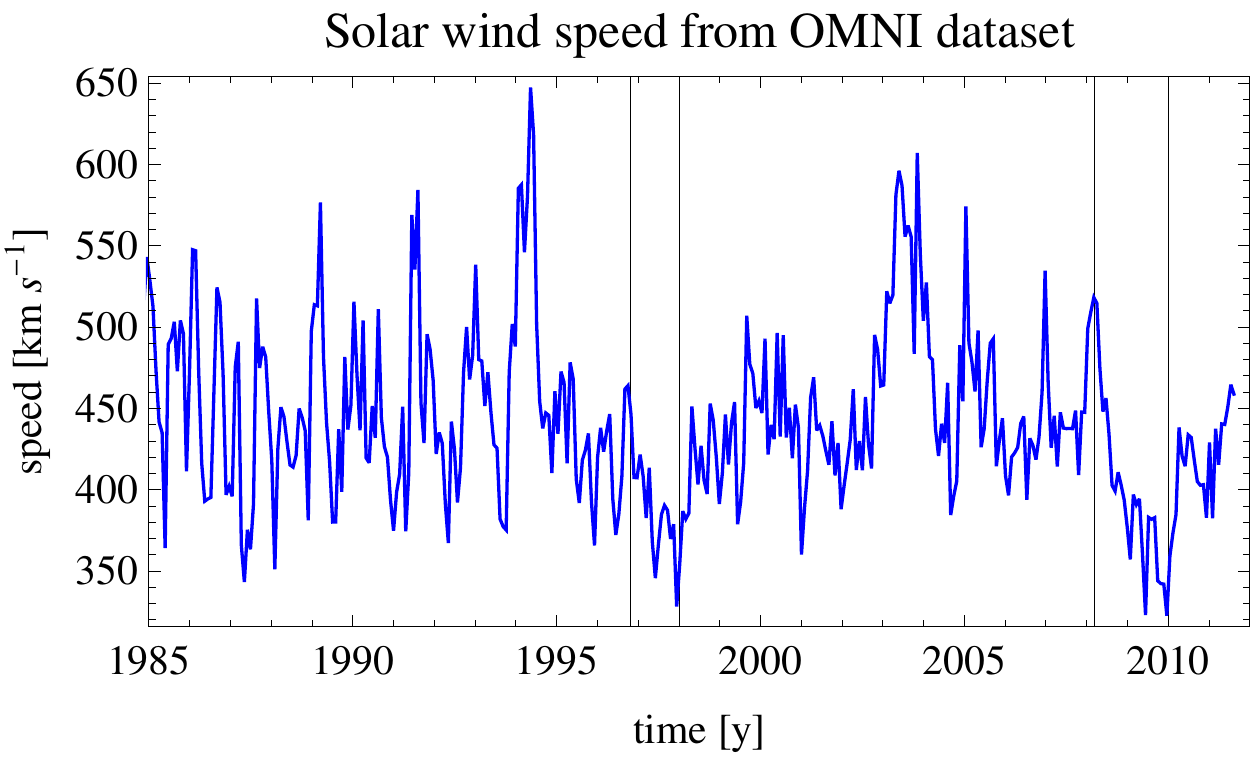}		
  	\caption{Carrington rotation averages of solar wind speed in the ecliptic plane, calculated from the hourly averages from the OMNI-2 database, with two times intervals marked by vertical lines indicating the short time intervals when the drop in speed is registered. They correspond to the minimum phase of solar cycle.}
 		\label{figOMNISpeed}
		\end{figure}

\begin{figure}
		\centering
		\includegraphics[scale=0.7]{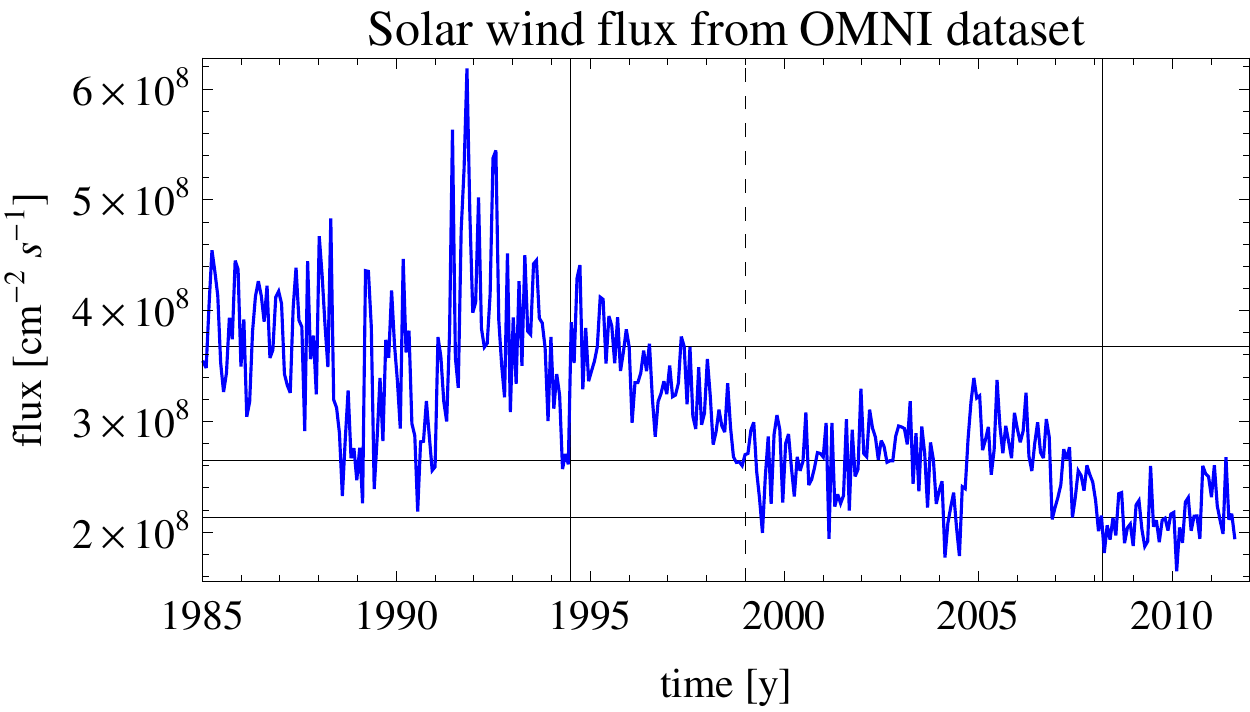}		
  	\caption{Carrington rotation averages of solar wind flux in the ecliptic plane adjusted to 1~AU, calculated from the CR-averages of in-ecliptic solar wind speed and density. The vertical lines show the time intervals when the slow long-term drop in the flux values occurs. The horizontal lines mark the average values in flux described in the text.}
 		\label{figOMNIFlux}
		\end{figure}

Here in our analysis, we average-out heliolongitude variations. We start from the hourly values of solar wind density and speed available in the OMNI-2 collection, and we construct a time series of Carrington rotation-averaged parameters of the solar wind with the grid points set precisely at halves of the CR intervals. Small deviations of the averaged times from the halves of the rotation periods are linearly interpolated. Thus, we develop an equally-spaced time series of the solar wind in-ecliptic densities adjusted to 1~AU and speeds, which are presented in Figure~\ref{figOMNIDens} and Figure~\ref{figOMNISpeed}, respectively. From these series we obtain a time series of solar wind flux calculated as the product of speed and density, shown in Figure~\ref{figOMNIFlux}.

The time interval shown in Figures~\ref{figOMNIDens}, \ref{figOMNISpeed}, \ref{figOMNIFlux} starts before the solar maximum in 1990 and includes the solar minima in 1995 and 2008, as well as the maximum in 2001. Neither the density nor speed seems to be correlated with solar activity. The density features a secular change \citep{mccomas_etal:08a}, which began just before the last solar maximum and leveled off shortly before the present minimum. The overall drop in the average solar wind density is on the order of 30\%.

During the last 20 years we can see 3 phases of the SW density changes: until 1998 the normal values phase, when the average number density was $\sim 8~\mathrm{cm}^{-3}$ (approximately equal to the average value since the beginning of observations), then between 1998 and 2002 a phase of a rapid decrease that seems to be correlated with the ascending phase of solar activity during SC~23, and finally a phase of low density, with an average number density reduced to $\sim 5.5~\mathrm{cm}^{-3}$, which lasts until present. These phases are marked with vertical lines in Figure~\ref{figOMNIDens}. The horizontal lines indicate the average values before and after the drop. 

The fluctuations of density are generally anticorrelated with speed fluctuations and within the low values phase smaller in magnitude than these in the phase before 1998. These can be associated with the persistence of coronal holes at equatorial latitudes, as convincingly illustrated by \citet{deToma:11a}. 

The changes in solar wind flux adjusted to 1~AU are the most pronounced of all discussed parameters. The steady and slow decrease that started after 1995 seems to continue to the present despite an 8-year plateau between 1999 and 2007 (see Figure~\ref{figOMNIFlux}). The changes in flux are of the order of $40\%$ from the average calculated from the data interval between 1985 and 1995 to the average calculated from the interval 2008.2 -- 2011.5. 

The three phases in the solar wind flux corresponding to the phases pointed out in the discussion of the density variations have different time boundaries; they are marked with horizontal and vertical lines in Figure~\ref{figOMNIFlux} in analogy with Figure~\ref{figOMNIDens}. The speed shows multi-timescale variations, but its average value seems to be basically constant in time apart from the last 2-3 years, when a small drop of $\sim 3\%$ is observed \citep{mccomas_etal:08a}. In Figure~\ref{figOMNISpeed} we indicate only the two time intervals with a small decrease in SW speed that are common in time with the lowest part of solar activity in 1997 and 2008.

\subsection{\textit{in~situ} out-of-ecliptic \textit{Ulysses} measurements}
\textit{Ulysses} was launched in October 1990 and has been the first and only spacecraft that traversed regions of the heliosphere close to solar poles and provided unique samples of solar wind. Its orbit was nearly polar, with aphelion of $\sim 5.5$~AU and perihelion $\sim 1.4$~AU, in a plane almost perpendicular to ecliptic and solar equator. The period of the orbit was about 6 years, with the so-called fast scan (from south to north pole through perihelion), which lasted $\sim1$~year and the slow scan (from north to south pole through aphelion), which lasted $\sim5$~years. 

\begin{figure}[t]
\centering
\includegraphics[scale=0.65]{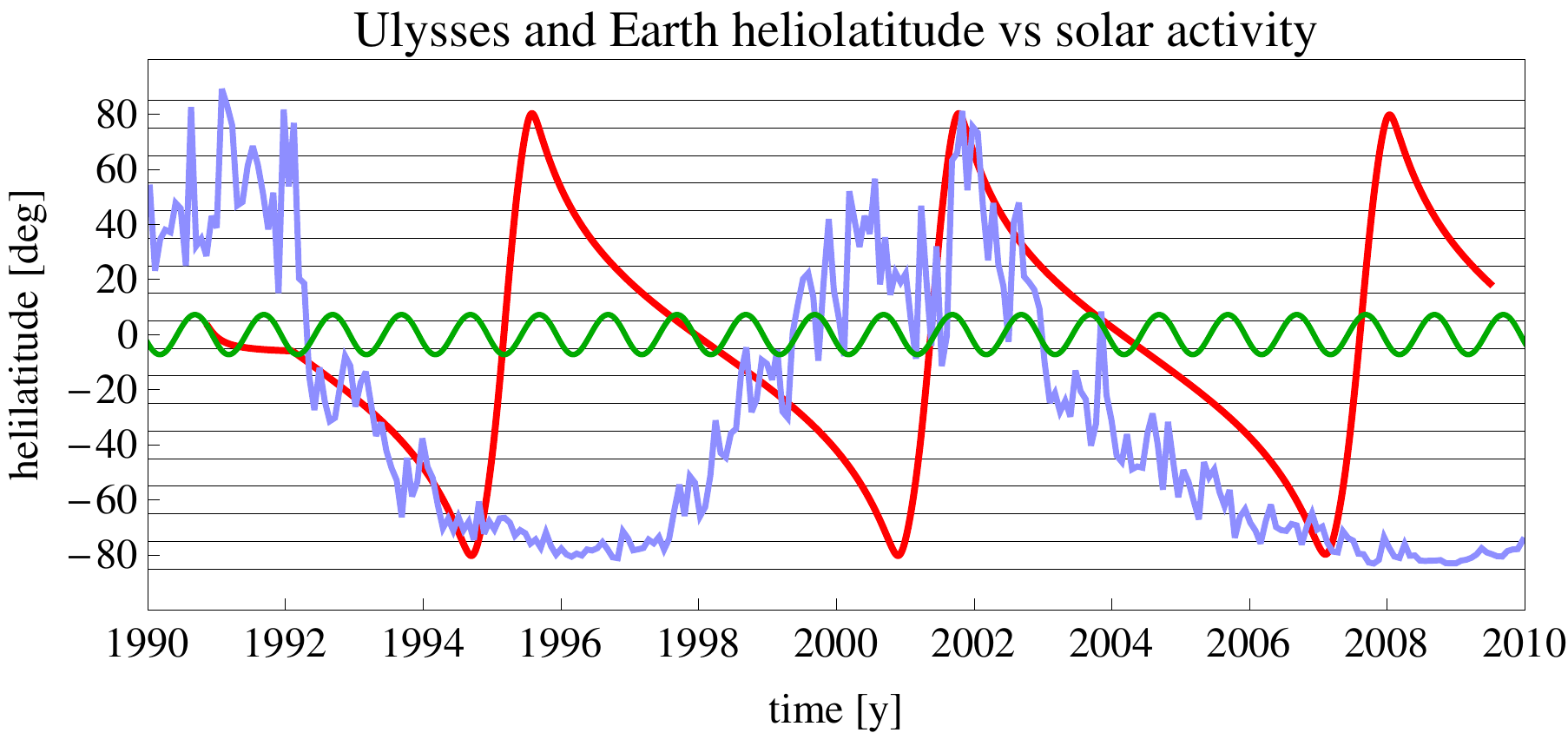}
\caption{Illustration of heliolatitude position of \textit{Ulysses} (red) and Earth (green) during the time span of \textit{Ulysses} mission (November 1990 - June 2009). The pale blue line is the F$_{10.7}$ solar radio flux \citep{covington:69, tapping:87}, superimposed to correlate variations in solar activity with \textit{Ulysses} heliolatitude during its more than 3 polar orbits.}
\label{figUlyssesComposite}
\end{figure}

\begin{figure}[t]
\centering
\includegraphics[scale=0.6]{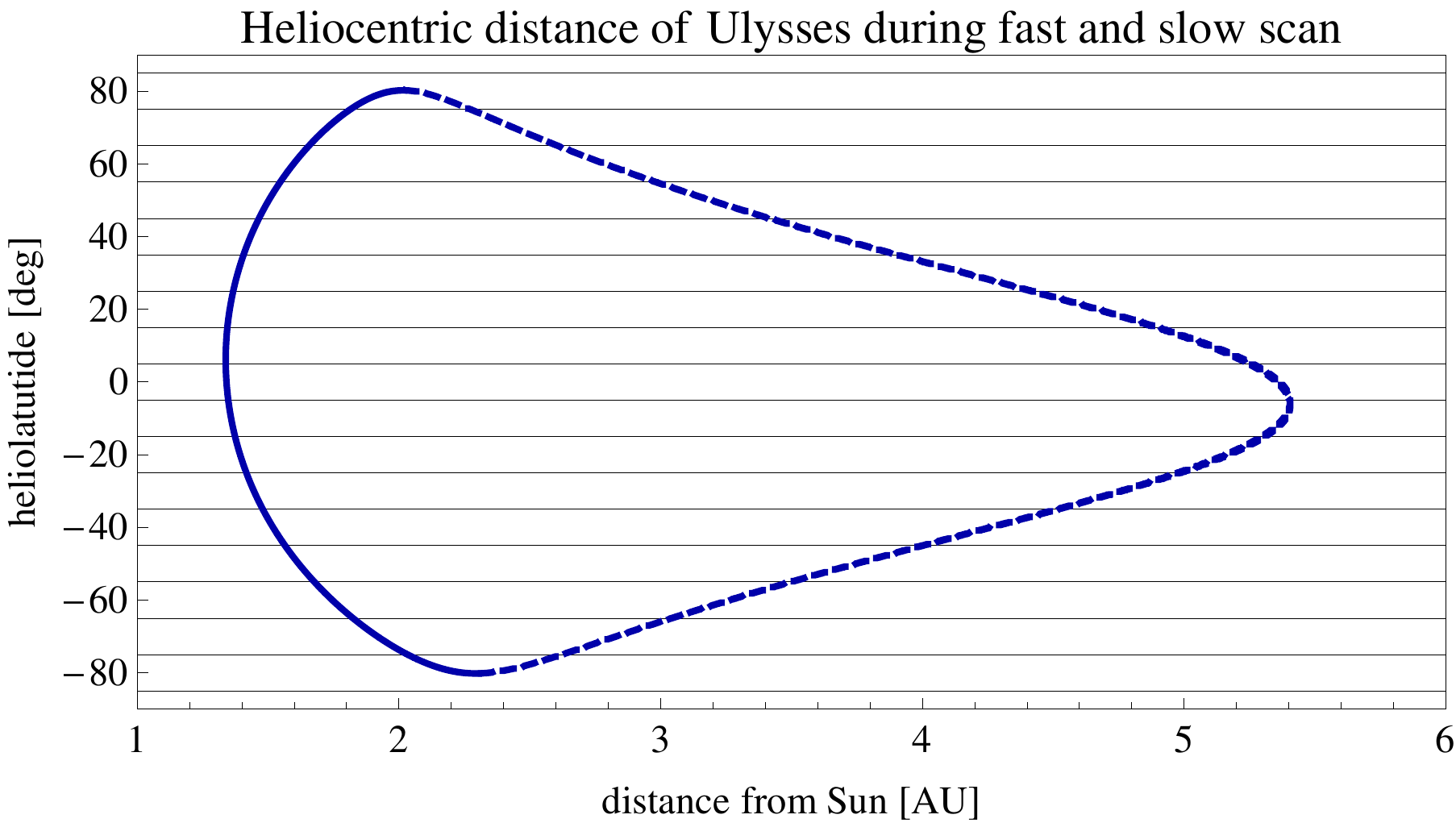}
\caption{\textit{Ulysses} heliocentric distance during one of its polar orbits. The solid line corresponds to the fast part of the scan, and the dashed line to the slow part of the scan. The horizontal lines mark the 10-degree heliolatitude bins used in the analysis.}
\label{figUlyRFast}
\end{figure}

The heliolatitude track of \textit{Ulysses} is shown in Figure~\ref{figUlyssesComposite}, with superimposed Earth heliolatitude position and a solar activity graph represented by the F$_{10.7}$~cm flux \citep{covington:69, tapping:87, svalgaard_hudson:10a}. The heliocentric distance of \textit{Ulysses} for one complete orbit is shown in Figure~\ref{figUlyRFast}. The spacecraft was launched during the solar maximum conditions and during its 20~years life it observed the Sun during the whole solar cycle and further until the last prolonged solar minimum.

The discoveries and findings from the plasma measurements by the \textit{Solar Wind Observations Over Poles of the Sun}~(SWOOPS) experiment on \textit{Ulysses} \citep{bame_etal:92a} can be found in many papers \citep[see e.g.][]{phillips_etal:95c, marsden_smith:97, mccomas_etal:00b}. They include the bimodal structure of solar wind \citep{mccomas_etal:98b}, which is fast and uniform at mid- and high-heliolatitudes and more variable, slower and denser at lower heliolatitudes during low activity of the solar cycle. During maximum of solar activity the solar wind is highly variable at all heliolatitudes, with flows of slow and fast wind interleaved \citep{mccomas_etal:03a}. 

Detailed studies of the fast solar wind parameters presented by \citet{mccomas_etal:00b} revealed latitudinal gradients in proton speed and density of flows from polar coronal holes. These gradients are equal to $\sim 1~ \mathrm{km}~\mathrm{s}^{-1} \mathrm{deg}^{-1}$ and $\sim -0.02 \mathrm{cm}^{-3} \mathrm{deg}^{-1}$, respectively. \textit{Ulysses} measurements also confirm the existence of the secular changes in SW parameters reported by in-ecliptic spacecraft. \citet{mccomas_etal:08a} showed that the global solar wind exhibits a slight reduction in speed ($\sim 3\%$), but a much greater one in proton density ($\sim 17\%$) and dynamic pressure ($\sim 22\%$). This result was demonstrated simultaneously at low and high latitudes, allowing these authors to conclude that the variations were truly global. It was also recorded that the band of the slow SW variability extended to a higher latitude during the last \textit{Ulysses} orbit \citep[see][]{mccomas_etal:06a} than during the first one. \citet{ebert_etal:09a} reported that above $\pm 36^\circ$ heliolatitude the spatial and radial variability in SW parameters remains consistent and relatively small, indicating that the fast solar wind plasma flows from the polar coronal holes are steady and uniform.

\begin{figure}[ht]
		\centering
		\includegraphics[scale=0.7]{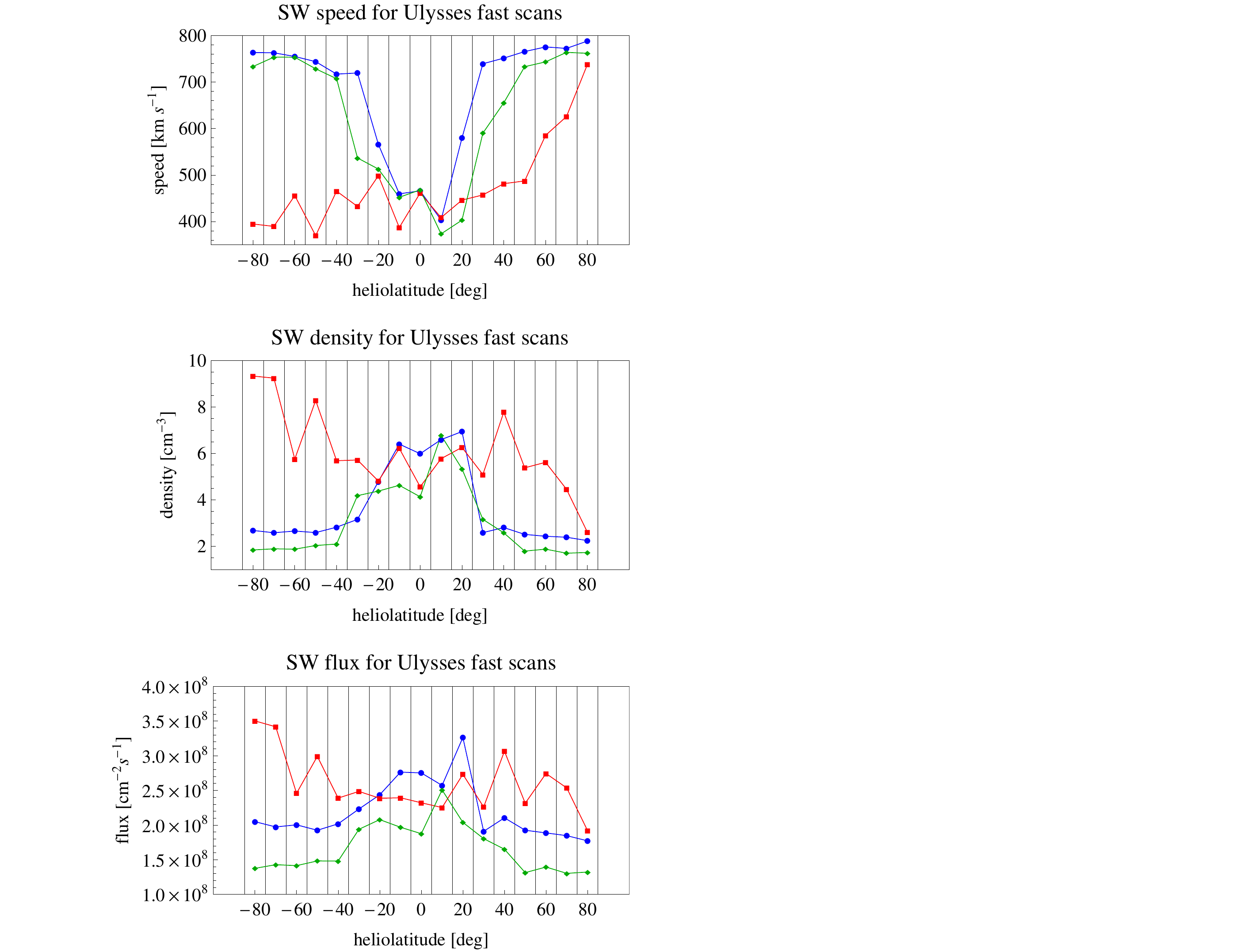}		
  	\caption{Solar wind speed (upper panel), adjusted density (middle panel), and adjusted flux (lower panel) as a function of heliolatitude for the first (blue), second (red) and third (green) \textit{Ulysses} fast heliolatitude scan. The parameters are averaged over the 10-degree heliolatitude bins marked with vertical lines.}
 		\label{figUlyDensSpeedFlux}
		\end{figure}

\textit{Ulysses} found that the heliolatitude structure of the SW during the two solar minima was largely similar (see Figure~\ref{figUlyDensSpeedFlux}), featuring an equatorial enhancement in density with the associated reduction in velocity (the slow wind region), and that during solar maximum the slow wind and fast wind from small coronal holes exist at all heliolatitudes (see also middle panel in Figure~\ref{figUlysses2IPSfast}). However, the region of slow wind seems to reach higher in heliolatitude during the last solar minimum than during the minimum of 1995, which is much less conspicuous in density (see Figure~\ref{figUlyDensSpeedFlux}). 
		
Here we study the latitudinal variation of solar wind parameters based mostly on the fast latitude scans to avoid possible convolution of heliolatitude, time and heliocentric distance effects. We leave the data from the slow scans, which covered almost 5 years each (i.e. almost a half of solar cycle) for verification of the derived model. Two of the 3 fast scans (the first and the third one) occurred during the descending phase of solar activity and the middle one during the solar maximum conditions (see Figure~\ref{figUlyssesComposite}). During the fast scans \textit{Ulysses} was close to the Sun (see Figure~\ref{figUlyRFast}), so possible distance related effects in the solar wind are small, in contrast to the slow scans, when distance related effect may be significant \citep{ebert_etal:09a}.

The evolution of solar wind speed and adjusted density and flux obtained by \textit{Ulysses} during the fast latitude scans are compiled in Figure~\ref{figUlyDensSpeedFlux}, where the parameter values are averaged over 10-degree bins in heliolatitude.

		\clearpage
		
\subsection{Remote-sensing observations of interplanetary scintillations}
\begin{figure}
\centering
\includegraphics[scale=0.6]{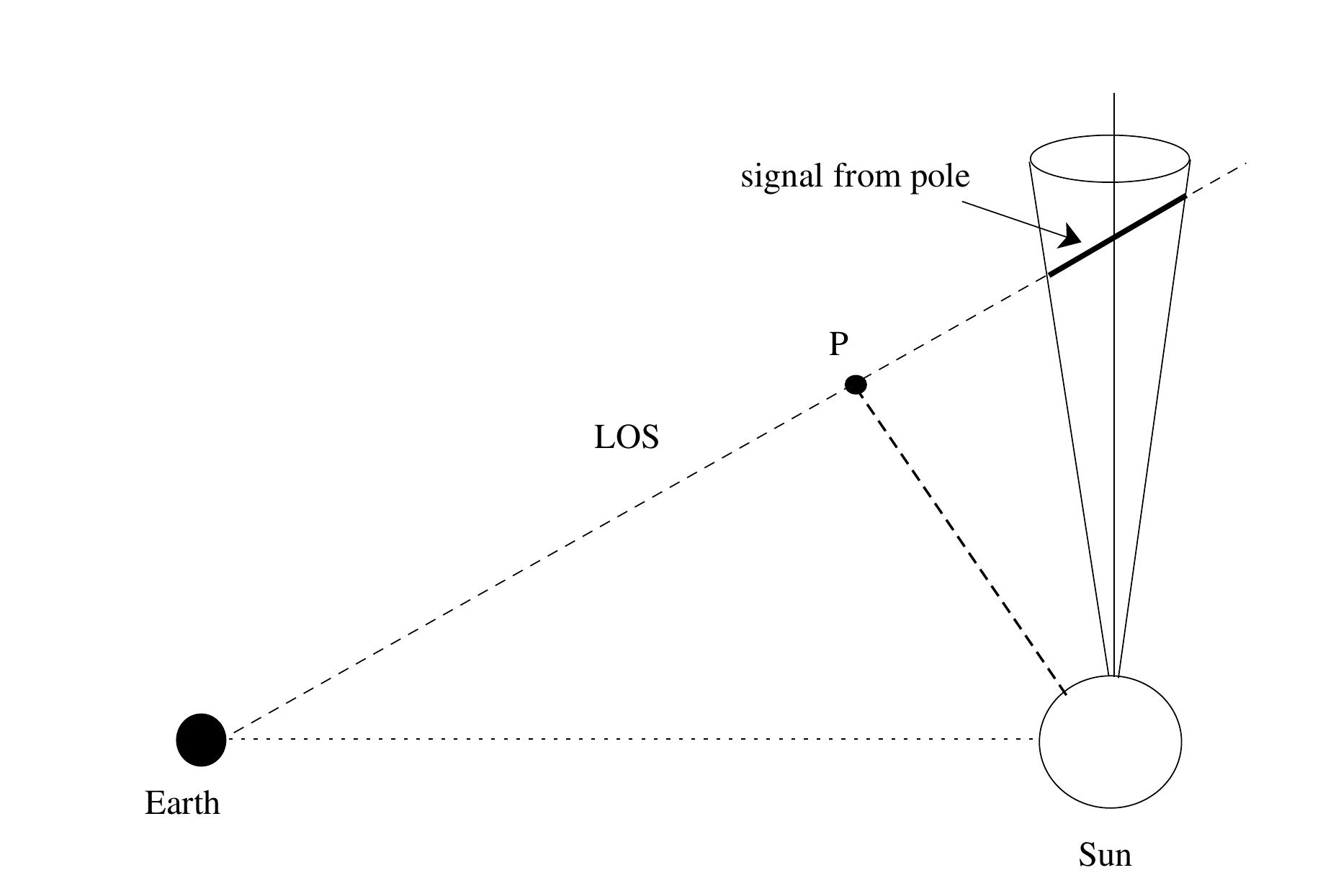}
\caption{Illustration of the geometry of the line of sight in remote-sensing observation of a polar region in the solar wind. The observer is close to the Earth in the ecliptic plane and aims its radio-telescope antenna at a target so that the LOS crosses a cone with a small opening angle centered at the pole. The signal is collected from the full length of the LOS, but the contributions from various parts are different \citep[for details of weighting function see][]{kojima_kakinuma:87a}. In the case of IPS observations, the strongest contribution to the signal is from the point nearest to the Sun along the LOS, marked with P, because the source function of the scintillation signal drops down with the square of solar distance. Note that the angular area of the polar region is quite small, so with an observations program that maps the entire sky only a small region in the map indeed includes the signal from polar regions.}
\label{figLOSScheme}
\end{figure}

Interplanetary scintillation is a phenomenon of producing diffraction patterns on an observer's plane by the interferencing radio waves from a remote compact radio-source (like quasar) that are scattered by electron density irregularities (fluctuations) in the solar wind \citep[see, e.g.,][]{hewish_etal:64a, coles_kaufman:78a, kojima_kakinuma:90a, kojima_etal:98,  kojima_etal:07a}. The scintillation signal is a sum of waves scattered along the line of sight (LOS) to the observed radio-source. Most of the scattering occurs at the closest distances to the Sun (the so-called P~point, see Figure~\ref{figLOSScheme}) along the LOS. This is because the absolute magnitude of the electron density fluctuations, which are approximately proportional to electron absolute density, rapidly decreases with solar distance \citep{coles_maagoe:72a}. The IPS observations are LOS integrated and to provide reliable information on the SW speed they have to be deconvolved. 

The magnitude of electron density fluctuations at a given solar distance is correlated with the magnitude of local solar wind speed. Thus, inferring the level of electron density fluctuations from IPS observations one can estimate the local speed of solar wind \citep{hewish_etal:64a, jackson_etal:97a, kojima_etal:98, jackson_etal:03a}. However, one needs a formula that links the electron density fluctuations $\delta n_e$ with solar wind speed $v$. Usually, a relation $\delta n_e \propto v^{\gamma}$ is used. The index $\gamma$ has to be established empirically. Such an analysis for various solar wind conditions was performed by \citet{asai_etal:98a}. 

Observations performed using a system of several radio antennas that are longitudinally distributed on the Earth allows to infer a very detailed information on the structure of solar wind both inside and outside the ecliptic plane \citep{coles_rickett:76a, tokumaru_etal:10a}. To that purpose best suitable seems the Computer Assisted Tomography technique \citep{jackson_etal:98a, kojima_etal:98, hick_jackson:01a, jackson_etal:03a, kojima_etal:07a}.

The accuracy of the LOS deconvolution depends on (1) geometrical considerations (e.g., the geographical latitude of the telescopes, the tilt of the ecliptic to the solar equator), (2) the number of observations available, and (3) the fidelity of the correlation formula between solar wind electron density fluctuations and speed. While correlating these quantities was possible early for the equatorial solar wind \citep{harmon:75a}, the out of ecliptic IPS measurements could only be calibrated once \textit{in~situ} data from \textit{Ulysses} became available \citep{kojima_etal:01}. Such a calibration should be repeated separately for each solar cycle because solar wind features secular changes, as discussed earlier in this paper. Still, even before the introduction of the CAT technique, the IPS observations suggested \citep[e.g.,][]{kojima_kakinuma:87a} that the solar wind structure varies with solar activity, with high speeds in the polar regions during low solar activity and the slow wind expanding to polar regions when the activity is high. This lends credibility to the IPS technique as a source of information on the global heliolatitude structure of solar wind speed.

An extensive program of solar wind IPS observations, initiated in 1980s in the Solar-Terrestrial Environmental Laboratory (STEL) at the Nagoya University in Japan \citep{kojima_kakinuma:90a}, resulted in a homogeneous data set that spans almost three solar cycles. This data set enables detailed studies of the evolution of solar wind speed profile with changes in solar activity \citep{kojima_kakinuma:87a, kojima_etal:99a, kojima_etal:01, fujiki_etal:03a, fujiki_etal:03b, fujiki_etal:03c, kojima_etal:07a, tokumaru_etal:09a, tokumaru_etal:10a}. 

In the analysis we used data from 1990 to 2011, obtained from the tomography processing of observations from 3 antennas (Toyokawa, Fuji and Sugadaira), and in addition from another antenna (Kiso) since 1994. The 4-antenna system was operated until 2005, when the Toyokawa antenna was closed \citep{tokumaru_etal:10a}; since then the system again was operated in a 3-antenna setup. We did not use the data collected in 2010, because the number of observations available was too small to obtain a reliable result on solar wind structure from the CAT analysis. Since 2011 the STEL IPS observations are again regular with the updated multi-station system.

The IPS data from STEL are typically collected on a daily basis during $\sim 11$ Carrington rotations per year: there is a break in winter because the antennas get covered with snow. Each day, 30-40 LOSs for selected scintillating radio-sources are observed. The lines of sight are projected on the source surface at 2.5 solar radii, which is used as a reference surface in the time-sequence tomography. 

The latitude coverage of the sky by the time-sequence IPS observations is not uniform and strongly correlated with the Sun position on sky, which changes during the year, and with the target distribution on sky. Relatively few of them are located near solar poles, because the polar regions are only a small portion of the sky (see Figure~\ref{figLOSScheme}). Additionally, the observations of the south pole are of lower quality than these of the north pole because of the low elevation of the Sun during winter in Japan. The original latitude coverage was improved owing to the new antenna added to the system in 1994 and by optimization of the choice of the targets.

Thus, the accuracy of the remote-sensing measurements of solar wind speed decreases with latitude because of geometry. The polar values are the most uncertain (and possibly biased) because the signal in the polar lines of sight is only partly formed in the actual polar region of space, which can be understood from the sketch presented in Figure~\ref{figLOSScheme}. 

\subsection{Comparison of IPS solar wind speed profiles with \textit{Ulysses} data}		

		\begin{figure}[ht]
		\centering
		\begin{tabular}{ccc}		
\includegraphics[scale=0.6]{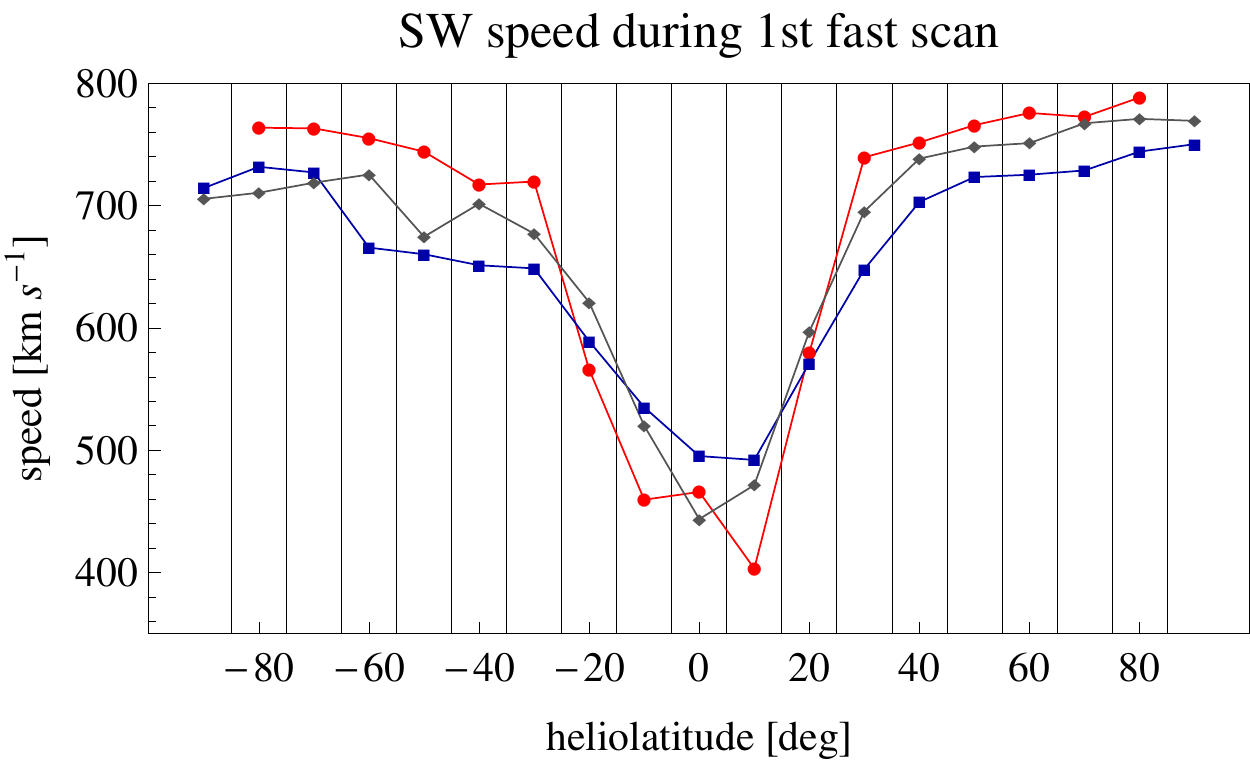}\\
\includegraphics[scale=0.6]{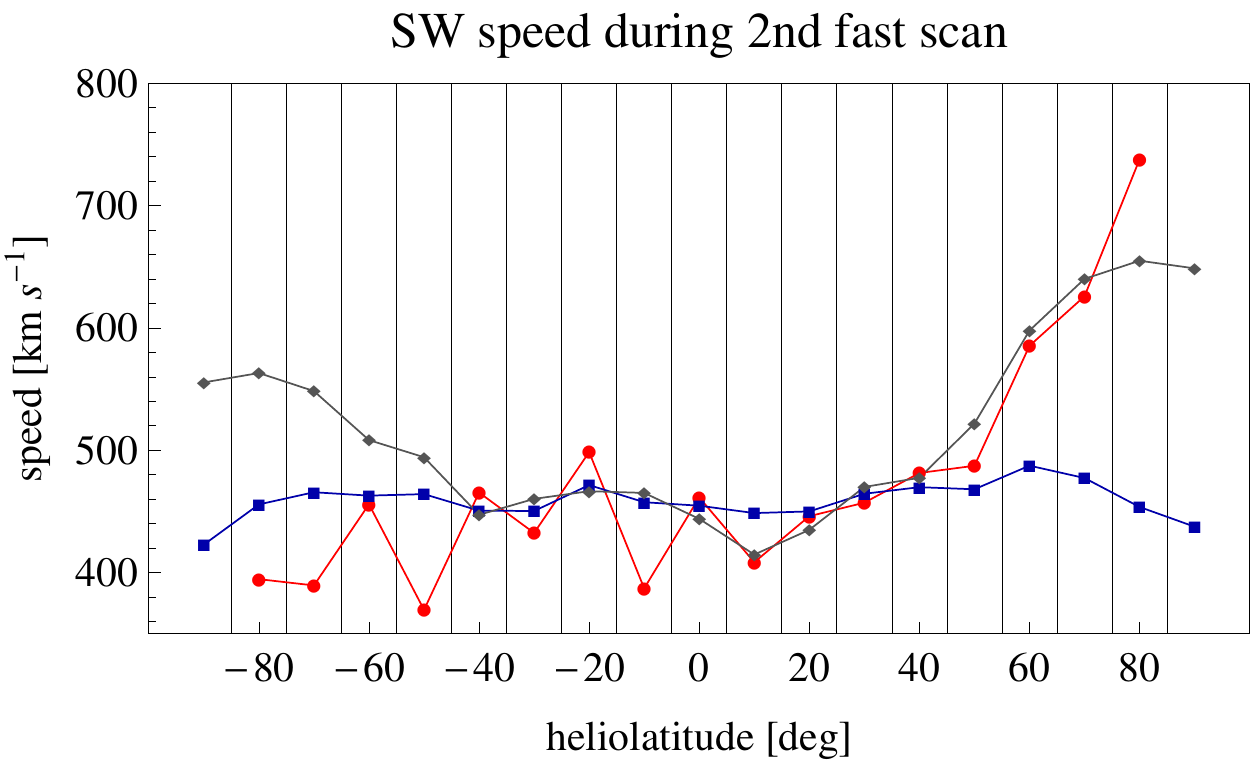}\\
\includegraphics[scale=0.6]{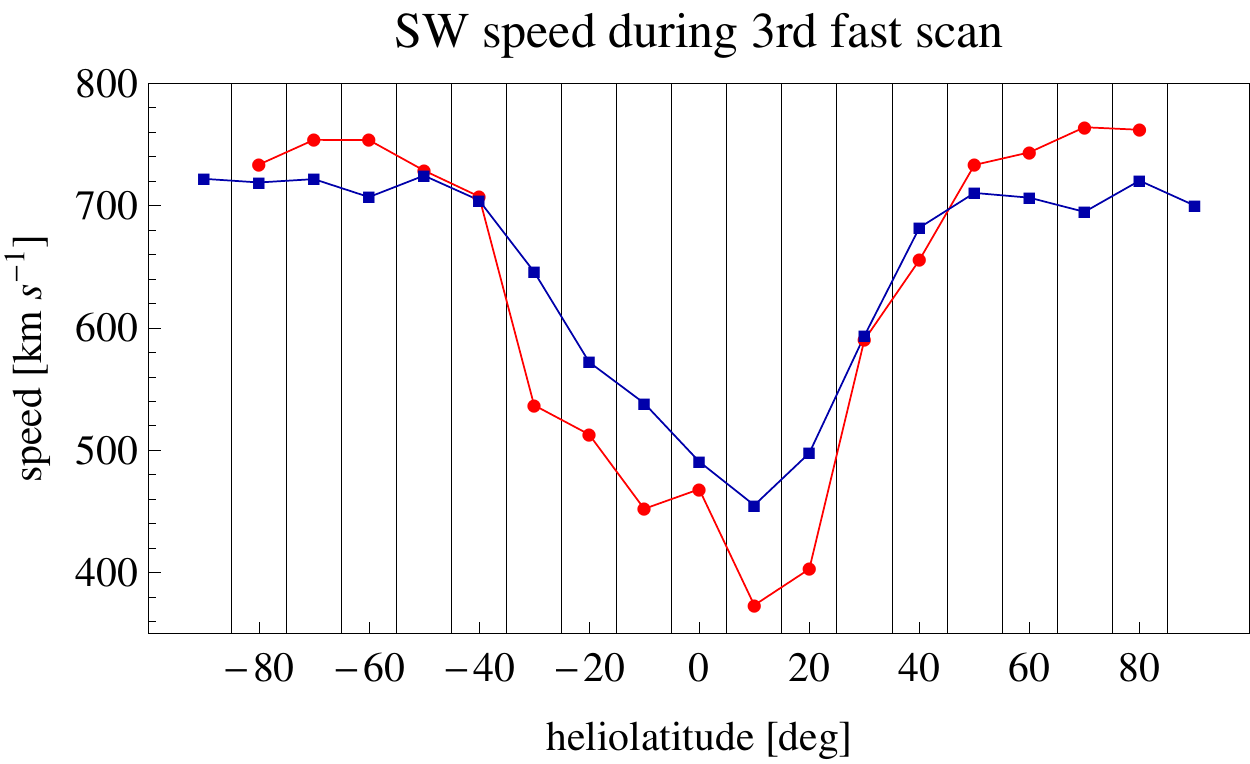}\\	
		\end{tabular}
		\caption{Solar wind speed profiles from \textit{Ulysses} measurements and IPS observations for the 3 fast heliolatitude scans. Red: \textit{Ulysses}, blue: IPS during the year of beginning of \textit{Ulysses} 1st and 2nd fast scan and the whole 3rd scan, gray: IPS during the year of the end of \textit{Ulysses} 1st and 2nd fast scan. Top panel: the first fast scan during solar minimum, middle panel: the second fast scan during solar maximum, bottom panel: the third fast scan during minimum.}
		\label{figUlysses2IPSfast}
		\end{figure}

To verify the results obtained from the IPS CAT analysis, we compared them with the data from the three \textit{Ulysses} fast latitude scans. The \textit{Ulysses} speed profiles used for this comparison were constructed from subsets of hourly averages available from the National Space Science Data Center of NASA (NSSDC), split into identical 10-degree heliolatitude bins and averaged. They are shown in Figure~\ref{figUlysses2IPSfast} in red. Since the acquisition of the \textit{Ulysses} profiles took one year each and the first and second scan straddle the break of calendar year, we show the IPS results for the years straddling the fast latitude scans; they are presented in blue and gray in Figure~\ref{figUlysses2IPSfast}. 

\begin{figure}
\centering
\includegraphics[scale=0.6]{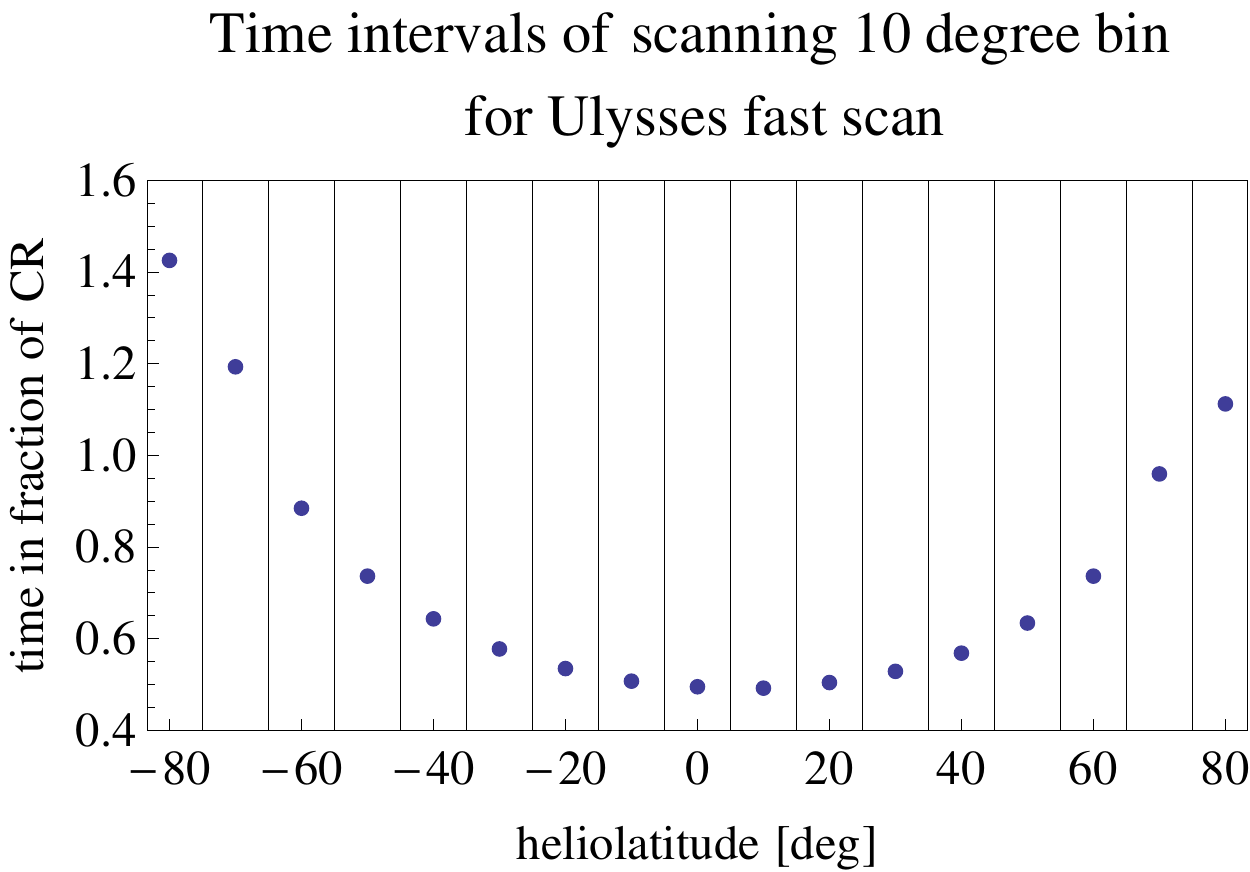}
\caption{Time intervals of scanning the 10 degree bins by \textit{Ulysses} during fast scans.}
\label{figUlyTimeFast}
\end{figure}

Generally the IPS and \textit{Ulysses} profiles agree quite well. The sawtooth feature in the \textit{Ulysses} profiles from the second fast scan and in the equatorial part of the first and third fast scans is due to the short time \textit{Ulysses} was sampling the 10-degree bins. The fast scans were performed within the perihelion part of the \textit{Ulysses} elliptical orbit, with the perihelion close to the solar equator plane. Hence, the angular speed of its motion was highest close to equator and traversing the 10-degree bin took it less than one solar rotation period (see Figure~\ref{figUlyTimeFast}). Thus the sawtooth is an effect of incomplete Carrington longitude coverage of the bimodal solar wind by \textit{Ulysses}, with slow wind interleaved with fast wind streams. 

Near the poles the angular speed was slower and it took more than 1~CR to scan the 10-degree bin. Thus, when the slow wind engulfed the whole space, the sawtooth effect expanded into the full heliolatitude span. By contrast, during the low-activity scans the solar wind speed at high latitude was stable, which resulted in the lack of the small-scale latitude variations in the CR-averages at high latitudes, despite the uneven heliolongitude coverage (see Figure~\ref{figUlyTimeFast}).

The IPS yearly averages do not show the short scale latitudinal variability of the solar wind speed seen in the \textit{Ulysses} data (the sawtooth feature) presented in Figure~\ref{figUlysses2IPSfast}, because this variability was smoothed by averaging over multiple full Carrington rotations, with the alternating fast/slow streams averaged out.

The difference between the blue and gray lines in the top and middle panels of Figure~\ref{figUlysses2IPSfast} is a measure of true variation of the latitudinal profile of solar wind speed during one year. \textit{Ulysses} was going from south to north during the fast latitude scans (see Figure~\ref{figUlyssesComposite}), so the south limb of the profile from \textit{Ulysses} ought to be closer to the southern limb of the blue profile obtained from the IPS analysis, while the north limb of the \textit{Ulysses} profile should agree better with the north limb of the gray IPS profile. And indeed, an almost perfect agreement is observed in the second panel, corresponding to solar maximum. In our opinion, this is a very interesting observation because it (1) shows how rapidly the latitude structure of solar wind varies during the maximum of solar activity, (2) confirms the credibility of IPS results for solar wind structure, and (3) confirms that interplanetary scintillation observations deconvolved by CAT algorithms reconstruct the SW structure in a very good agreement with \textit{in~situ} measurements. 

Originally, the interpretation of the speed profiles obtained from \textit{Ulysses} was not clear, it was pondered whether the north-hemisphere increase in solar wind speed was a long-standing feature of the solar wind or was it just a time-variability of the wind at the north pole. Similarly, the question arose whether IPS was able to credibly reproduce the solar wind profiles given the fact that some of the profiles obtained approximately at the time of the fast scan seemed to disagree with the \textit{in~situ} data (especially the first fast scan). The challenging question was which data were more reliable for the yearly profile: those from \textit{Ulysses} that measured features only from the point of solar surface it came from, or the yearly profiles from IPS observations that reveal the full-surface variations in time? 

The \textit{Ulysses} data are challenging to interpret in this respect. On one hand the fast scans give information on the solar structure from the south to north pole almost in one year, but the scans are so fast that the longitudinal variation is convolved with the latitudinal structure (see Figure~\ref{figUlyTimeFast}) and on the other hand, the slow scans take so long (nearly half of solar cycle) that the parameters obtained for different heliolatitude bands may differ not only because of a heliolatitude structuring of solar wind, but also because of possible time variations of this structure. Without an independent insight one is unable to separate between the two. 

Furthermore, the \textit{Ulysses} heliocentric distances during the fast and slow scans are different (see Figure~\ref{figUlyRFast}). During the fast scans the distance does not change very much (from 2.2~AU close at the poles to 1.4~AU in the ecliptic), so possible distance related effects are not pronounced, but during the slow scans, when the distance changes are much greater, one can not deconvolve the variation connected with the distance and those related to heliolatitude and/or time without a credible MHD model of the solar wind behavior between the Sun and the spacecraft. The IPS data provide us with the much wanted information on all heliolatitudes simultaneously over one year.  

\begin{figure}[t]
\centering
\includegraphics[scale=0.6]{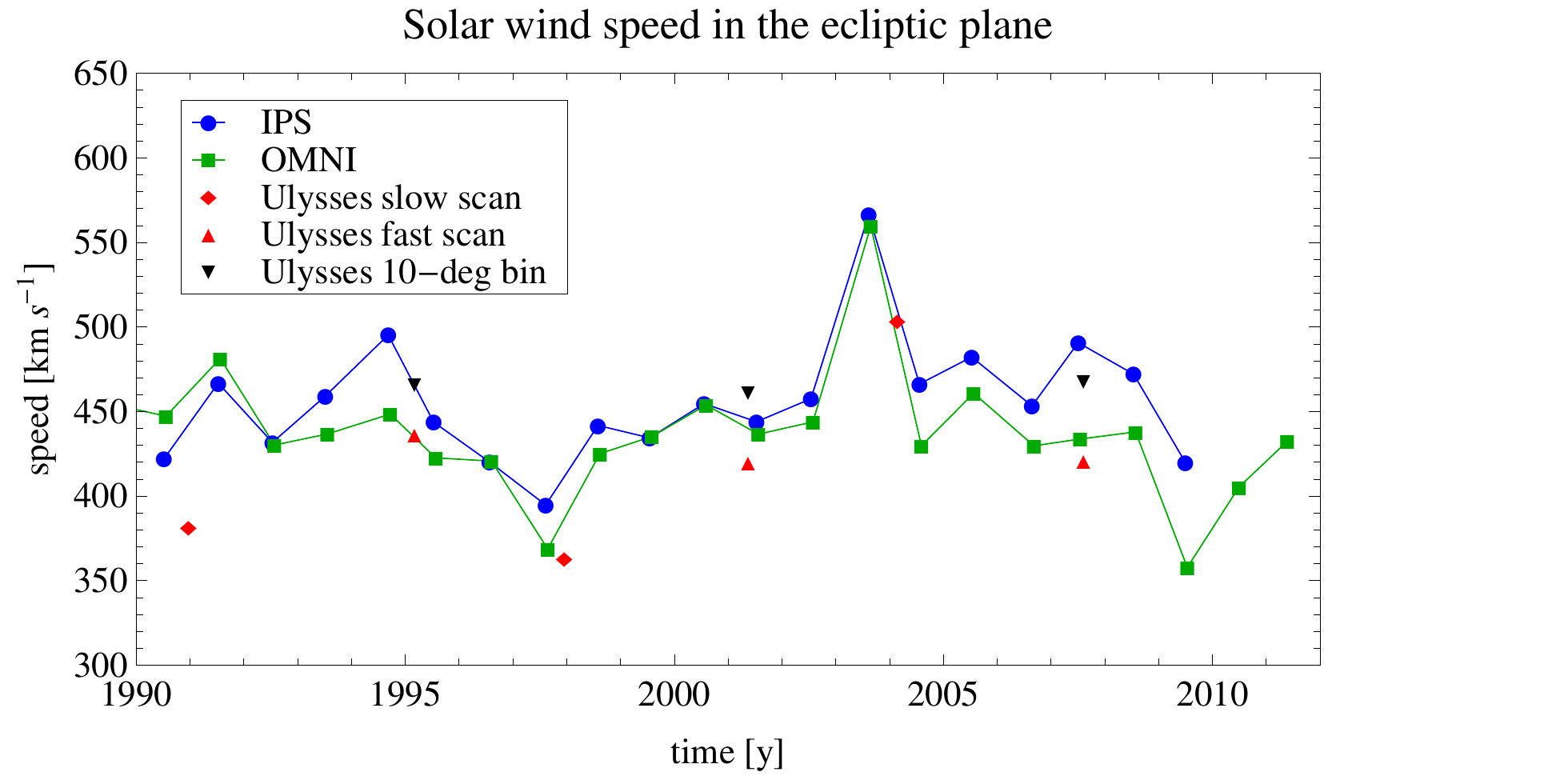}
\caption{Comparison of yearly solar wind speed values in the ecliptic plane from IPS and OMNI-2 database with \textit{Ulysses} fast scan data from the 10-degree bin around solar equator (black triangles; it is the $0^\circ$ bin value from the 10-degree profiles) and Carrington rotation average around the $0^\circ$ latitude for fast scans (red triangles) and 2 Carrington rotation average for slow scans (red diamonds).}
\label{figEclipticIPSOMNI}
\end{figure}

\begin{figure}[ht]
\centering
\includegraphics[scale=0.5]{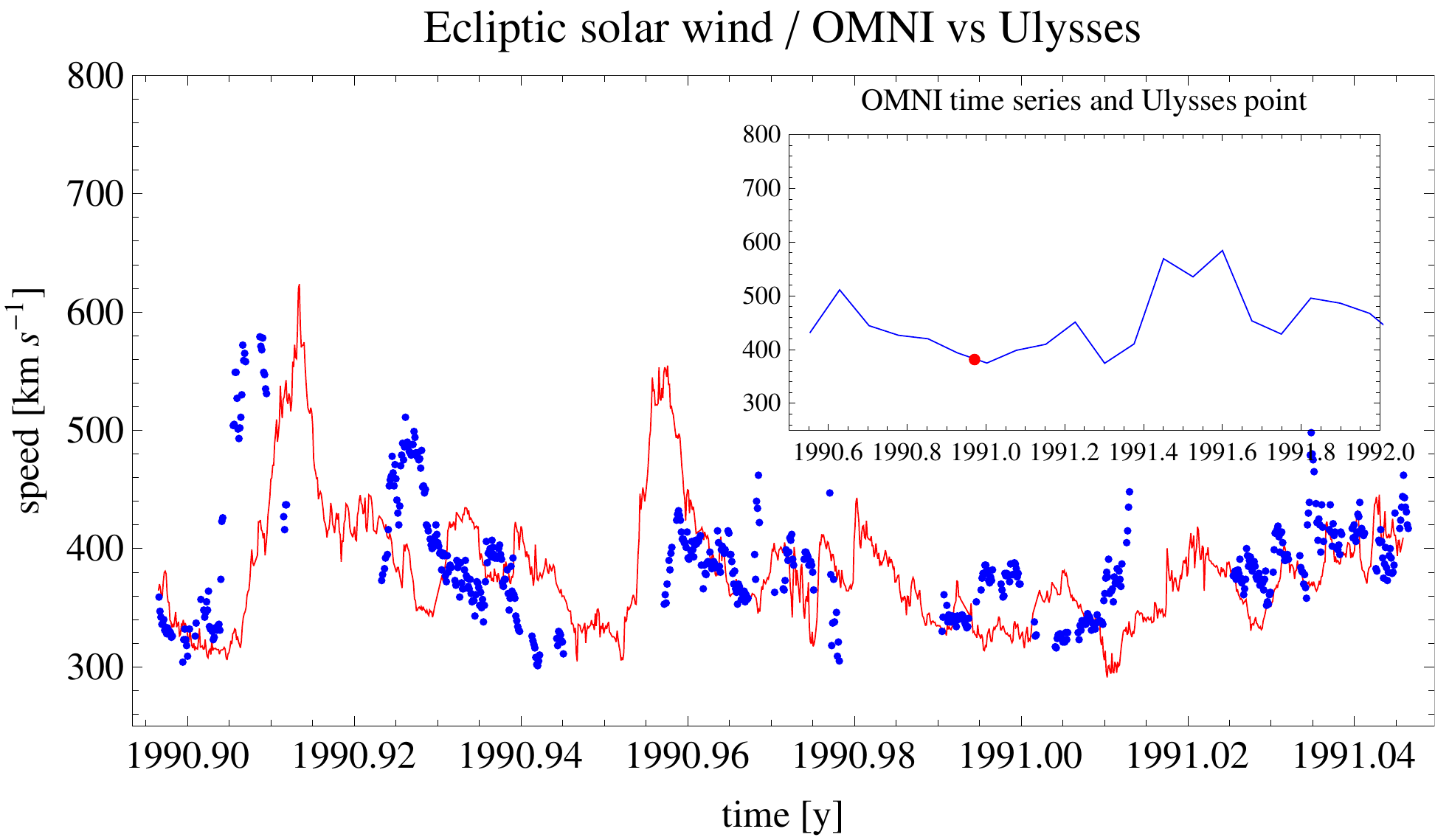}
\caption{Comparison of hourly averages of solar wind speed from the OMNI-2 collection (dots) and \textit{Ulysses} measurements (solid line) obtained during two Carrington rotations just after its launch. The inset presents a time series of CR-averages of solar wind speed from the OMNI-2 collection to illustrate that the average from the Carrington rotation in question happened to be the minimum over a longer time interval.}
\label{figUly1slowEcl}
\end{figure}

The IPS data are in a very good agreement with OMNI-2 in the ecliptic (see the blue and green lines of the corresponding yearly averages in Figure~\ref{figEclipticIPSOMNI}). Also the agreement between the \textit{Ulysses} measurements obtained from the two Carrington rotations during the passages of $0^\circ$ heliolatitude at aphelion and the OMNI-2 data is almost perfect (cf red diamonds in Figure~\ref{figEclipticIPSOMNI}), which is clearly seen for the passages in 1998 and 2004. The apparent discrepancy in 1991 is a result of a fluctuation, as explained in Figure~\ref{figUly1slowEcl}, which presents the \textit{Ulysses} hourly time series compared with the corresponding hourly time series from OMNI-2 for the 2~CRs when \textit{Ulysses} was in the ecliptic plane. The two time series are in excellent agreement and the big difference between the \textit{Ulysses} CR-average and the yearly-average from the OMNI-2 data exists because the CR-averaged speed during this Carrington rotation in question was minimum within the 12 rotations included in the yearly average, as illustrated in the inset in Figure~\ref{figUly1slowEcl}.

The agreement between the \textit{Ulysses} measurements and the OMNI-2 data set during the fast-scan passages is not straightforward to estimate because \textit{Ulysses} was passing the ecliptic latitude bins in a time shorter than a half of the Carrington rotation (see Figure~\ref{figUlyTimeFast}) and at heliolongitudes different from Earth's. In this case averaging over a full Carrington rotation (shown by red triangles in Figure~\ref{figEclipticIPSOMNI}) averages out the heliolongitude variation of the solar wind, but likely biases the result due to a possible presence of heliolatitudinal gradients. On the other hand, averaging over a 10-degree heliolatitudinal bin around the solar equator (black triangles in Figure~\ref{figEclipticIPSOMNI}) mostly eliminates the latitudinal gradients, but does not average out the heliolongitudinal modulation. In reality, however, the difference between the two averages turned out to be small, similar in magnitude to the differences between the yearly averages of the OMNI-2 and IPS solar wind speed time series. 

Summing up this section we conclude that the IPS solar wind speed profiles provide a reliable insight into the solar wind structure and its evolution with solar activity phase. They agree both with OMNI-2 in ecliptic and with \textit{Ulysses} out of ecliptic for the time intervals when they can be compared directly.

\section{Data processing and model description}
Our goal is to retrieve the solar wind speed and density structure and its evolution in time and heliolatitude, beginning from the maximum of SC~22, through the minimum  and maximum of SC~23 (in 1996 and 2001, respectively) until the most recent prolonged minimum. We use all relevant and complementary data sets both in and out of the ecliptic plane, to construct a homogeneous set of solar wind parameters, which can be further used for calculation of ionization losses of ENAs, modeling of heliosphere and global interpretation of the Lyman-$\alpha$ helioglow etc.

Our procedure takes as baseline absolute calibration of the OMNI-2 data set both in speed and density in the ecliptic plane and the absolute calibration of \textit{Ulysses} measurements for density and speed out of ecliptic and interplanetary scintillation observations, interpreted by the tomography modeling for speed out of ecliptic. Up to now, no global, continuous measurements of solar wind density as a function of heliographic latitude have been available and this quantity has to be obtained using indirect methods \citep[see][]{bzowski_etal:12b}.

Tests revealed that an appropriate balance between the latitudinal resolution of the coverage and fidelity of the results is obtained at a subdivision of the data into  10-degree heliolatitude bins. Concerning the time resolution, ideal would be Carrington rotation averages. Regrettably, such a high resolution seems to be hard to achieve for two reasons. First, the time coverage in the data from IPS has gaps that typically occur during almost four months at the beginning of each year, which would induce an artificial 1-year periodicity in the data. Second, the fast latitude scans by \textit{Ulysses} lasted about 12 months and hence differentiating between the time and latitude effects in its measurements is challenging. Thus, the full and reliable latitude structure of solar wind can currently be obtained only at a time scale of 1~year and this will be the time resolution of the model we present. Furthermore, the typical time of solar wind proton travel to the termination shock (TS) is about 1~year for $\sim 1$~keV protons so the adopted time resolution is reasonable for the heliosphere modeling.  

\subsection{Solar wind speed profiles}
For our analysis we take the solar wind speed available from 1990 to 2011 mapped at the source surface on a grid of $11 \times 360 \times 180$ records per year, which corresponds to a series of 11 Carrington rotations. The data are organized in heliolatitude from $89.5^\circ$ North to $89.5^\circ$ South in the so-called Carrington maps of solar surface \citep[see][]{ulrich_boyden:06a}. 

A comparison of the tomography-derived solar wind speed with the \textit{in~situ} measurements by \textit{Ulysses} (see Figure~\ref{figUlysses2IPSfast}) showed that the accuracy of the tomographic results depends on the number of IPS observations available for a given rotation. Intervals with a small number of data points clearly tend to underestimate the speed. Consequently, we removed from the data the Carrington rotations with the total number of points less than 30\,000. Rotations with small numbers of available observations typically happen at the beginning and at the end of the year and at the edges of data gaps. The selection of data by the total number of points per rotation constrained the data mainly to the summer and autumn months, when all latitudes are fully sampled.

\begin{figure}[t]
		\centering
		\begin{tabular}{ccc}		
\includegraphics[scale=0.35]{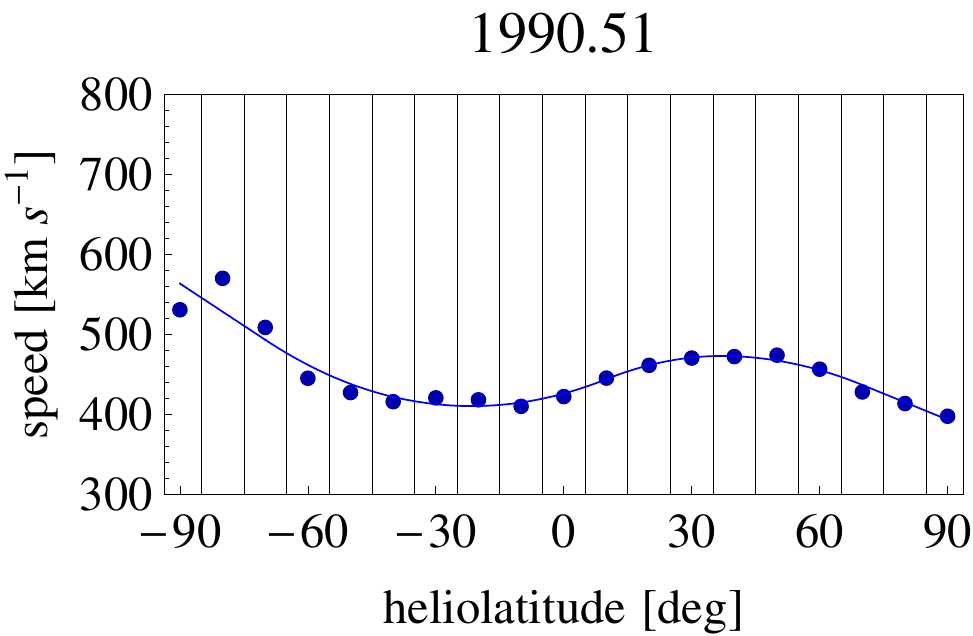}&\includegraphics[scale=0.35]{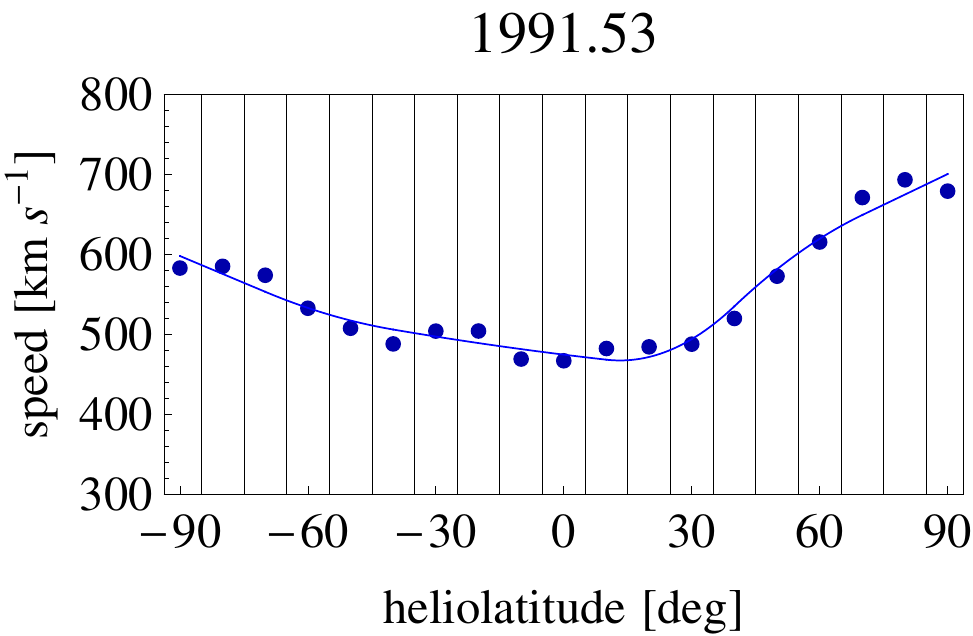}&\includegraphics[scale=0.35]{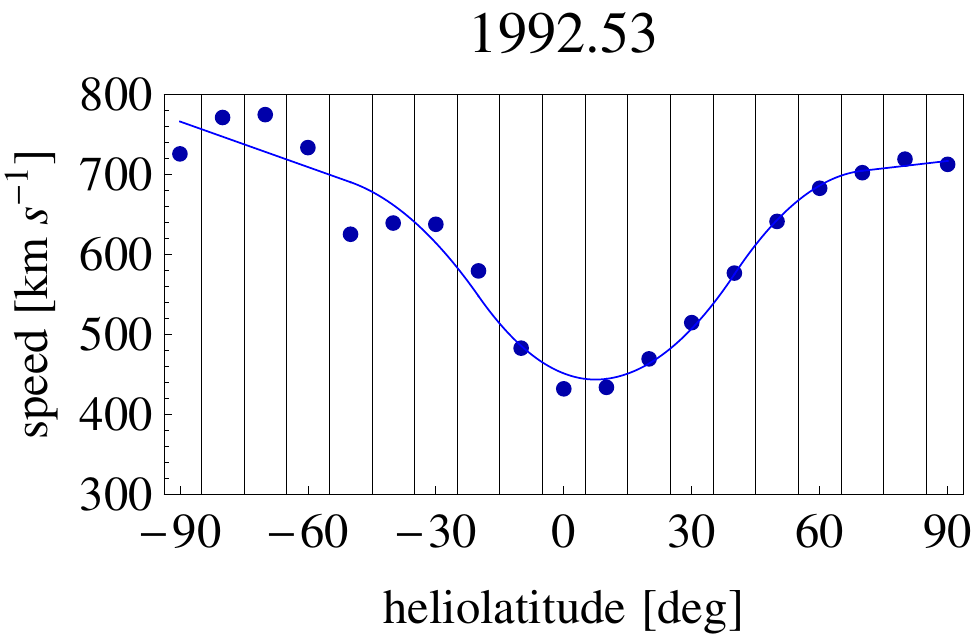}\\	\includegraphics[scale=0.35]{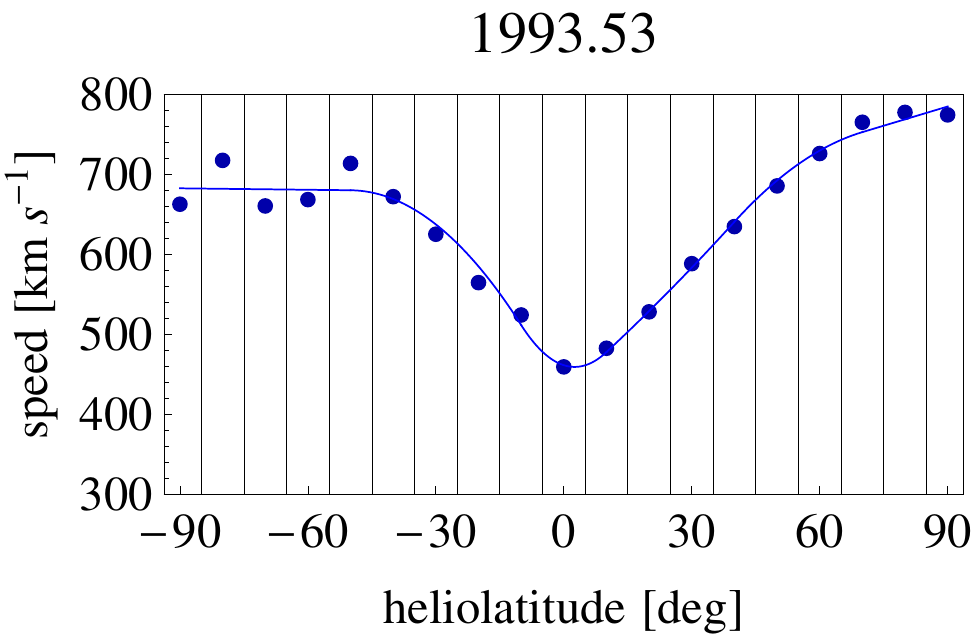}&\includegraphics[scale=0.35]{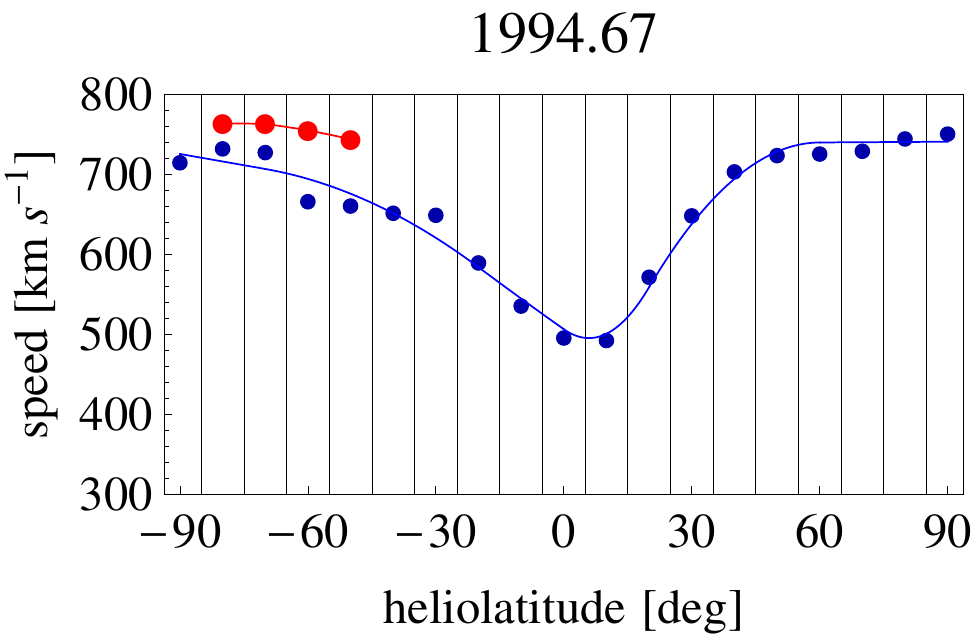}&\includegraphics[scale=0.35]{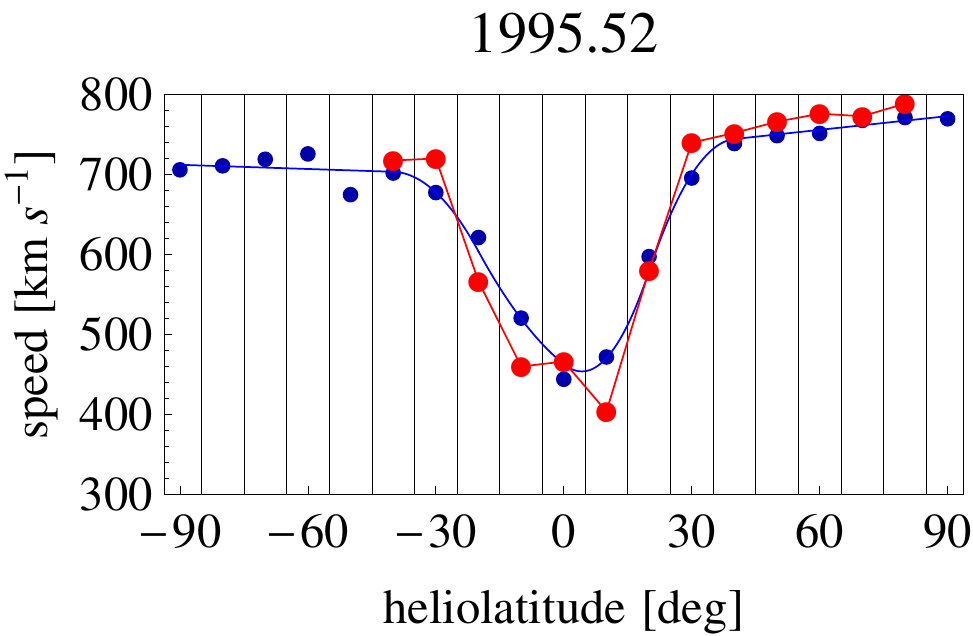}\\	\includegraphics[scale=0.35]{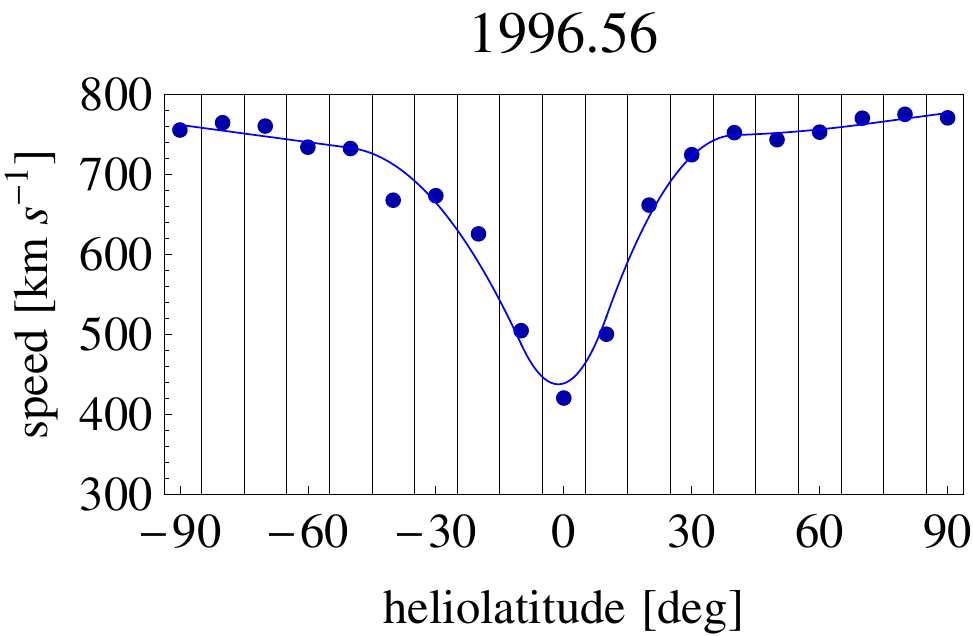}&\includegraphics[scale=0.35]{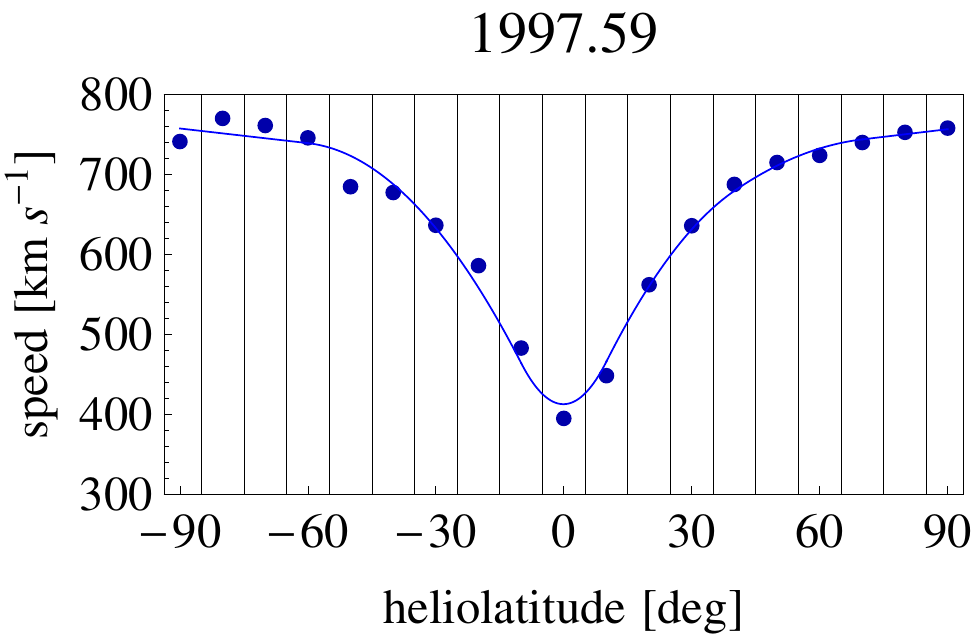}&\includegraphics[scale=0.35]{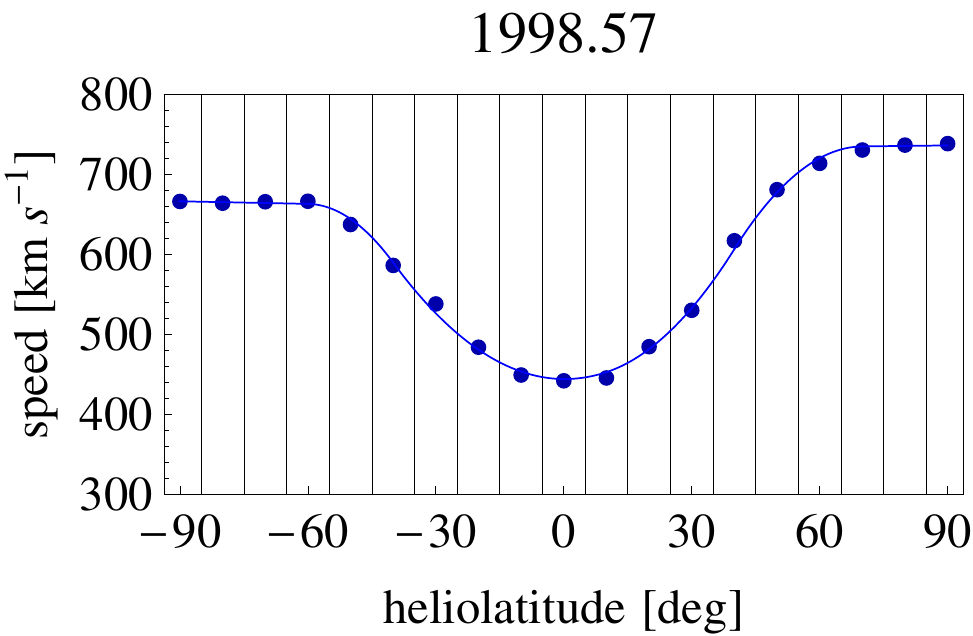}\\	\includegraphics[scale=0.35]{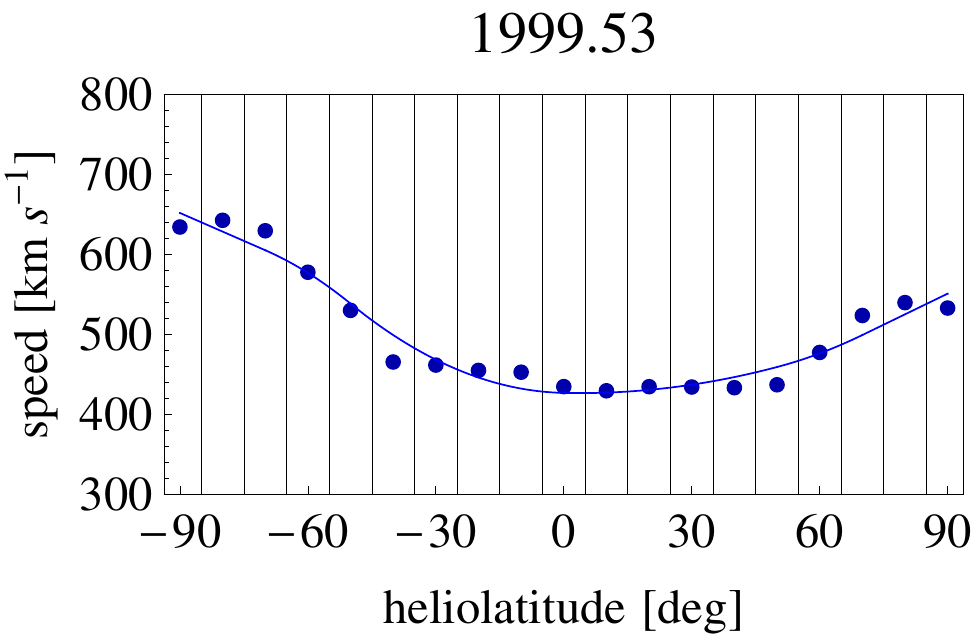}&\includegraphics[scale=0.35]{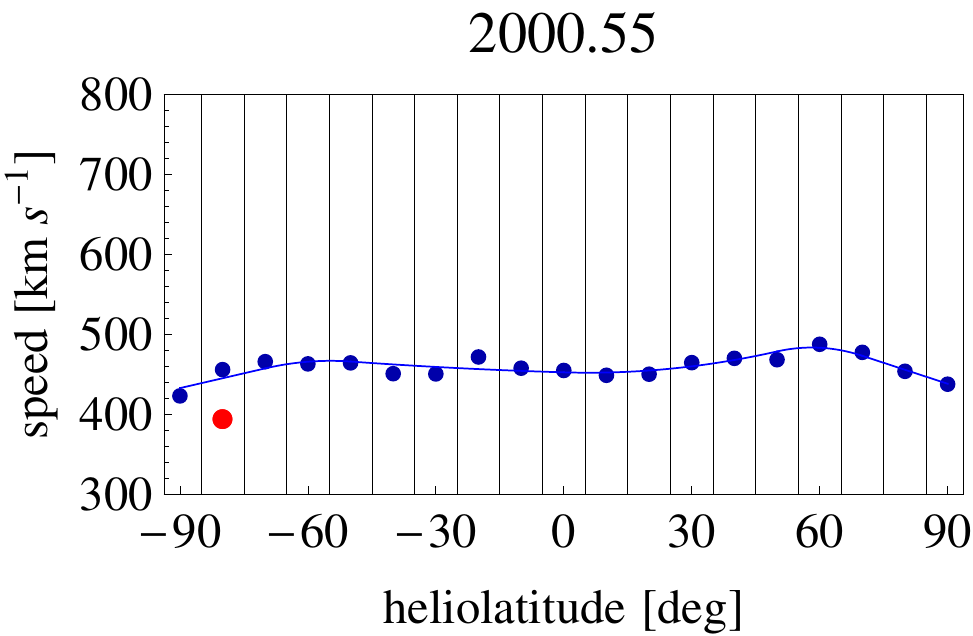}&\includegraphics[scale=0.35]{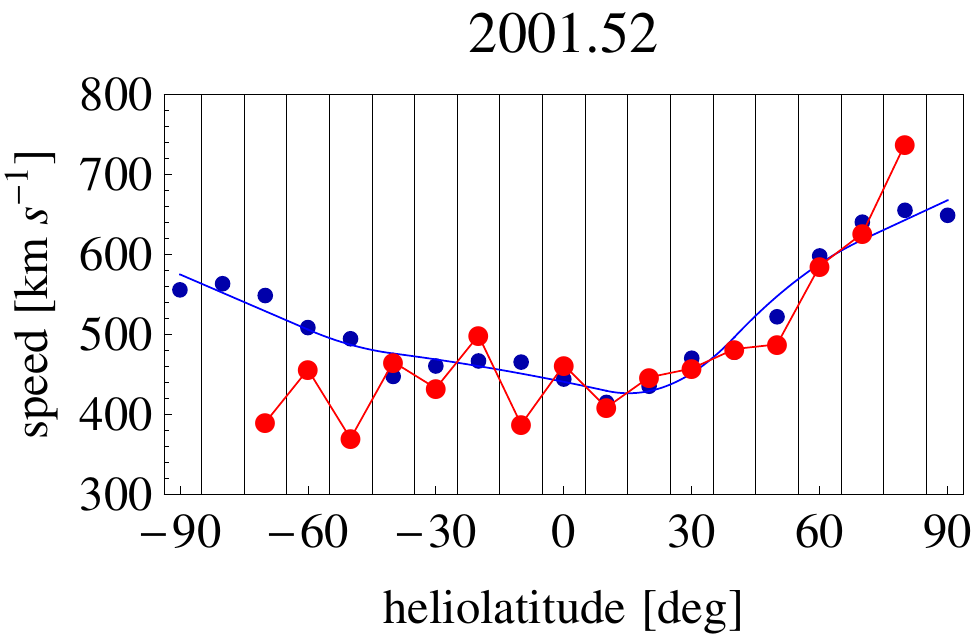}\\	\includegraphics[scale=0.35]{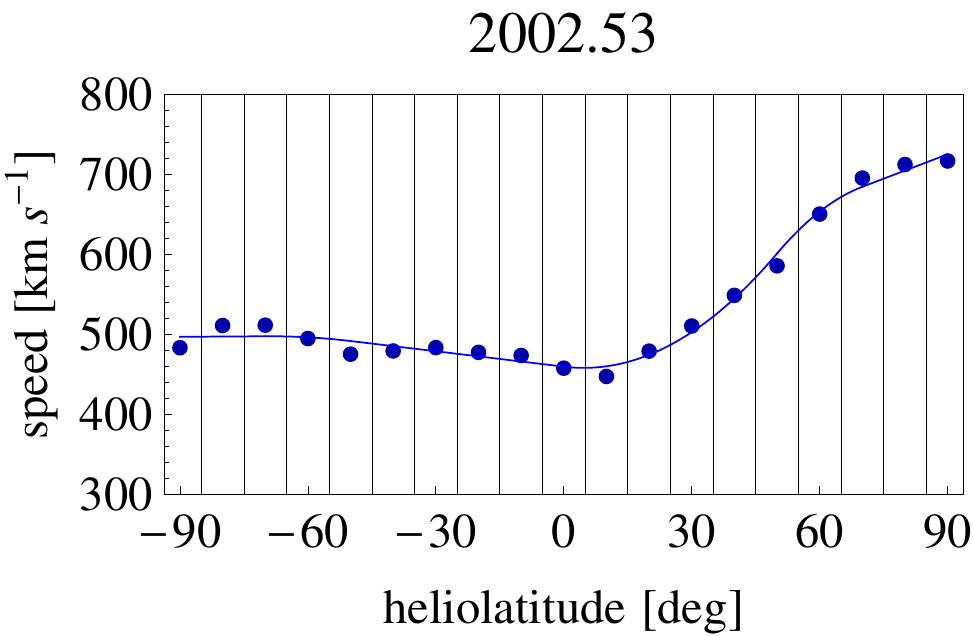}&\includegraphics[scale=0.35]{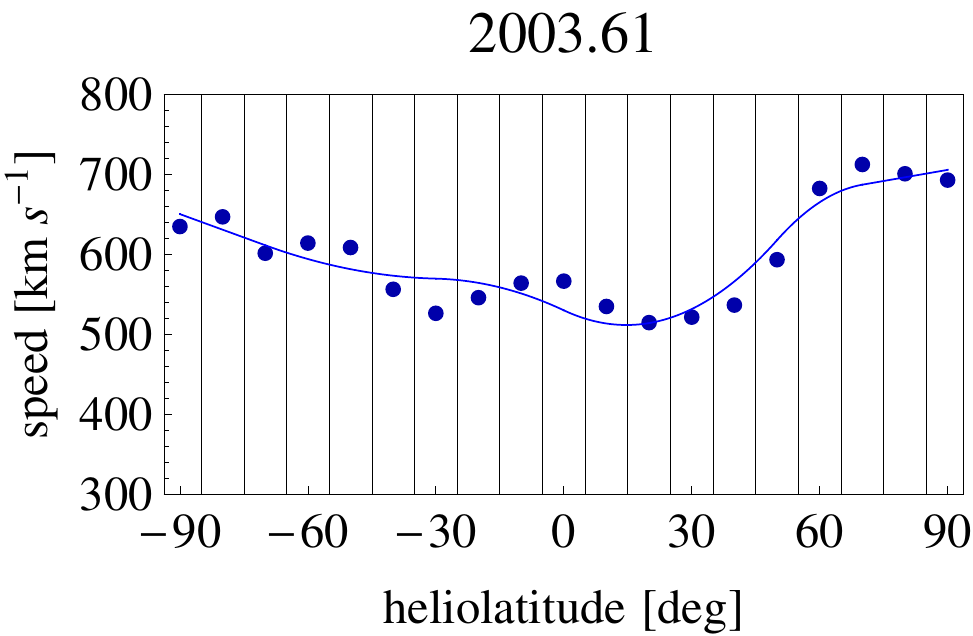}&\includegraphics[scale=0.35]{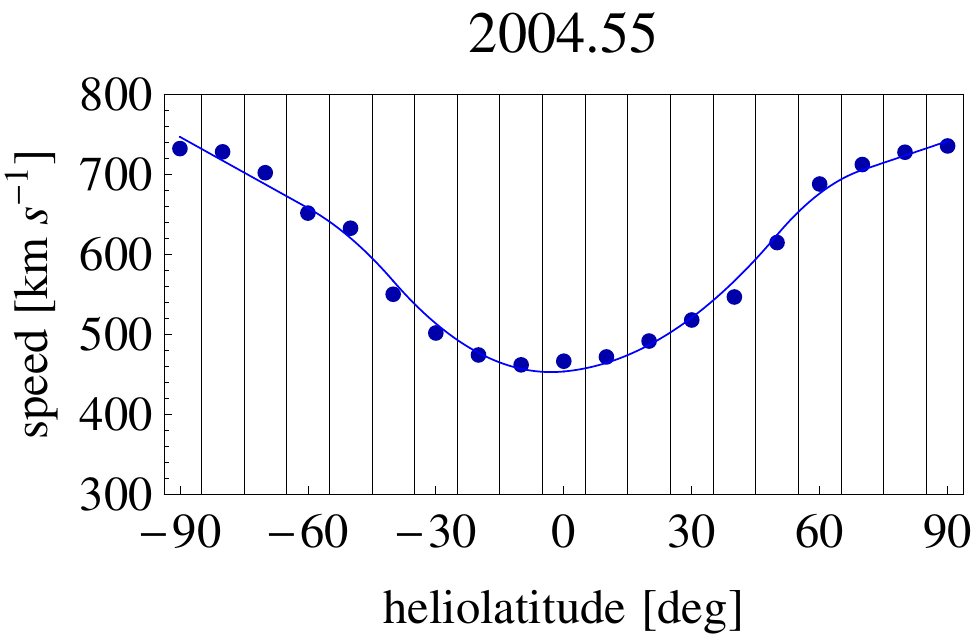}\\	\includegraphics[scale=0.35]{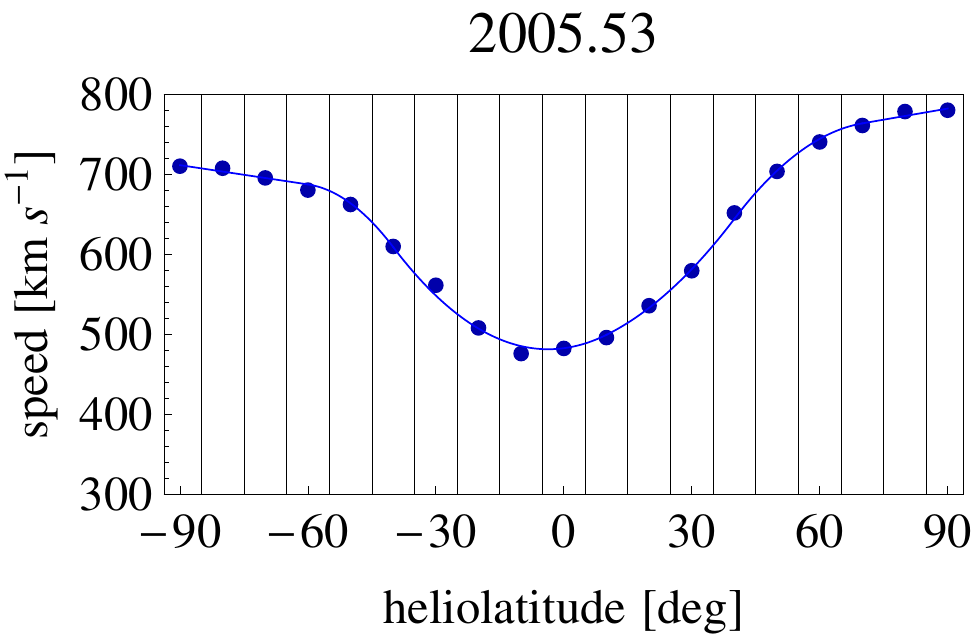}&\includegraphics[scale=0.35]{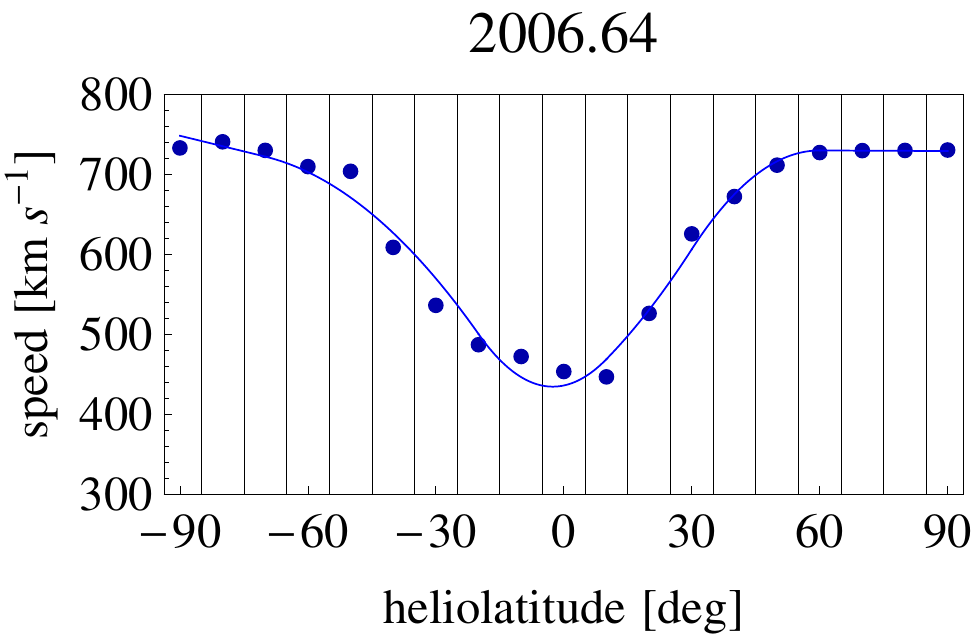}&\includegraphics[scale=0.35]{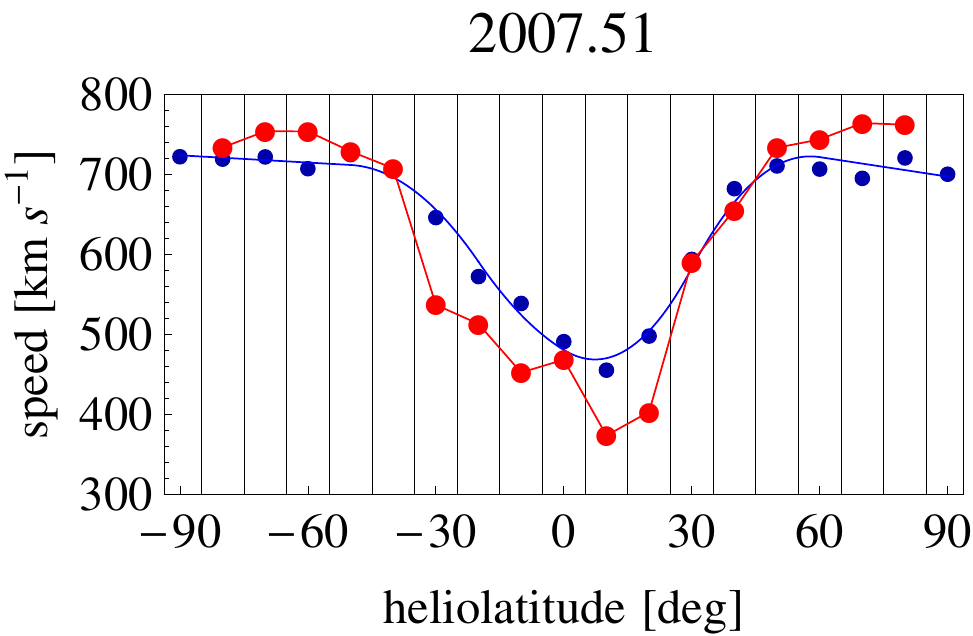}\\		\includegraphics[scale=0.35]{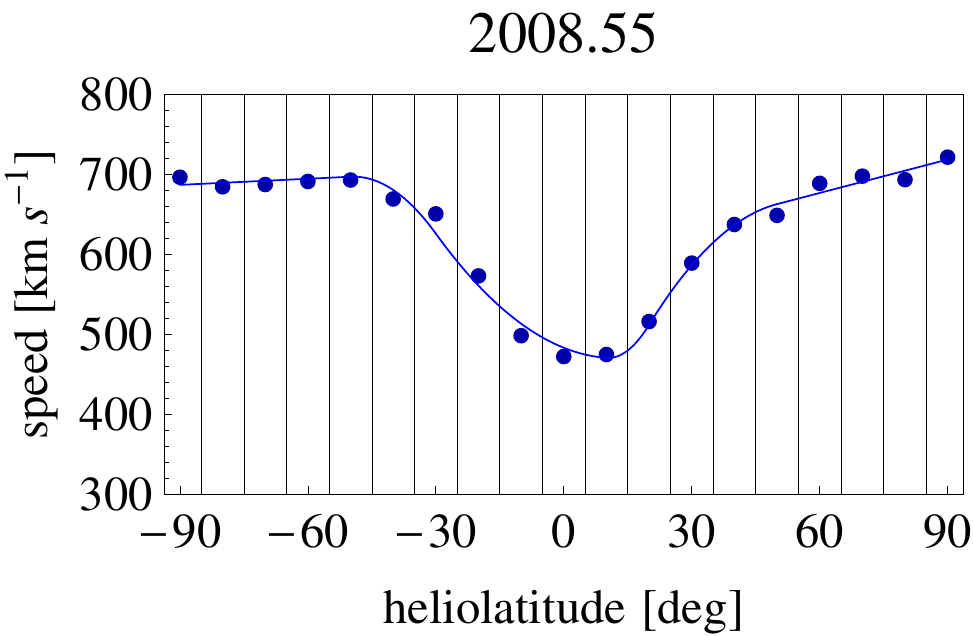}&\includegraphics[scale=0.35]{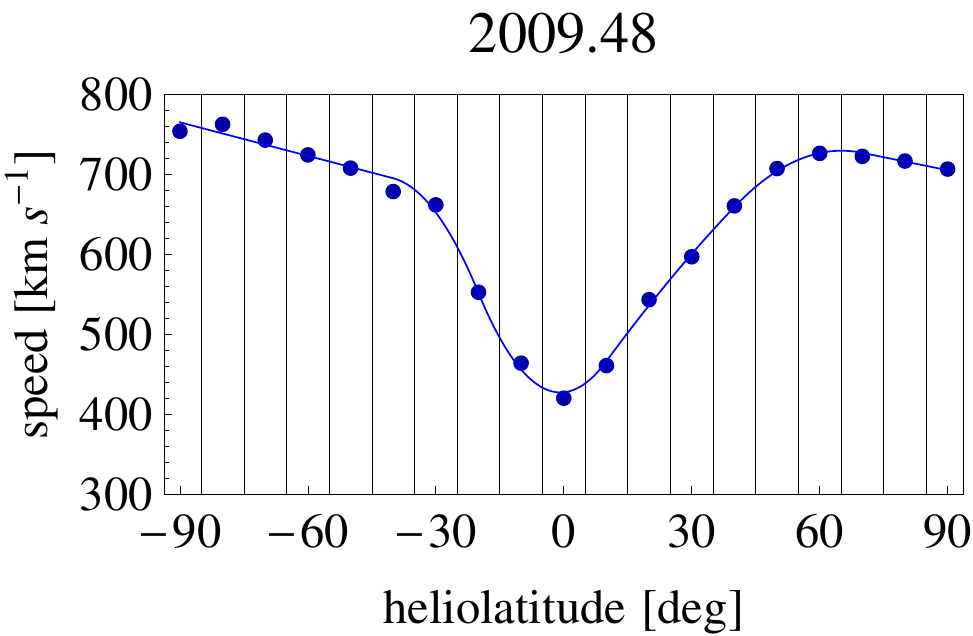}&\includegraphics[scale=0.35]{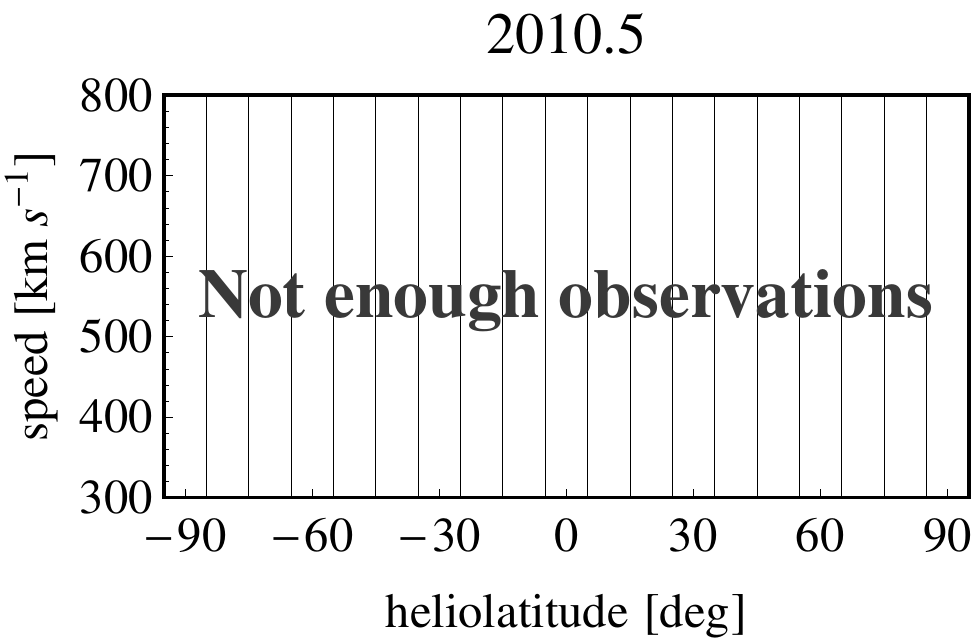}\\
\includegraphics[scale=0.35]{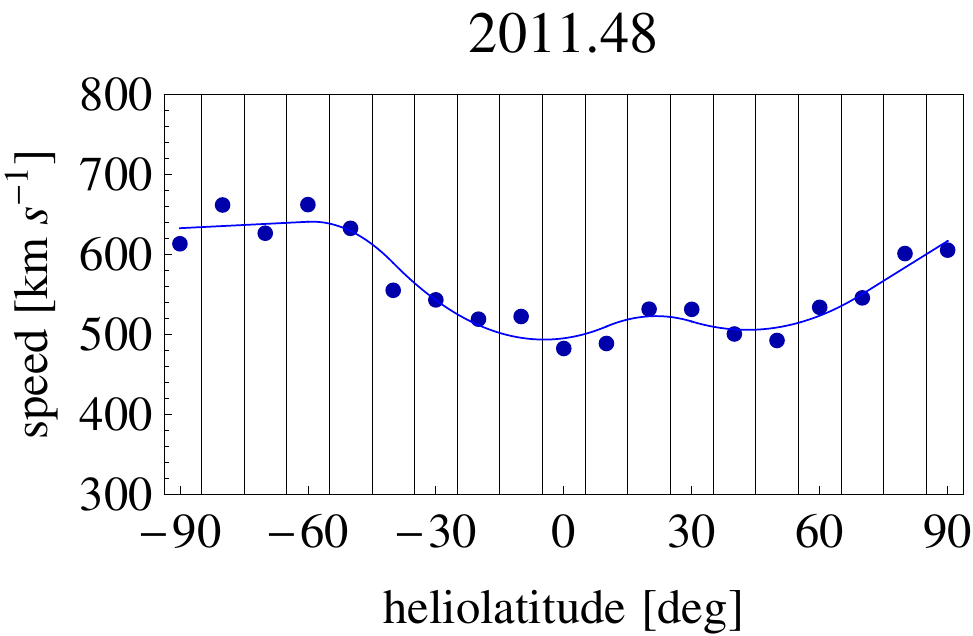}& & \\
		\end{tabular}
		\caption{Heliolatitude yearly profiles of solar wind speed. The dots represent the IPS yearly data obtained from the CAT analysis \citep{tokumaru_etal:10a} and the line shows the smoothing function defined in Equation~\ref{eqModel4and2} for each year separately. The red points and lines superimposed on the profiles (in analogy with Figure~\ref{figUlysses2IPSfast}) are parts of the \textit{Ulysses} fast scan latitude profiles for years 1994, 1995, 2000, 2001 and 2007 averaged over 10-degree heliolatitude bins. The average times for the profiles are indicated in the panel headers.}
		\label{figIPSprofiles}
		\end{figure}

The selected subset of data was split into years, and within each yearly subset into 19 heliolatitude bins, equally spaced from $-90^\circ$ to $90^\circ$. They cover the second half of Solar Cycle 22 and full Solar Cycle 23. We performed the two-step calculation: first Carrington rotation averaged values of heliolatitudinal profiles and next the yearly averages calculated from them. They are shown in Figure~\ref{figIPSprofiles}. It is worth noting that the bin-averaged solar wind profiles for specific Carrington rotations have very similar shapes to the corresponding yearly profiles with some scatter, which suggests that the general latitude structure is stable over a year and changes only on a time scale comparable with the scale of solar activity variations.  

Because of the lack of direct data we had to use the speed profiles to infer density profiles, as discussed below. Therefore, we decided to construct a function that retrieves the SW speed profile for arbitrary time and heliolatitude and smooths the remnant variations in the IPS speed profiles, so that they do not bias the inferred densities. An additional benefit of such an approach is the ability to determine latitudinal boundaries of the fast wind outside the solar maximum phase of solar activity.

		\begin{figure}[t]
		\centering
		\includegraphics[scale=0.6]{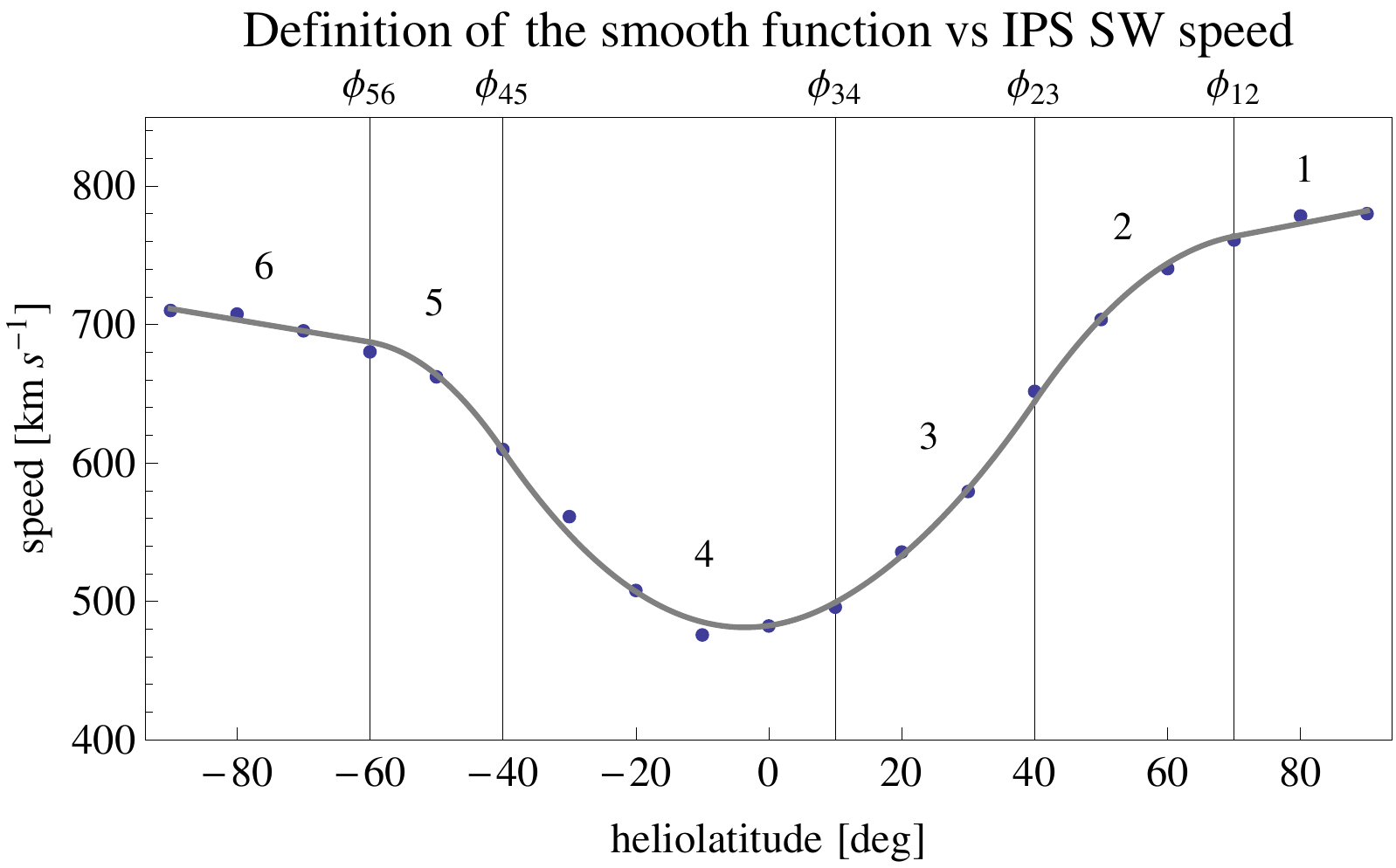}
		\caption{Comparison of the solar wind speed profiles in 2005 obtained directly from IPS (dots) with the model with the smoothing function defined in Equation~\ref{eqModel4and2Def} and Equation~\ref{eqModel4and2} applied (solid line). The vertical lines indicate the boundary bins between different sections of the model function. In 2005 they are: -60, -40, 10, 40, 70 bins. The numbers above the lines indicate the sections of the model function.}
		\label{figModel4and2}
		\end{figure}

We smoothed the speed profiles by fitting an approximating piecewise function defined as: 
	\begin{equation}
	f \left( \phi \right) = \left\{ 
	\begin{array}{ll}
	a_3+b_3 \phi +c_3 \phi ^2, & \phi_{34} < \phi \leq \phi_{23} \\
	a_4+b_4 \phi +c_4 \phi ^2, & \phi_{45} < \phi \leq \phi_{34} \\
	a_2+b_2 \phi +c_2 \phi ^2, & \phi_{23} < \phi \leq \phi_{12} \\
	a_5+b_5 \phi +c_5 \phi ^2, & \phi_{56} < \phi \leq \phi_{45} \\
	a_1+b_1 \phi, & \phi \geq \phi_{12} \\
	a_6+b_6 \phi, & \phi \leq \phi_{56} \\
	\end{array}
	\right.
	\label{eqModel4and2Def}
	\end{equation}
and
	\begin{equation}
	\phi_{56} < \phi_{45} < \phi_{34} < \phi_{23} < \phi_{12}
	\label{eqPhiOrder}
	\end{equation}
with the additional requirement that the sections of the function connect smoothly, i.e. the first derivative of $f \left( \phi \right)$ is continuous at all heliolatitudes. 

This model approximates the SW speed as a function of heliolatitude by linear relations in the polar caps and 4 parabolae at midlatitudes and in the equatorial band, which all transition smoothly between the neighboring sections. A sketch of the function definition is presented in Figure~\ref{figModel4and2}, where the splitting into sections is shown on an example profile of the IPS data. 

The ordering in Equation~\ref{eqModel4and2Def} is important because those parts of the profile for which the number of observations is the highest and which are at the north hemisphere are the most reliable. In this way, first of all we fit the function to part 3, that means to the north--near ecliptic latitudes, next to part 4, the south--near ecliptic latitudes, next to the part for the mid-latitudes and at the end to the polar regions, but with a higher weight for the north hemisphere because the number of observations per year is greater in the north hemisphere than in the south. 

Under these assumptions the function comes the following form:
	\begin{equation}
	f \left( \phi \right) = \left\{ 
	\begin{array}{ll}
	A_3+B_3 \phi +C_3 \phi ^2, & \phi_{34} < \phi \leq \phi_{23} \\
	A_4+B_4 \phi +C_4 \phi ^2, & \phi_{45} < \phi \leq \phi_{34} \\
	A_2+B_2 \phi +C_2 \phi ^2, & \phi_{23} < \phi \leq \phi_{12} \\
	A_5+B_5 \phi +C_5 \phi ^2, & \phi_{56} < \phi \leq \phi_{45} \\
	A_1+B_1 \phi, & \phi \geq \phi_{12} \\
	A_6+B_6 \phi, & \phi \leq \phi_{56} \\
	\end{array}
	\right.
	\label{eqModel4and2}
	\end{equation}
where the formulae for the coefficients $A_i$, $B_i$ and $C_i$ are defined in Table~\ref{tabABC}. The number of free parameters is reduced from 16 in Equation~\ref{eqModel4and2Def} to the following 6: $a_3, b_4, c_4, b_1, b_6, c_3$. The boundaries $\phi_{12}, \phi_{23}, \phi_{34}, \phi_{45}, \phi_{56}$ separate the heliolatitude pieces of the fitted function (see Figure~\ref{figModel4and2}) and are chosen separately for each of the yearly profiles so that the fit residuals are minimized. In Table~\ref{tabModel4and2} we present the values of the parameters of the smooth function for each yearly profile. The residuals of the fits are typically 4\% and do not exceed 10\%. 

\begin{sidewaystable}
\caption{Formulae for the coefficients from Equation~\ref{eqModel4and2}}
\label{tabABC}
\begin{tabular*}{\textwidth}{c|c|c|c} \hline 
i & $A_\mathrm{i}$ & $B_\mathrm{i}$ & $C_\mathrm{i}$ \\ \hline 
1 & $\frac{1}{2} \left(2 a_3-b_1 \phi_{12}-b_1 \phi _{23}+b_4\left(\phi_{12}+\phi _{23}\right)+2 c_3 \phi_{12} \phi _{23}-2\left(c_3-c_4\right) \left(\phi_{12}+\phi _{23}\right) \phi_{34}\right)$ & $b_1$ & - \\ \hline

2 & $\frac{2 a_3 \left(\phi _{12}-\phi _{23}\right)+\phi _{23}^2 \left(b_1-b_4-2 c_3 \phi_{12}+2 c_3 \phi _{34}-2 c_4 \phi _{34}\right)}{2 \left(\phi _{12}-\phi_{23}\right)}$ & $\frac{b_4 \phi _{12}-b_1 \phi _{23}+2 \phi _{12} \left(c_3 \left(\phi _{23}-\phi_{34}\right)+c_4 \phi _{34}\right)}{\phi _{12}-\phi _{23}}$ & $\frac{b_1-b_4-2 c_3 \phi _{23}+2 \left(c_3-c_4\right) \phi _{34}}{2 \left(\phi_{12}-\phi _{23}\right)} $\\ \hline

3 & $a_3$ & $ b_4+2 \left(c_4-c_3\right) \phi _{34} $ & $c_3$ \\ \hline

4 & $a_3+\left(c_4-c_3\right) \phi _{34}^2$ & $b_4$ & $c_4$ \\ \hline

5 & $\frac{2 \left(\phi _{45}-\phi _{56}\right) \left(a_3+\left(c_4-c_3\right) \phi_{34}^2\right)+b_4 \phi _{45}^2-b_6 \phi _{45}^2+2 c_4 \phi _{56} \phi_{45}^2}{2\left(\phi _{45}-\phi _{56}\right)}$ & $ \frac{ \left(b_6 \phi _{45}-\phi_{56} \left(b_4+2 c_4 \phi_{45}\right)\right)}{\phi _{45}-\phi _{56}} $ & $\frac{b_4-b_6+2 c_4 \phi _{45}}{2 \phi _{45}-2 \phi _{56}}$ \\ \hline

6 & $ \frac{1}{2} \left(2 a_3-b_6 \phi _{45}-b_6 \phi _{56}+b_4 \left(\phi_{45}+\phi_{56}\right)+2 \left(c_4-c_3\right) \phi _{34}^2+2 c_4 \phi_{45} \phi_{56}\right)$ & $b_6$ &  - \\ \hline
\end{tabular*}
\end{sidewaystable}

\begin{table}
\caption{Values of the free parameters of the function defined in Equation~\ref{eqModel4and2Def} and Table~\ref{tabABC} for all years.}
\label{tabModel4and2}
\begin{tabular}{cllllll}      \hline                
year & $a_3$ & $b_4$ & $c_4$ & $b_1$ & $b_6$ &  $c_3$ \\  \hline
1990 & 418.566 & ~1.45736~~ & ~0.0337921~~ & -2.17352~~ & -3.51384~~ & -0.0392348 \\
1991 & 483.579 & -0.66172~~ & ~0.00308865~ & ~2.55869~~ & -2.23751~~ & ~0.0941952 \\
1992 & 449.381 & -2.06557~~ & ~0.138359~~~ & ~0.614671~ & -1.89668~~ & ~0.121885~ \\
1993 & 429.457 & -1.69448~~ & ~0.333417~~~ & ~1.60133~~ & -0.0603855 & ~0.0151053 \\
1994 & 506.33~ & -3.77307~~ & ~0.00505828~ & ~0.0313874 & -0.93796~~ & ~0.319795~ \\
1995 & 462.637 & -4.29315~~ & ~0.137468~~~ & ~0.573251~ & -0.17428~~ & ~0.505656~ \\
1996 & 347.645 & ~1.71087~~ & ~0.662762~~~ & ~0.717709~ & -0.719594~ & -0.246573~ \\
1997 & 348.65~ & ~0.0918378 & ~0.523877~~~ & ~0.636597~ & -0.6109~~~ & -0.115214~ \\
1998 & 446.188 & -0.0377569 & ~0.0900351~~ & ~0.0607183 & -0.106951~ & ~0.112234~ \\
1999 & 426.63~ & -0.115668~ & ~0.0423373~~ & ~2.58806~~ & -2.32994~~ & ~0.0152633 \\
2000 & 452.346 & -0.158137~ & ~0.00223068~ & -1.74742~~ & ~1.23412~~ & ~0.0136059 \\
2001 & 452.326 & -1.07672~~ & -0.00476791~ & ~2.47585~~ & -2.29808~~ & ~0.113553~ \\
2002 & 459.426 & -0.647343~ & -0.000129429 & ~2.0322~~~ & ~0.0351119 & ~0.0695614 \\
2003 & 529.919 & -2.50005~~ & -0.0393349~~ & ~0.927344~ & -1.94706~~ & ~0.0850843 \\
2004 & 453.411 & ~0.494147~ & ~0.0831141~~ & ~1.80231~~ & -2.96517~~ & ~0.058379~ \\
2005 & 480.172 & ~0.705444~ & ~0.0966926~~ & ~0.925081~ & -0.803117~ & ~0.0732095 \\
2006 & 421.171 & ~1.1226~~~ & ~0.217231~~~ & -0.0220226 & -1.3036~~~ & ~0.0698785 \\
2007 & 480.365 & -3.25418~~ & ~0.112857~~~ & -0.811889~ & -0.302357~ & ~0.222071~ \\
2008 & 519.668 & -2.13859~~ & ~0.0872578~~ & ~1.39421~~ & ~0.255003~ & ~0.452983~ \\
2009 & 391.003 & ~0.523058~ & ~0.336663~~~ & -1.08514~~ & -1.39841~~ & -0.0276554 \\ 
2011 & 477.692 & ~0.704998~ & ~0.0765034~~ & ~3.32028~~ & ~0.267478~ & -0.0960392 \\  \hline
\end{tabular}
\end{table}

\begin{table}
\caption{Boundaries between sections of the smoothing function defined in Equation~\ref{eqModel4and2Def}.}
\label{tabPhi}
\begin{tabular}{clllll}      \hline                
year & $\phi_{56}$ & $\phi_{45}$ & $\phi_{34}$ & $\phi_{23}$ & $\phi_{12}$  \\  \hline
 1990 & -70 & -40 & 10 & 40 & 70 \\ 
 1991 & -70 & -40 & 10 & 40 & 70 \\ 
 1992 & -50 & -20 & 10 & 40 & 70 \\ 
 1993 & -50 & -10 & 10 & 40 & 70 \\
 1994 & -70 & -20 & 0~ & 20 & 60 \\
 1995 & -40 & -20 & 0~ & 20 & 40 \\
 1996 & -50 & -10 & 10 & 40 & 70 \\
 1997 & -60 & -10 & 10 & 40 & 70 \\
 1998 & -60 & -40 & 10 & 40 & 70 \\
 1999 & -70 & -50 & 0~ & 50 & 70 \\ 
 2000 & -70 & -50 & 0~ & 50 & 70 \\
 2001 & -60 & -40 & 10 & 40 & 70 \\
 2002 & -70 & -50 & 0~ & 50 & 70 \\
 2003 & -70 & -30 & 0~ & 50 & 70 \\
 2004 & -60 & -40 & 0~ & 50 & 70 \\
 2005 & -60 & -40 & 10 & 40 & 70 \\
 2006 & -70 & -20 & 10 & 30 & 60 \\
 2007 & -50 & -20 & 0~ & 30 & 60 \\
 2008 & -50 & -30 & 10 & 20 & 50 \\
 2009 & -40 & -20 & 10 & 40 & 70 \\ 
 2011 & -60 & -40 & 10 & 30 & 70 \\  \hline
\end{tabular}
\end{table}

\begin{figure}
\centering\includegraphics[scale=0.6]{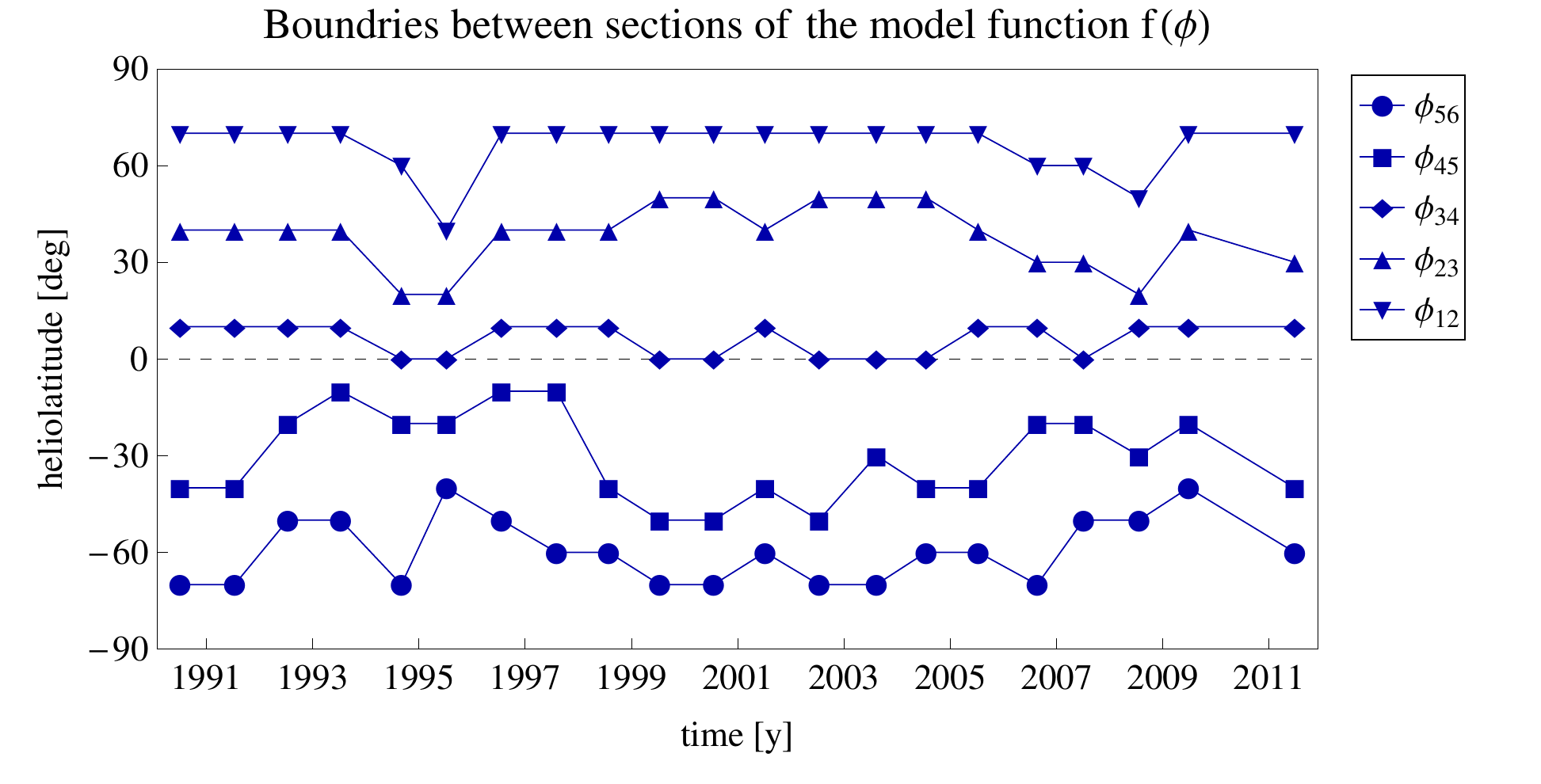}
\caption{Boundaries between the sections of the model function defined in Equation~\ref{eqModel4and2Def} and Equation~\ref{eqModel4and2}, fitted to the yearly solar wind heliolatitude profiles obtained from the IPS observations. See Table~\ref{tabPhi} for numerical values.}
\label{figButterflyPlot}
\end{figure}

Table~\ref{tabPhi} presents the heliolatitude boundaries adopted for each yearly profile of the IPS SW speed. They are also shown in Figure~\ref{figButterflyPlot}, which illustrates how the heliographic latitude bands change with time and solar cycle phases and how the ranges of solar wind regimes evolve. During the interval 1998--2006 the low-latitude bands of the slow and variable SW expand to midlatitudes. The behavior of the boundaries fitted for the south hemisphere is somewhat different, which is probably due to a lower quality of the IPS speed profiles in the south hemisphere because of the observations conditions discussed above.

\begin{figure}[ht]
		\centering	
		\begin{tabular}{cc}
    \includegraphics[scale=0.4]{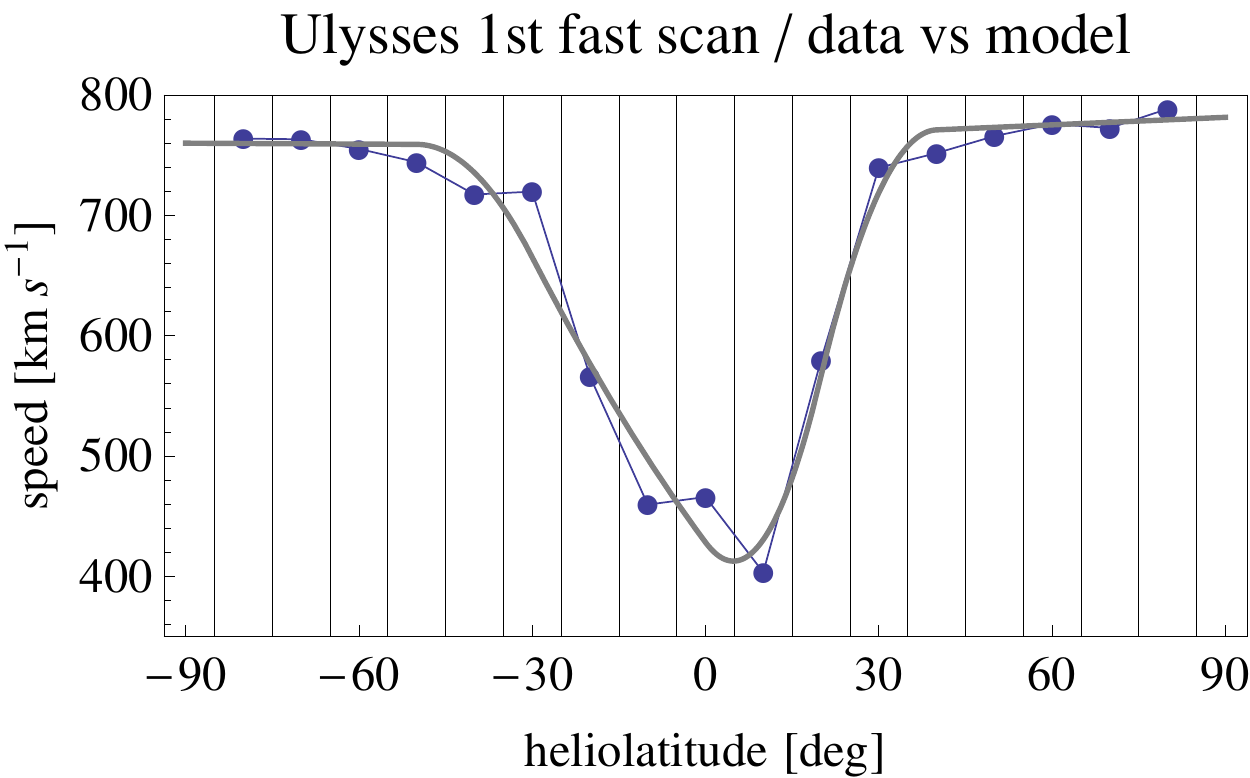} & \includegraphics[scale=0.4]{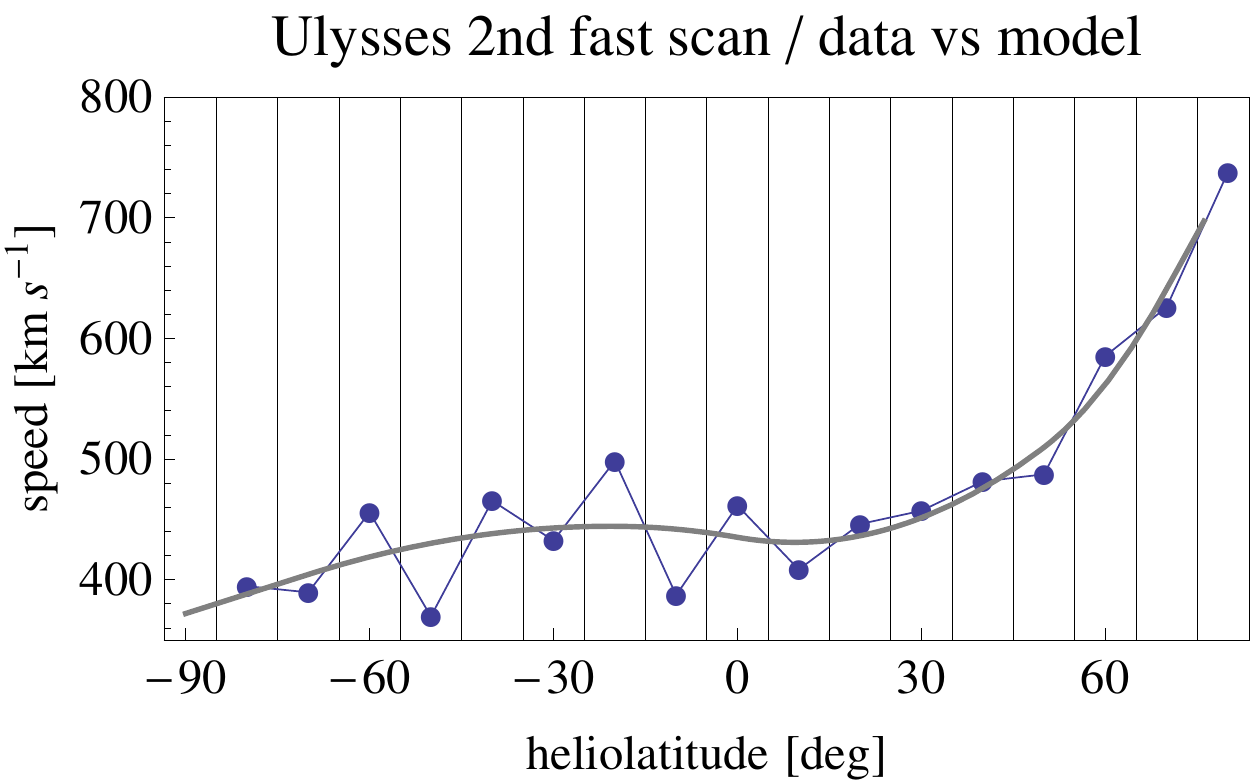}\\
    \includegraphics[scale=0.4]{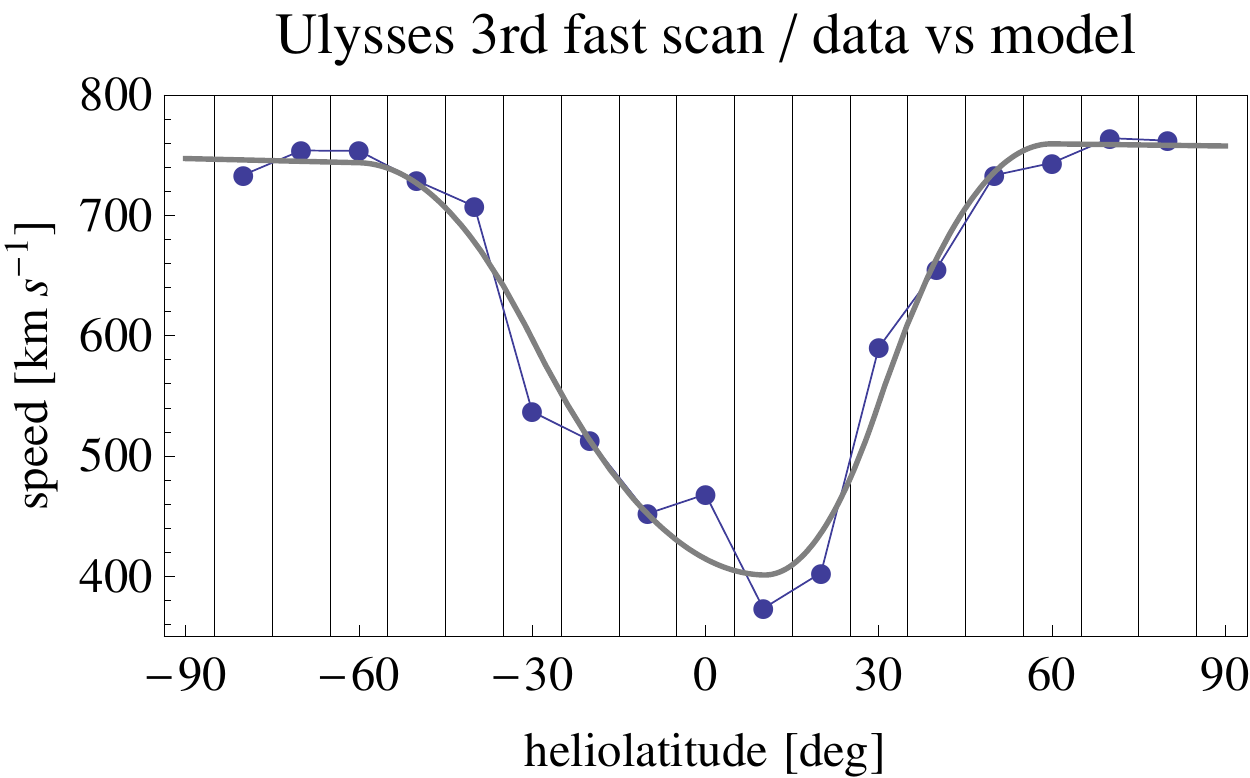} & \\
    \end{tabular}
  	\caption{Results of application of the smoothing function defined in Equation~\ref{eqModel4and2} to solar wind speed data obtained from the \textit{Ulysses} fast scans, averaged over 10-degree heliolatitudes bins. The boundaries between the model function sections are: $\left(-50, -30, 0, 20, 40\right)$ for the first fast scan, $\left(-70, -20, 0, 50, 70\right)$ for the second slow scan and $\left(-60, -30, 10, 30, 60\right)$ for the third fast \textit{Ulysses} scan. Blue dots and lines: \textit{Ulysses} original data, gray solid lines: results of the smoothing.}
		\label{figModel4and2forUlysses}
		\end{figure}

This model function was used to smooth the yearly profiles obtained from the IPS tomography data. We use them in further analysis keeping the 10-degree resolution in latitude. As it is seen in Figure~\ref{figIPSprofiles}, the smoothing procedure works very well for all years, with slightly higher residuals at higher latitudes. The model can be used equally well for the solar minimum and maximum conditions and it can be applied to both IPS and \textit{Ulysses} data (see Figure~\ref{figModel4and2forUlysses}).

		\clearpage
		
\begin{figure}
		\centering
		\begin{tabular}{cc}		
\includegraphics[scale=0.33]{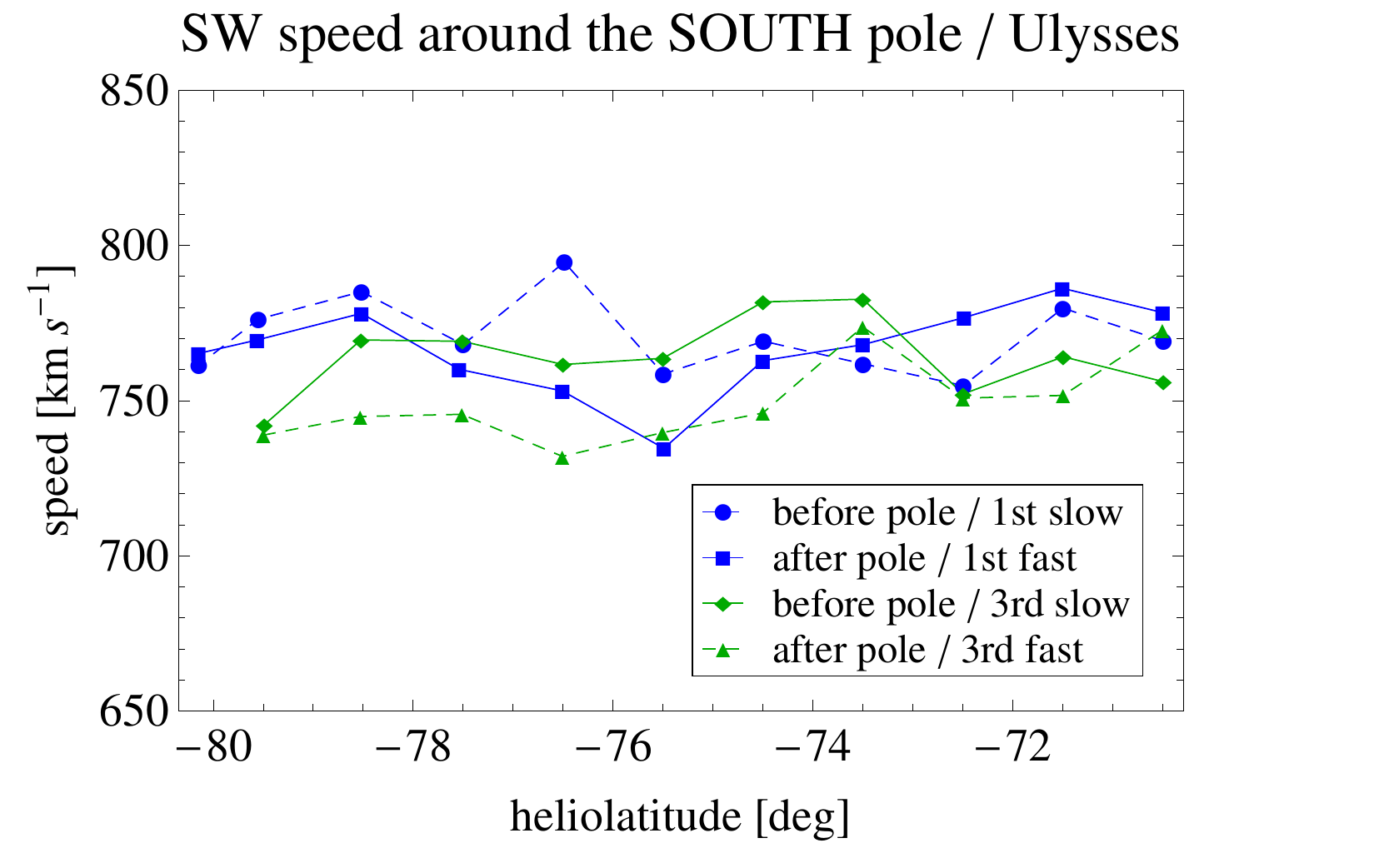}&\includegraphics[scale=0.33]{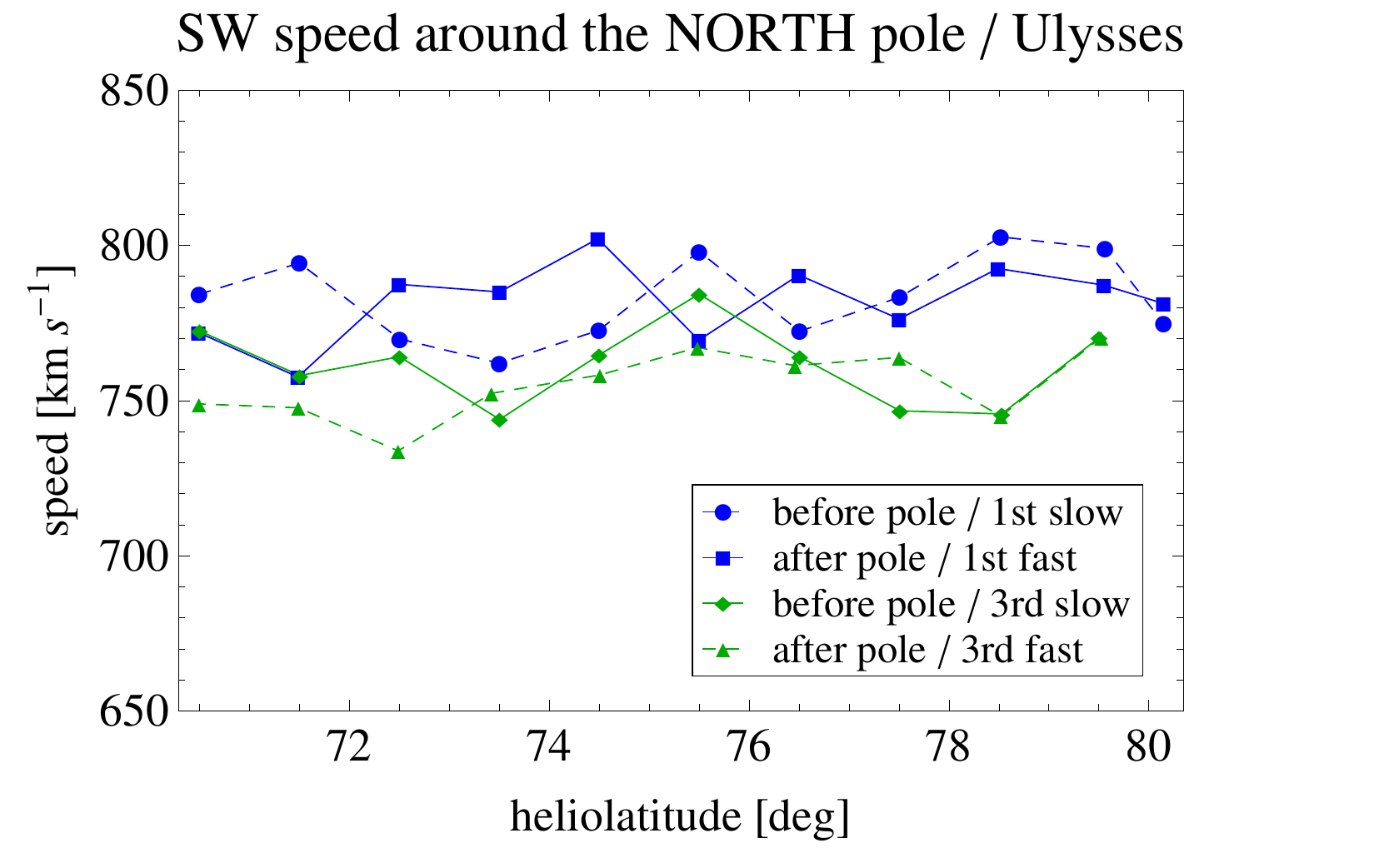}\\	
\includegraphics[scale=0.33]{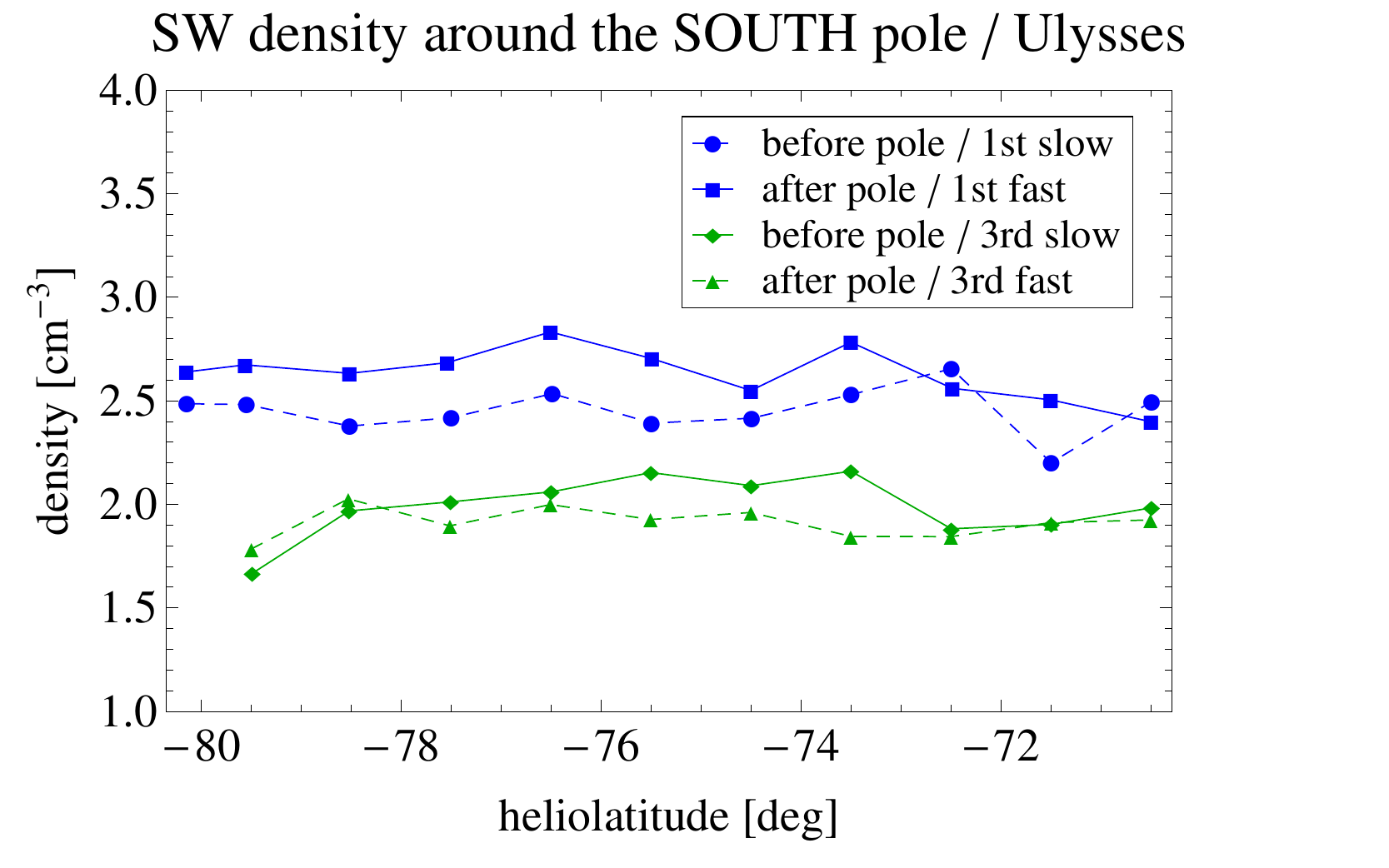}&\includegraphics[scale=0.33]{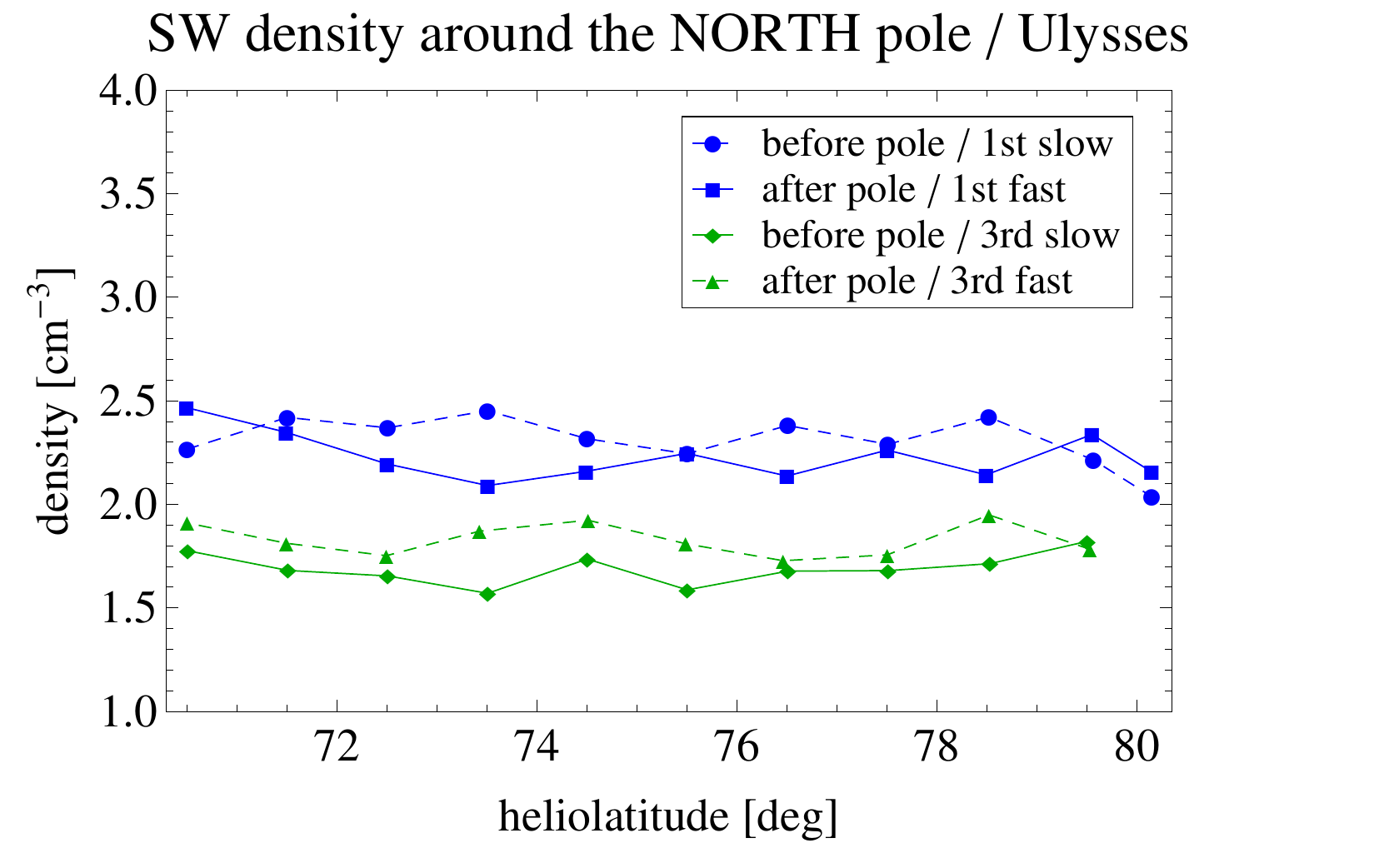}\\
		\end{tabular}
		\caption{Solar wind speed (upper row) and density (lower row) variations near the polar regions during the solar minimum conditions from \textit{Ulysses} measurements. Blue lines represent the beginning and end of the first fast scan, green lines the beginning and end of the third fast scan. Solid lines are for \textit{Ulysses} approaching to the pole and the dashed lines are for \textit{Ulysses} receding from the pole. The dots represent the 1-degree latitude averages.}
		\label{figUlyssesPoles}
		\end{figure}

The linear behavior of the fast solar wind as a function of heliolatitude was discovered by \citet{mccomas_etal:02a} based on \textit{in~situ} measurements by \textit{Ulysses}. In Figure~\ref{figUlyssesPoles} we show the behavior of SW speed for polar regions for the first and third \textit{Ulysses} polar orbit. There is a slight drop in speed at the north hemisphere between the first and third scan, while the drop in density is well seen for both hemispheres.	

\subsection{Solar wind density profiles from density-speed correlation}
There have been no direct solar wind density measurements apart from the \textit{Ulysses} \textit{in~situ} data and in lack of remote-sensing observations (e.g. density deconvolved from remote-sensing observations of Lyman-$\alpha$ helioglow from SWAN/SOHO) the density has to be estimated using indirect approximate methods. The SW density adjusted to 1 AU can be approximately inferred from the speed profiles, but the correlation between speed and density for different heliolatitudes is challenging to obtain. Some insight was provided by \citet{ebert_etal:09a}, but in their approach the data from the fast and slow scans were treated collectively and it was hard to deconvolve the temporal and spatial effects (see Figure~\ref{figUlyRFast} for the correlation between the solar distance and heliolatitude). 

\begin{figure}[ht]
\centering
\includegraphics[scale=0.6]{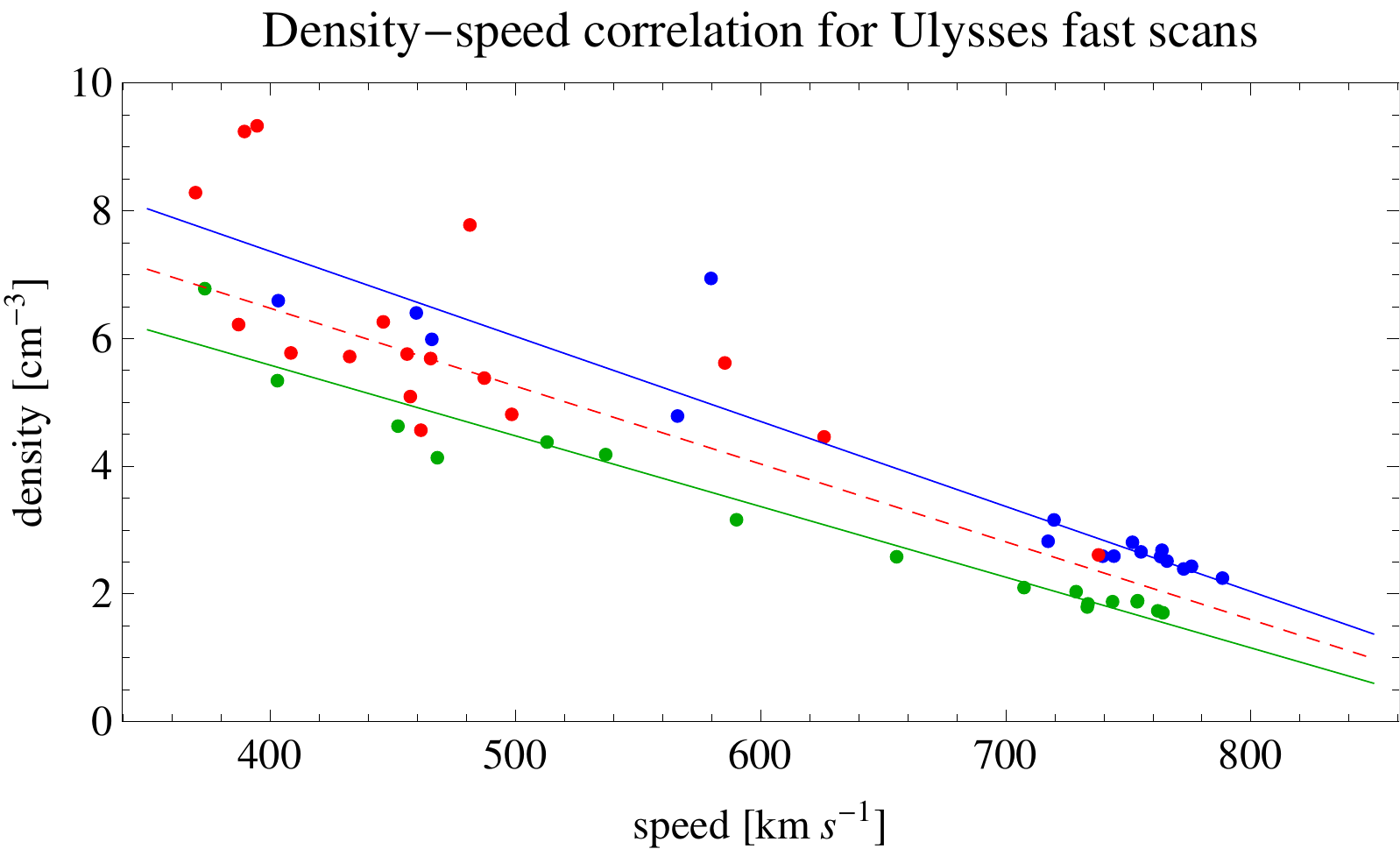}
\caption{Correlation between the solar wind adjusted density and speed obtained from the \textit{Ulysses} fast latitude scans. Blue corresponds to the first scan (see Figure~\ref{figUlyssesComposite}), green to the third scan and red to the second scan, performed during the solar maximum conditions. The dots correspond to the density-speed pairs averaged over the 10-degree heliolatitude bins, the blue and green lines are the linear correlations specified in Equation~\ref{eqUlyDensCorr}. The dotted red line is the density-speed relation proposed for the transition interval close to the solar maximum in 2002, calculated as an arithmetic mean of the correlation relations obtained from the first and third latitude scans. }
\label{figUlyDensSpeedCorr}
\end{figure}	

\begin{figure}
\centering
\includegraphics[scale=0.6]{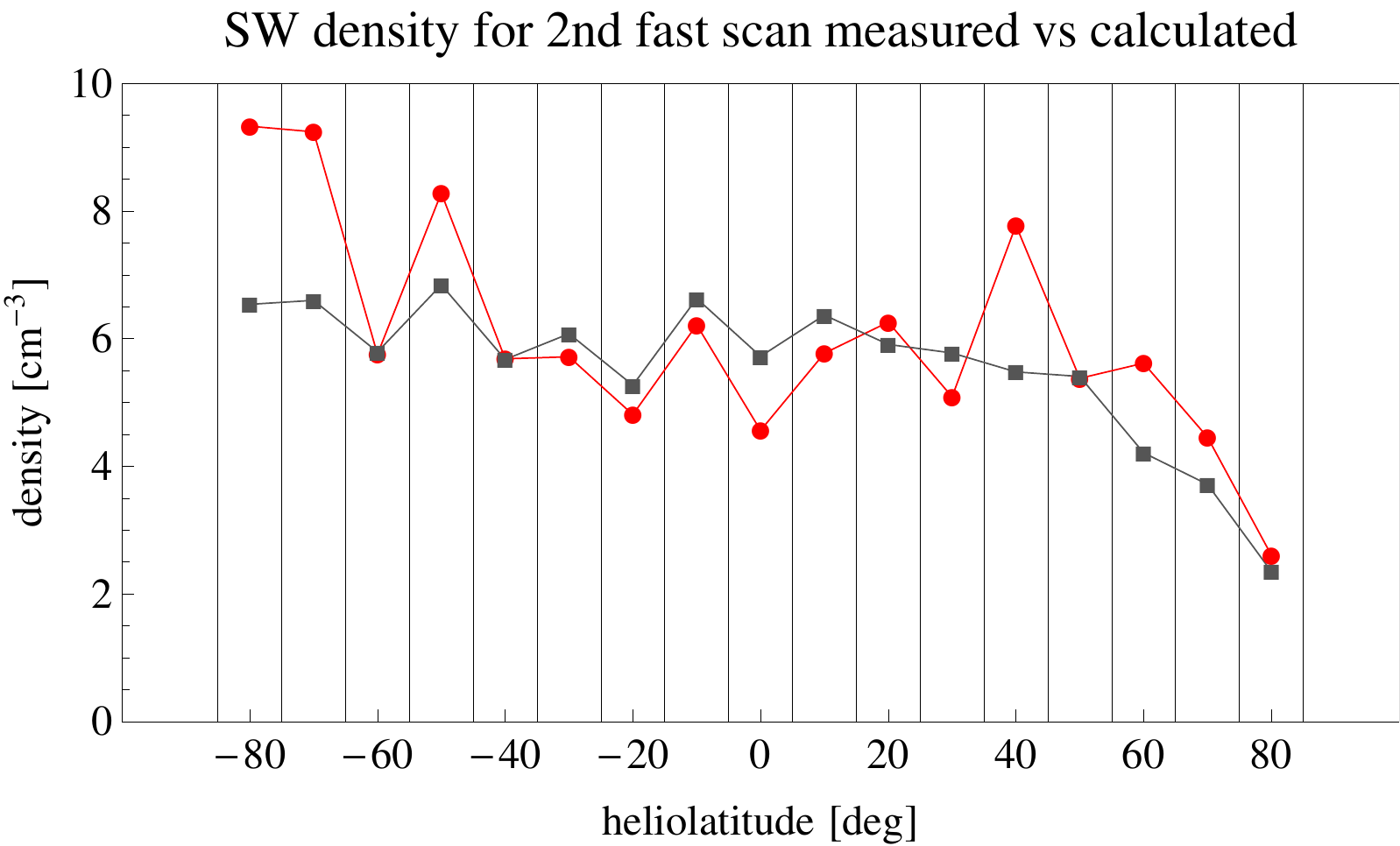}
\caption{Comparison of adjusted solar wind density from \textit{in~situ} measurement by \textit{Ulysses} during the second fast scan during solar maximum (red) with the density calculated from the Equation~\ref{eqUlyDensCorr} and Equation~\ref{eqDensFormSecond} adopted to speed values from \textit{Ulysses} data from second fast scan (gray).}
\label{figUlyDens2}
\end{figure}

Here we decided to use an interim solution and retrieve solar wind density from correlations between the solar wind speed and density obtained specially for this project from the \textit{Ulysses} fast latitude scans. During solar minimum the solar wind speed and density are correlated in heliolatitude but the correlations seem slightly different between the first and third latitude scans because of the secular changes observed in the solar wind \citep{mccomas_etal:08a}. The correlations for the fast scans are illustrated in Figure~\ref{figUlyDensSpeedCorr}. We assume the following linear relation between the SW  proton density $n_{\mathrm{p}}$ and speed $v_{\mathrm{p}}$:
\begin{equation}
\label{eqUlyDensCorr}
n_{\mathrm{p, Ulysses}} \left(v_{\mathrm{p}}\right) = a_{scan} + b_{scan}\, v_{\mathrm{p}}
\end{equation}
where $a_{scan}, b_{scan}$ are fit separately for the speed and density values averaged over 10-degree bins using the ordinary least squares bisector method \citep[see OLS bisector in][]{isobe_etal:90a}, which allows for uncertainty in both ordinate and abscissa, separately for the first and third latitude scans. For the first scan (blue line and points in Figure~\ref{figUlyDensSpeedCorr}) we obtain:
\begin{equation}
\label{eqDensFormFirst}
a_{first} = 12.69 \pm 1.17, \quad b_{first} = -0.01332 \pm 0.00154,
\end{equation}
and for the third scan (green line and dots in Figure~\ref{figUlyDensSpeedCorr}) the correlation formula parameters are:
\begin{equation}
\label{eqDensFormThird}
a_{third} = 10.01 \pm 0.65, \quad b_{third} = -0.01107 \pm 0.00094.
\end{equation} 
Thus, the slopes are almost identical and the main difference between the two formulae is in the intercept, which reflects the overall secular decrease in solar wind density between the two solar minima.

The relation between density and speed for the second scan, which occurred during solar maximum, is hard to establish directly because there are few points with high speed values. In this case the spatial and temporal effects seem to be convolved (as discussed earlier). Therefore, we propose to use a very simple relation based on the assumption that the density--speed correlation does not change with solar activity phase and use an arithmetic mean of the relations for the first and third scans as an valid for the middle time period: 
\begin{equation}
\label{eqDensFormSecond}
a_{second} = \left(a_{first} + a_{third} \right)/2 \quad \mathrm{and} \quad b_{second} = \left(b_{first} + b_{third} \right)/2.
\end{equation} 
Relation from Equation~\ref{eqDensFormSecond} is shown in Figure~\ref{figUlyDensSpeedCorr} as the red broken line. It seems to reconstruct the SW density reasonably well, as shown in Figure~\ref{figUlyDens2}, where a comparison of the density values actually measured during the second fast latitude scan and calculated from the adopted correlation formula is presented.

Since we have three different correlation formulae, we have to specify the time intervals of the applicability of them. We connect them with the changes in SW density in time and adopt the following rules: the formula from the first scan applies to the interval before 1998, because before this time no long-term density changes were observed; the second relation to the interval from 1998 until 2002, when the density decrease was the most visible; and the relation from the third scan for the interval since 2002, when the density seems to have changed its temporal gradient. 

The correlation formulae mentioned seem to have a purely statistical character and are only good to reproduce the large-scale relation between bin-averaged speed and density. They do not allow to reliably reproduce density short-scale variations within the equatorial solar wind because, as we verified, there is no significant correlation between long-term averages of density and speed in this regime.

\begin{figure*}[t]
		\centering
		\begin{tabular}{ccc}		
\includegraphics[scale=0.35]{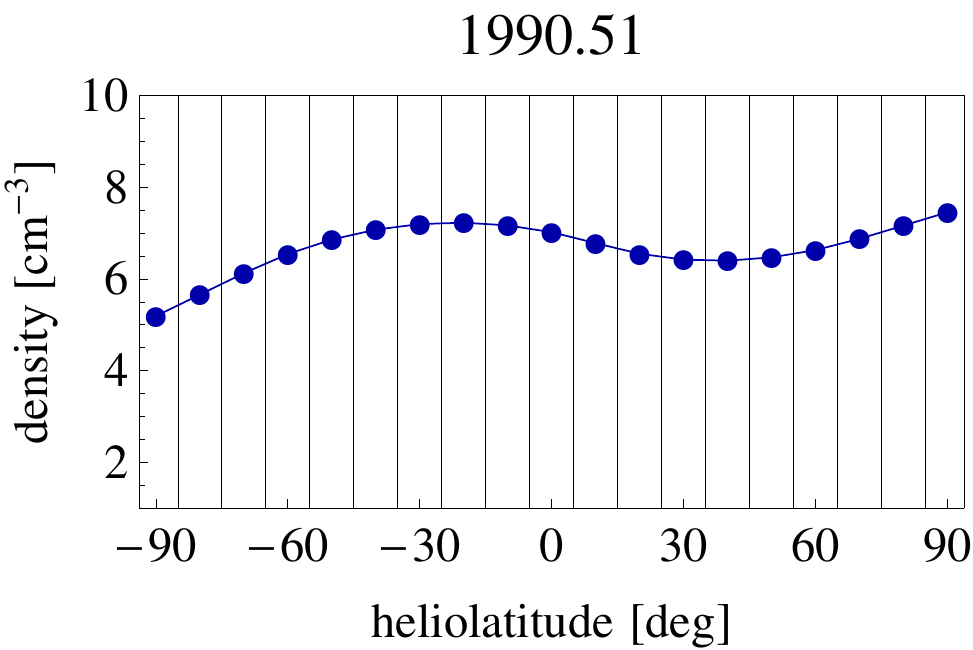}&\includegraphics[scale=0.35]{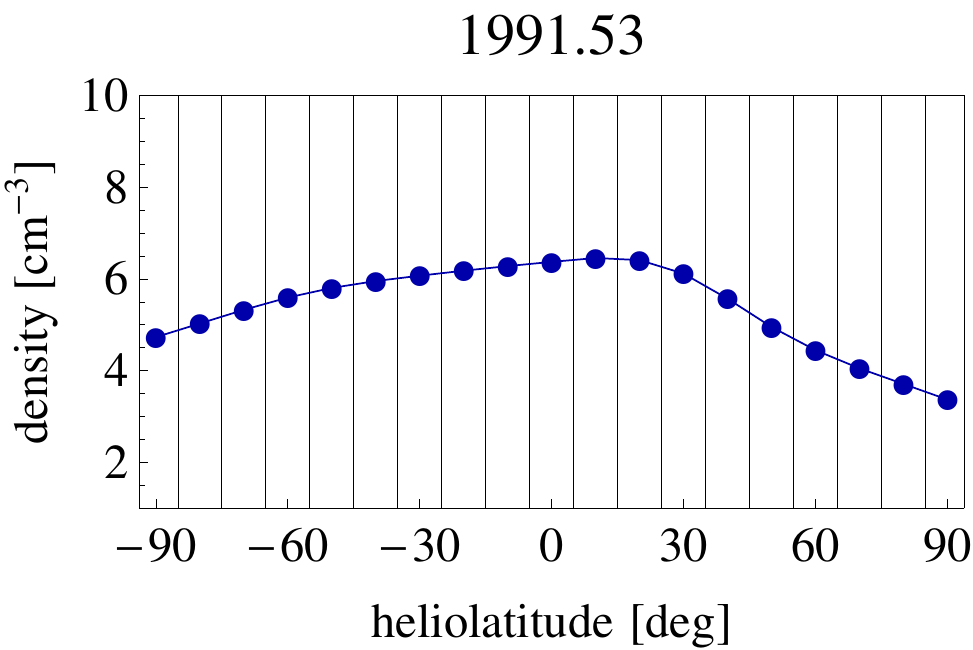}&\includegraphics[scale=0.35]{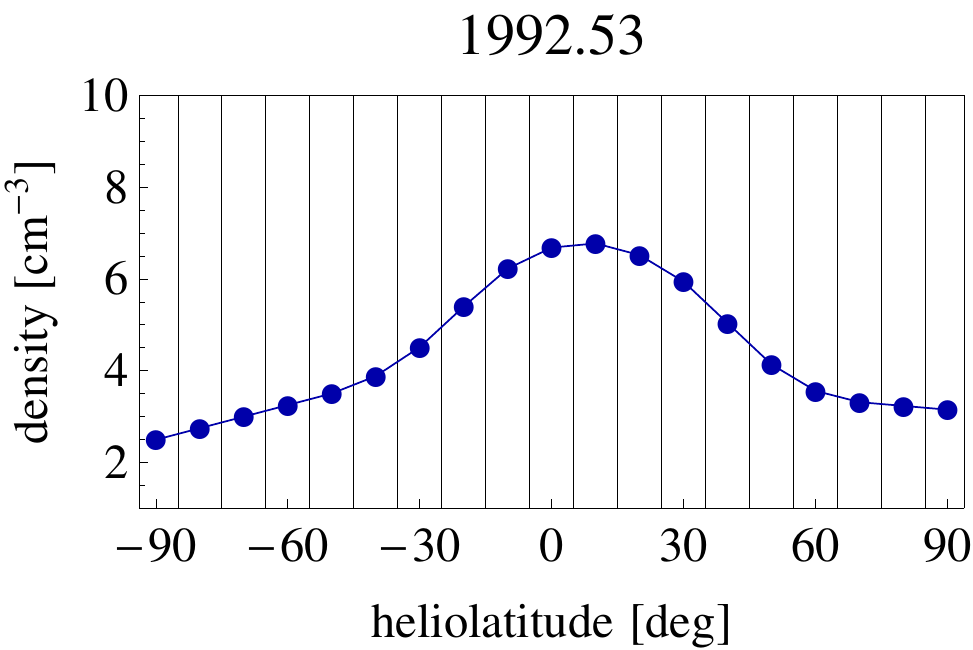}\\	\includegraphics[scale=0.35]{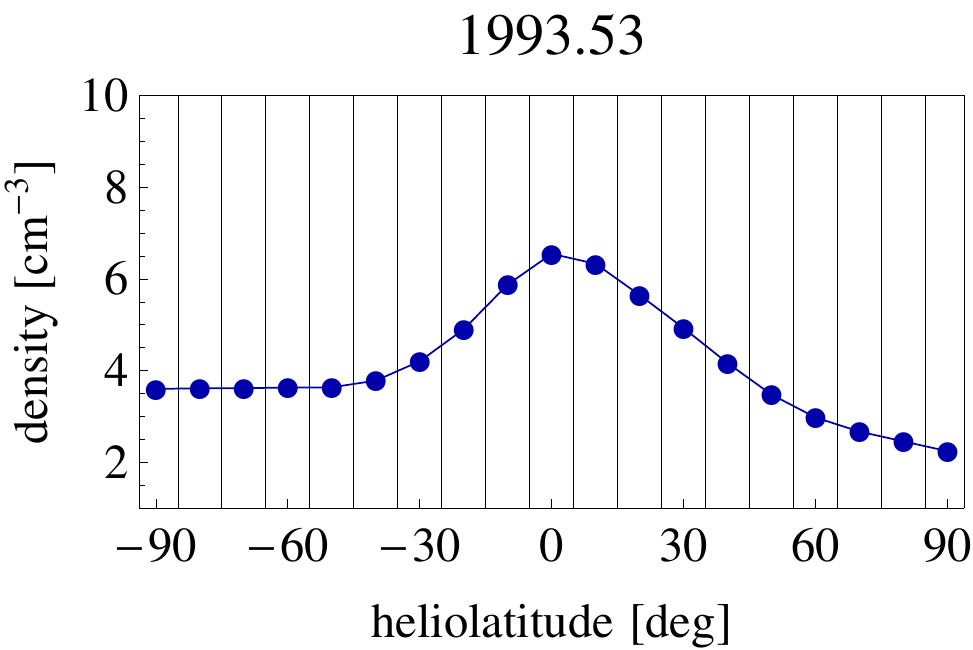}&\includegraphics[scale=0.35]{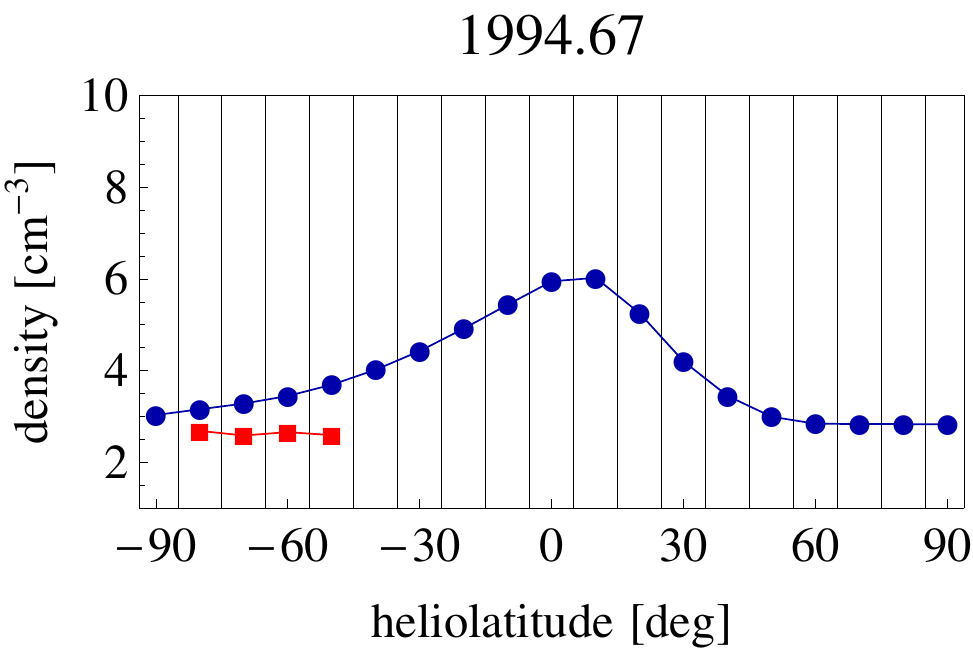}&\includegraphics[scale=0.35]{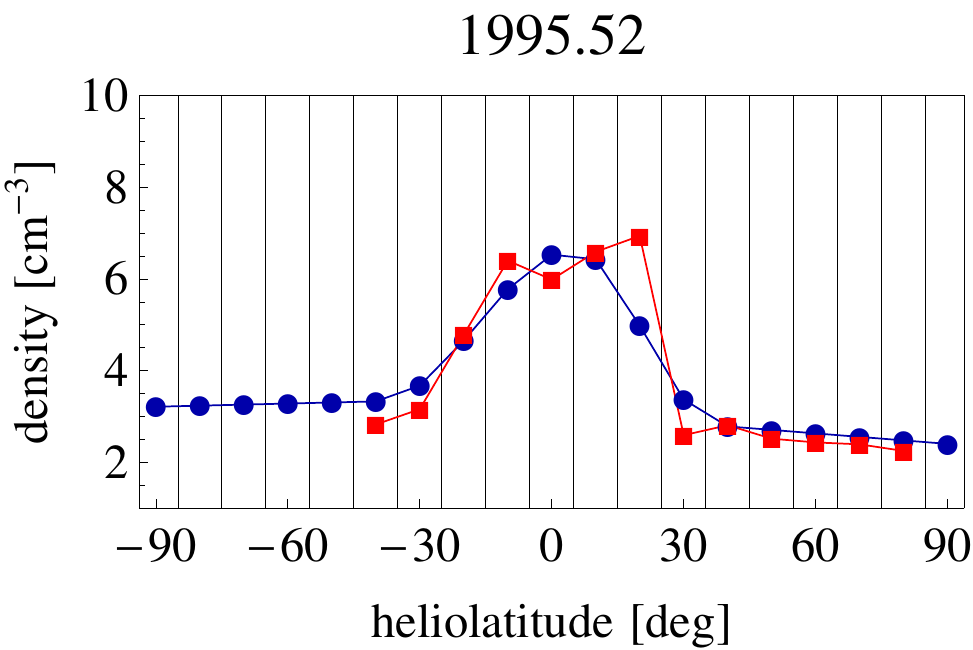}\\	\includegraphics[scale=0.35]{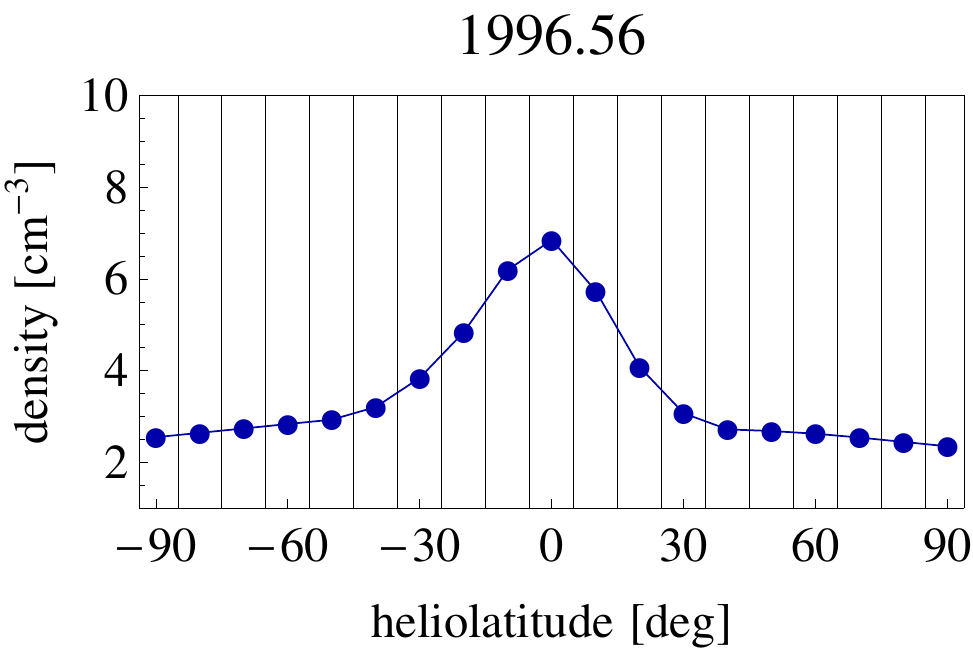}&\includegraphics[scale=0.35]{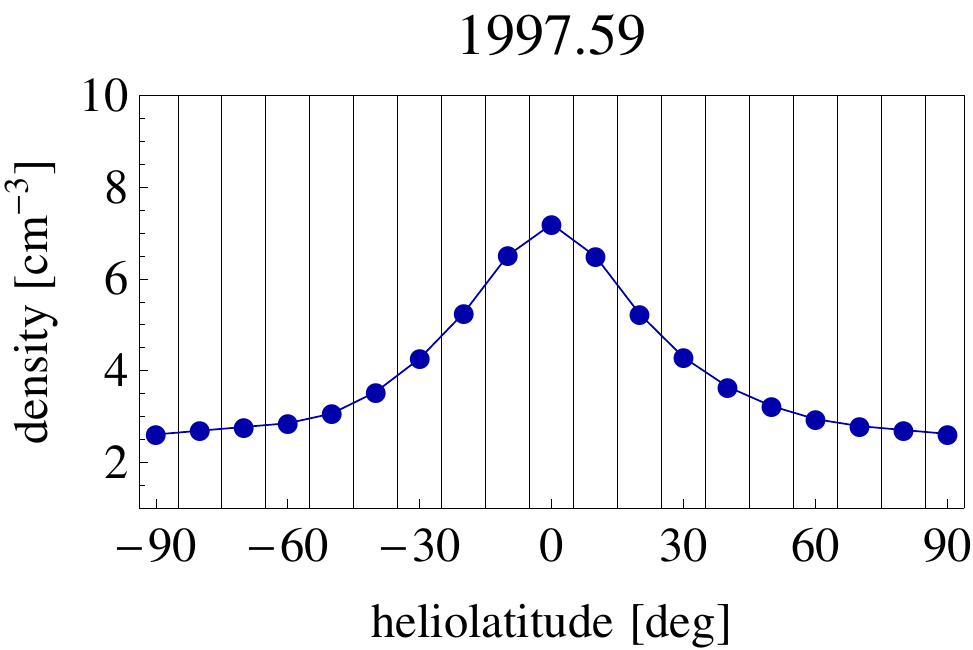}&\includegraphics[scale=0.35]{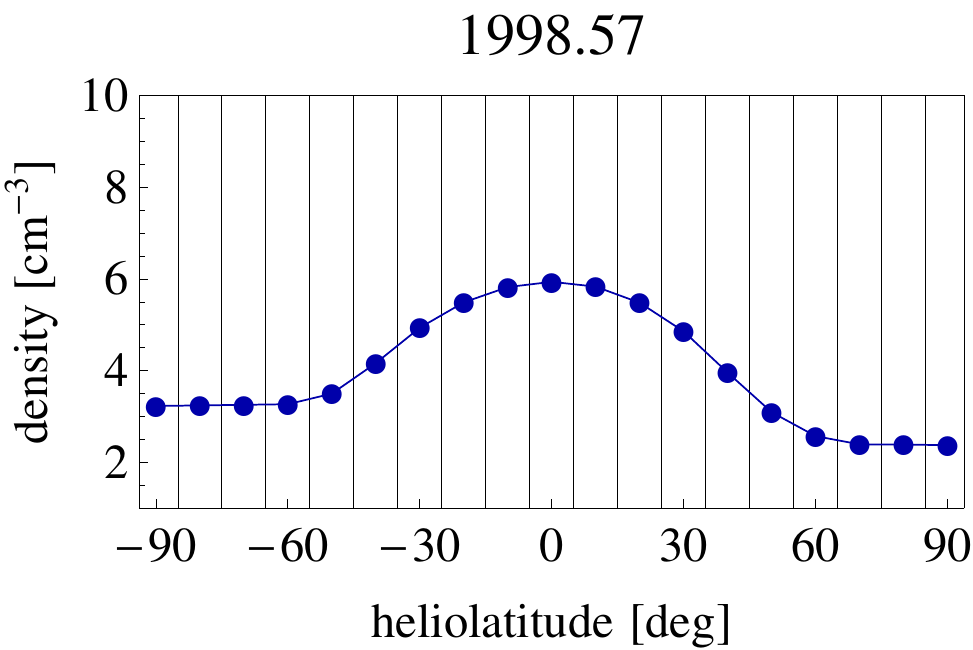}\\	\includegraphics[scale=0.35]{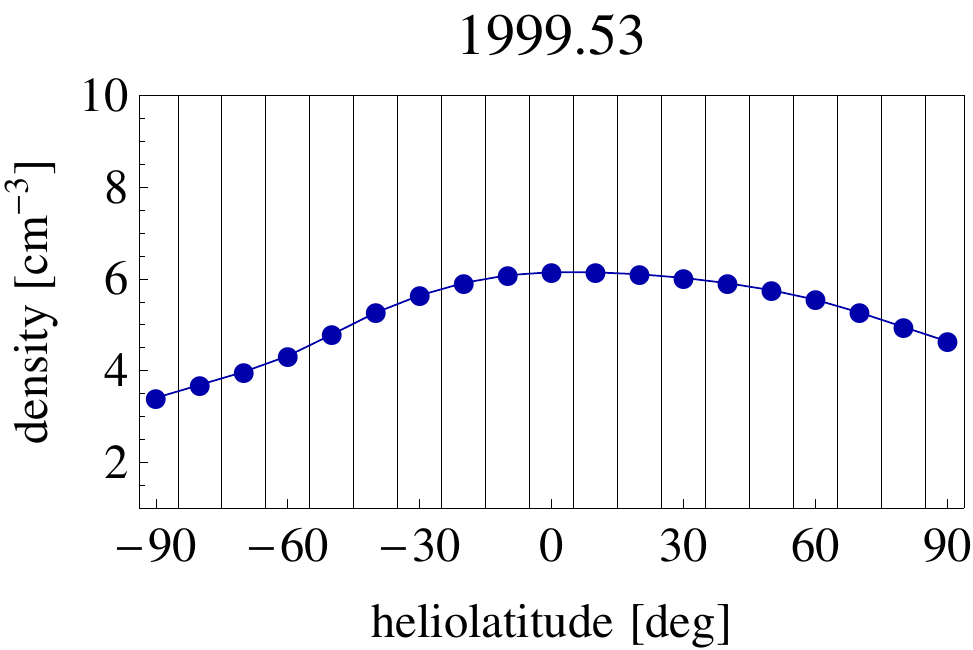}&\includegraphics[scale=0.35]{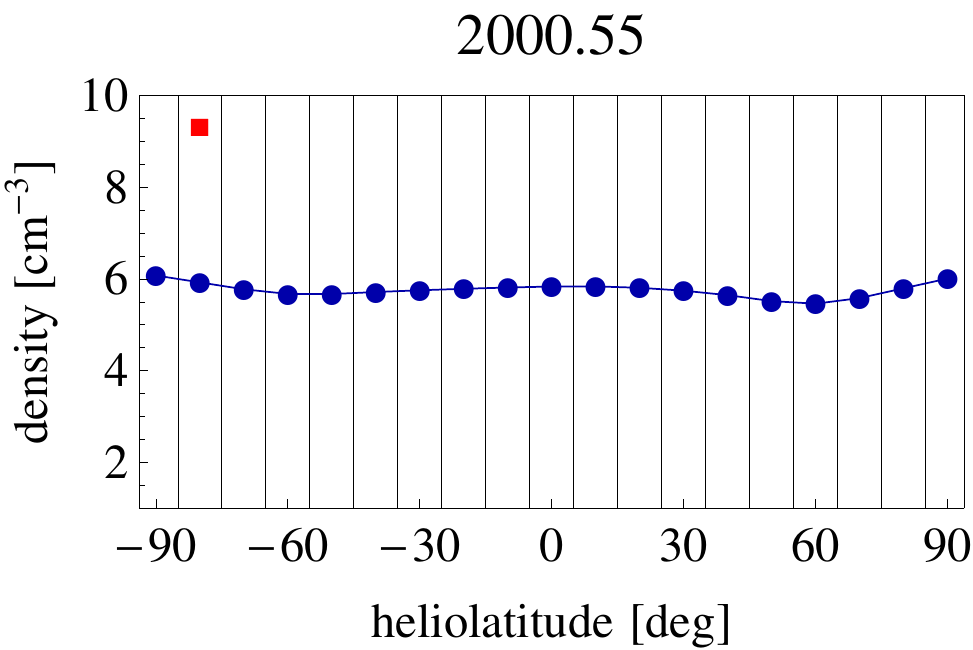}&\includegraphics[scale=0.35]{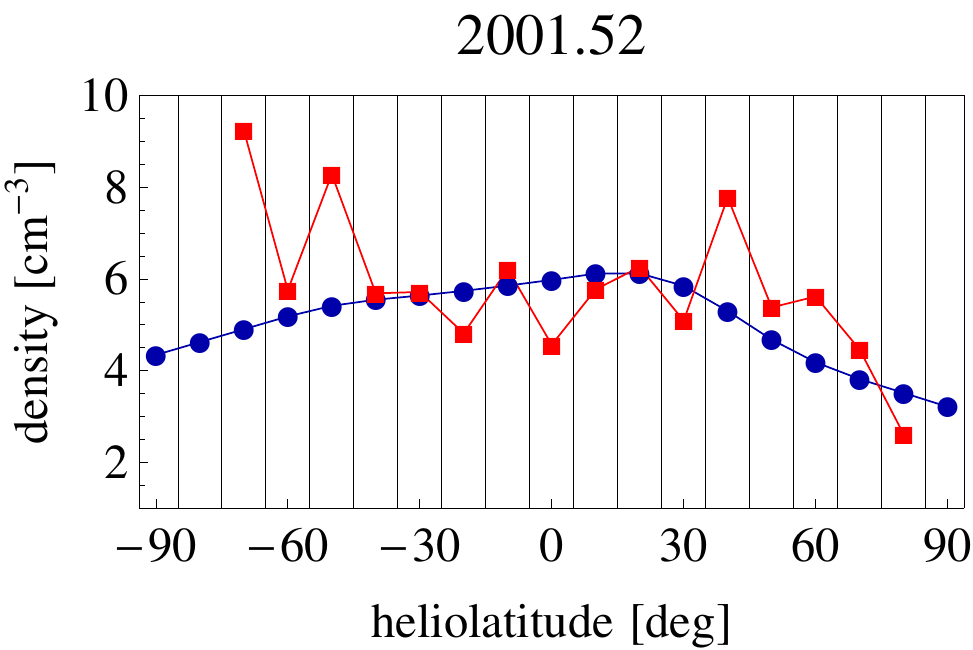}\\	\includegraphics[scale=0.35]{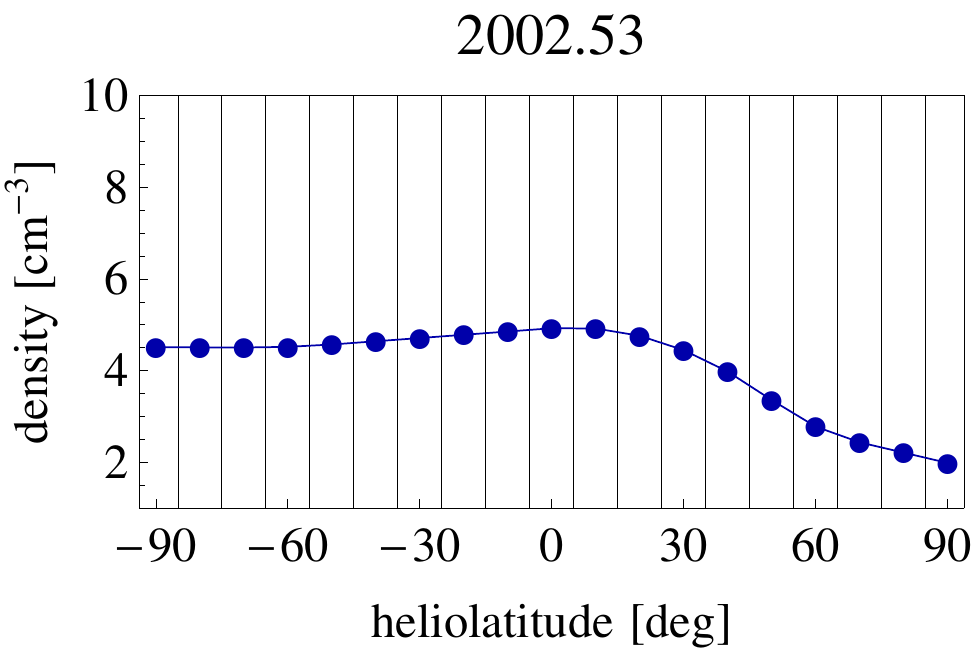}&\includegraphics[scale=0.35]{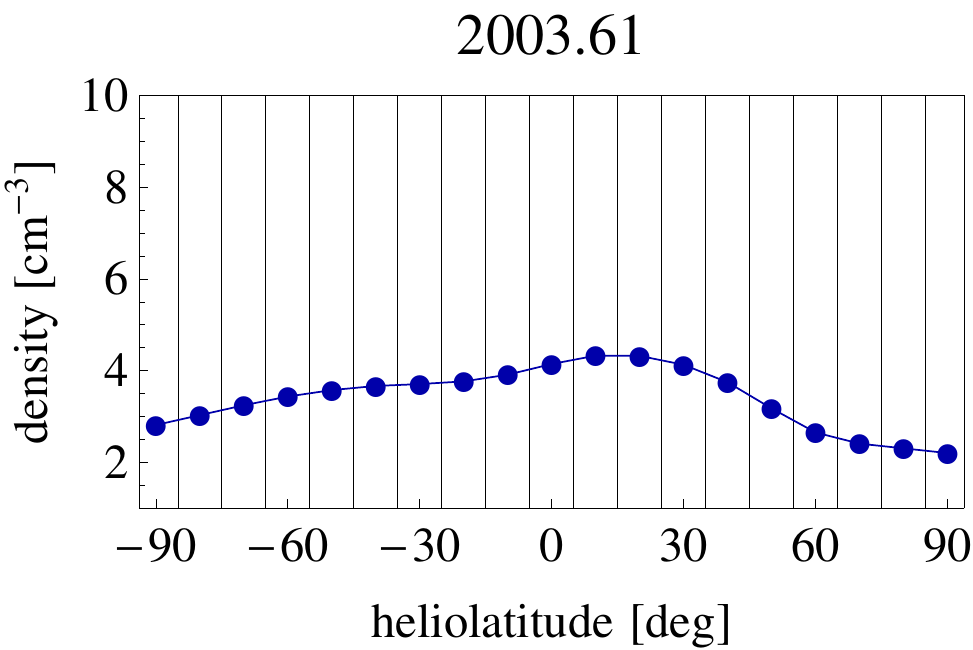}&\includegraphics[scale=0.35]{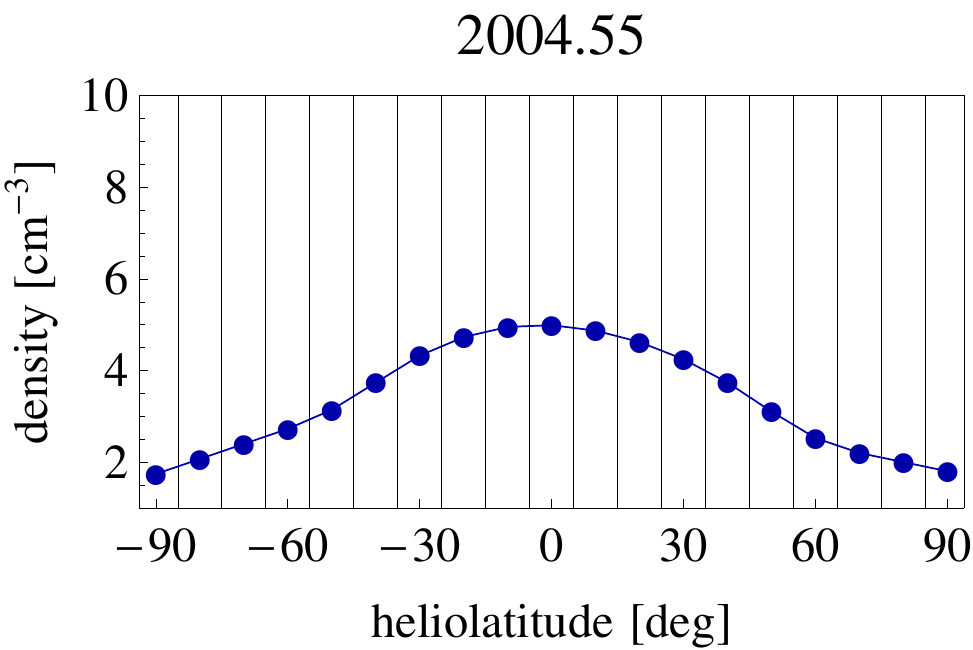}\\	\includegraphics[scale=0.35]{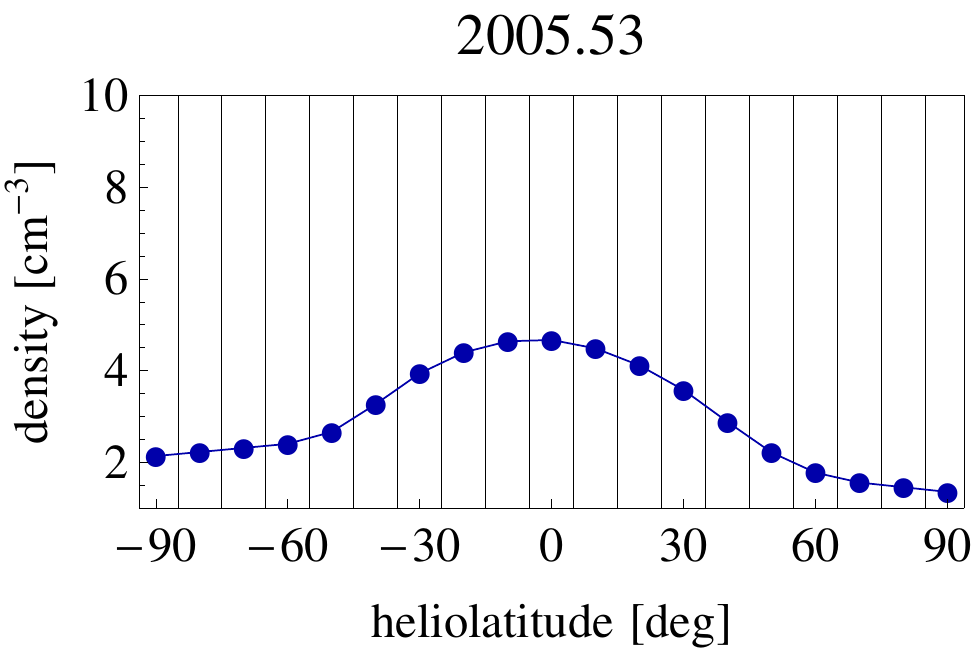}&\includegraphics[scale=0.35]{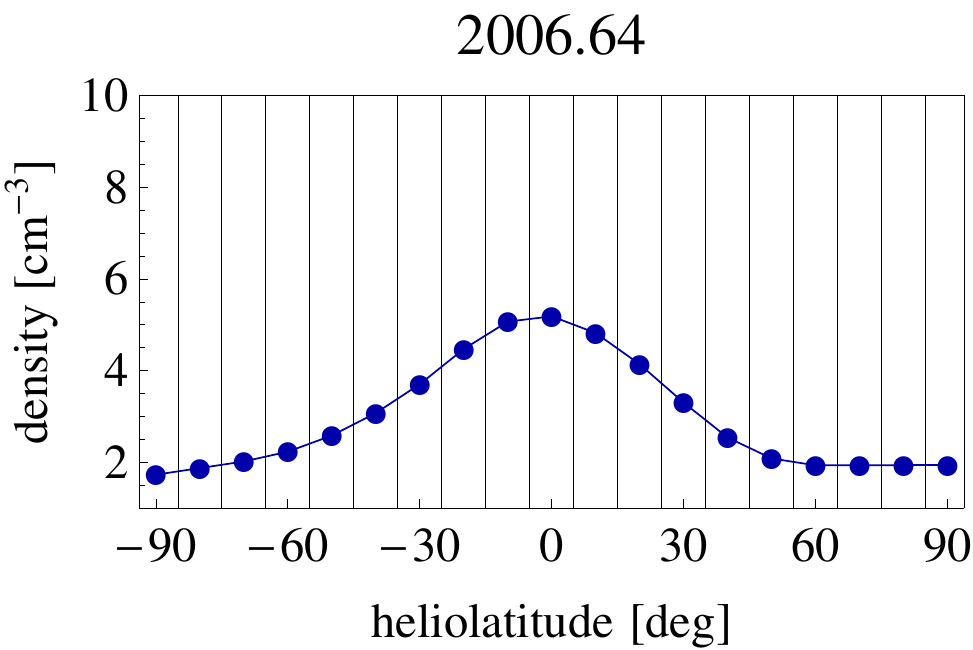}&\includegraphics[scale=0.35]{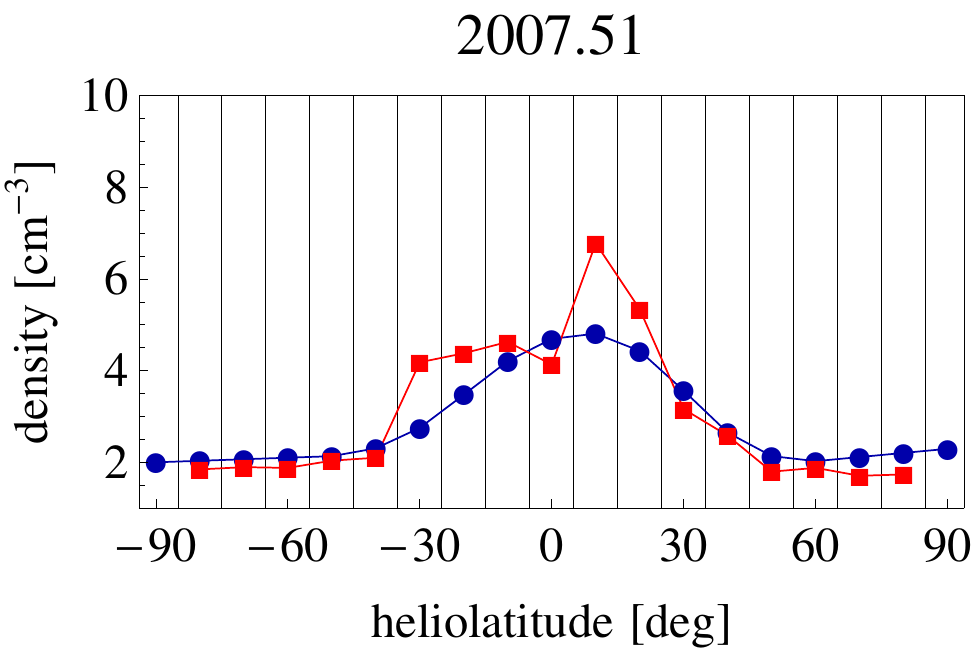}\\	
\includegraphics[scale=0.35]{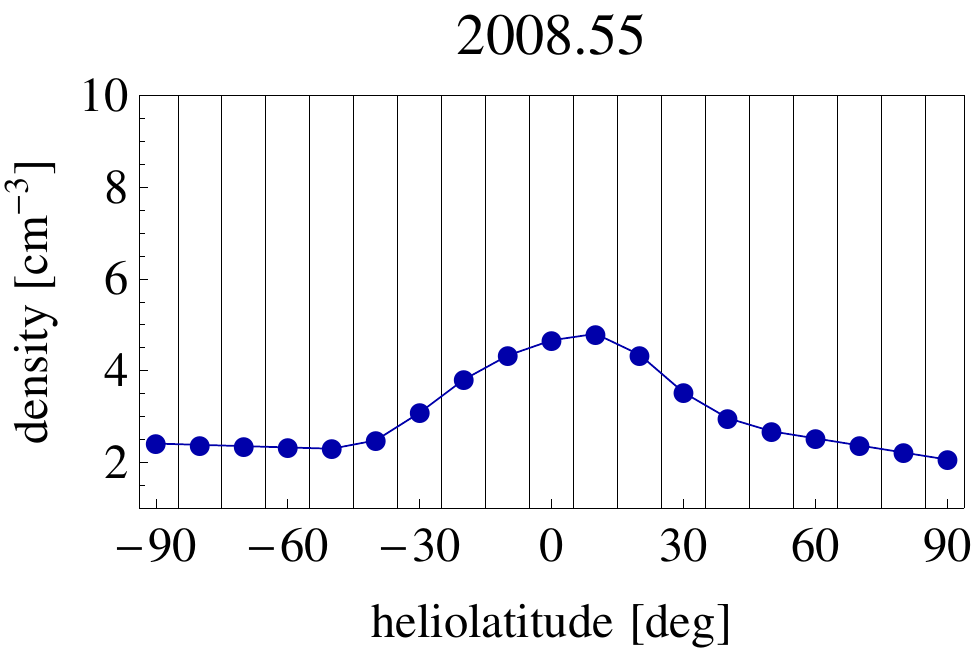}&\includegraphics[scale=0.35]{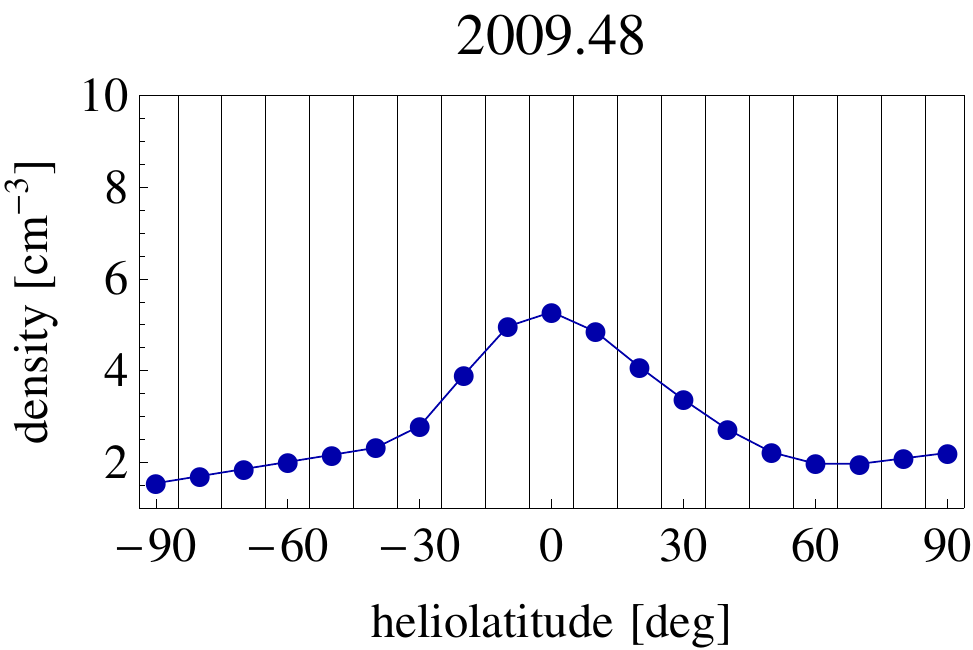}&\includegraphics[scale=0.35]{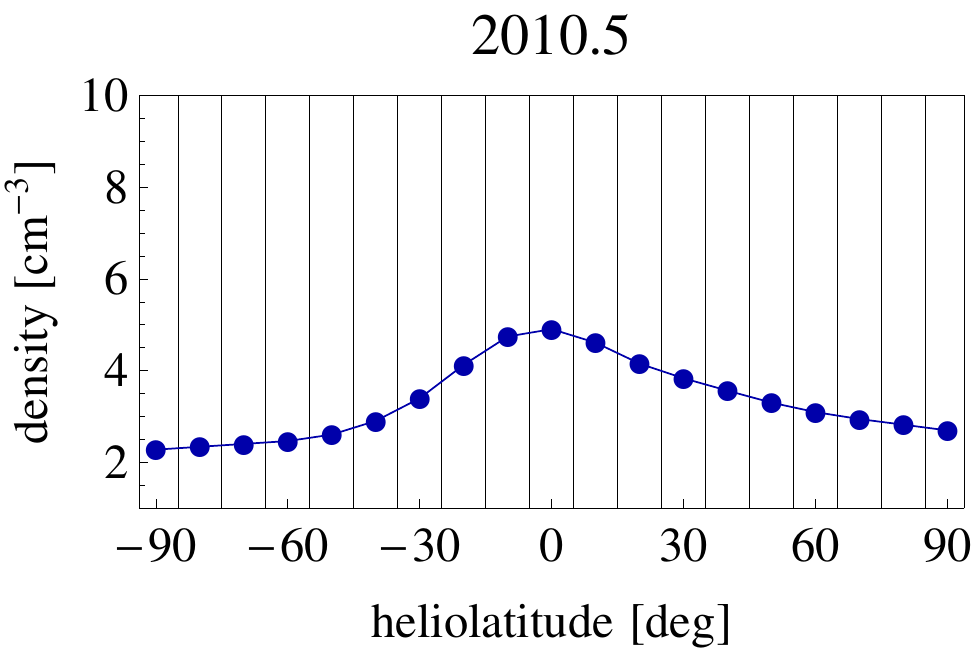}\\
\includegraphics[scale=0.35]{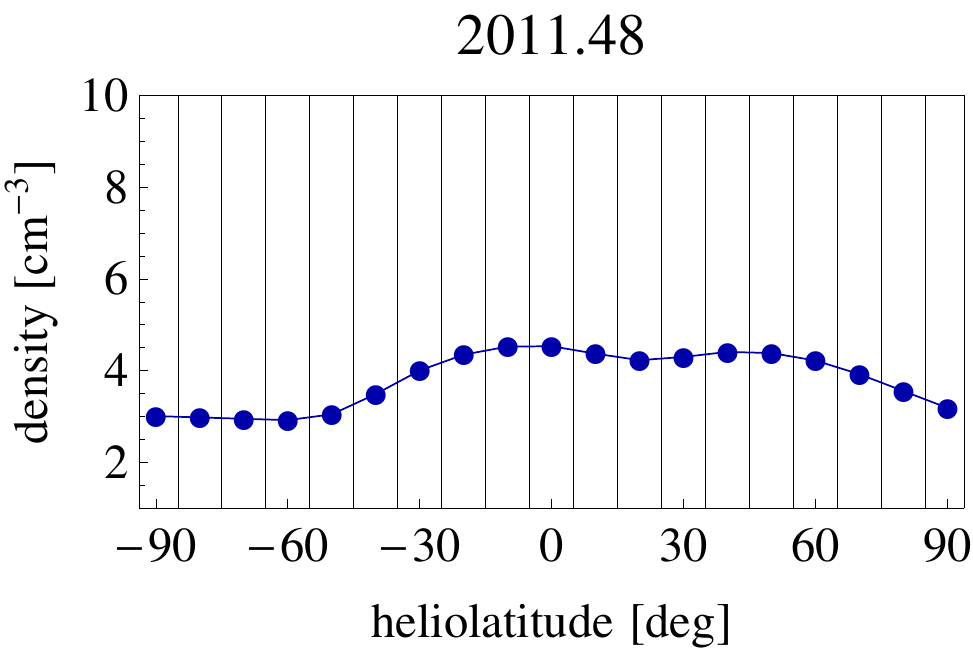}&&\\
		\end{tabular}
		\caption{Heliolatitude yearly profiles of solar wind density obtained from the density-speed correlation discussed in the text adopted to the smooth speed profiles obtained in the previous subsection. The red dots and lines show the corresponding parts of the \textit{Ulysses} fast scan profiles, similarly as in Figure~\ref{figIPSprofiles}. The average times for the profiles are indicated in the panel headers.}
		\label{figResDens}
		\end{figure*}

We calculate the yearly profiles of solar wind density as a function of heliolatitude by applying Equation~\ref{eqUlyDensCorr} to the smoothed speed profiles reported in the preceding section. The results are shown in Figure~\ref{figResDens}. Despite the limitations of the correlation formulae the agreement with \textit{Ulysses} profiles is quite good, especially for polar regions during the third fast scan and the northern limb during the first fast scan (see Figure~\ref{figCompareModelWithUlysses}). There is a slight difference between \textit{Ulysses} measurements and model density results for solar maximum in the south polar region in 2000 and 2001, which might be due the following reasons: the profile presents the yearly average value and \textit{Ulysses} the current conditions on the Sun; the data are for solar maximum, when the conditions on the Sun change dynamically in short time scales; the yearly value from IPS is calculated only from about $\sim~8$~CRs without winter months and \textit{Ulysses} sampled the highest south latitudes just in winter months; and the correlation formula adopted for the solar maximum is retrieved from the solar minimum conditions and thus may be not fully adequate.

\subsection{Interpolation in time}
The last step to retrieve the solar wind structure (both in speed and density) as a function of heliolatitude and time is a linear interpolation to nodes at halves of Carrington rotations. To be consistent with the direct measurement in the ecliptic plane we replace the equatorial bin obtained from the presented analysis with the CR-averaged time series from OMNI-2, linearly interpolated to halves of CRs. The $\pm 10^\circ$ bins are replaced with values linearly interpolated between the $\pm 20^\circ$ bins (respectively) and the equatorial bin. Because the data from \textit{Ulysses} are available only to $\sim 80^\circ$ heliolatitude and the IPS data at $\pm 90^\circ$ bins are scarce and thus not fully reliable, we calculate the polar values from a parabolic interpolation between the $\pm 70^\circ$ and $\pm 80^\circ$ bins. We fit a parabolic curve to the points from bins $\pm 70^\circ$ and $\pm 80^\circ$ and their mirror reflection around the appropriate pole and we calculate the values for the $\pm 90^\circ$ heliolatitude from the fits.

As a result of such a treatment, we have utilized all available information on the equatorial band of solar wind traversed by the Earth during its yearly orbital motion, and away from the equatorial band, where such detailed information is not available, we have a smooth transition into the region of the low time-resolution model. At the poles, because of the extra smoothing procedure, we avoid a conical sharp peak, having latitudinal gradient precisely zero.

The one-year gap for 2010, when data on solar wind speed from IPS analysis are not sufficient to retrieve solar wind speed, we fill in by the average value calculated from the straddling years 2009 and 2011.

		\clearpage
	
\section{Results}
\begin{figure}
		\centering	
		\includegraphics[scale=0.65]{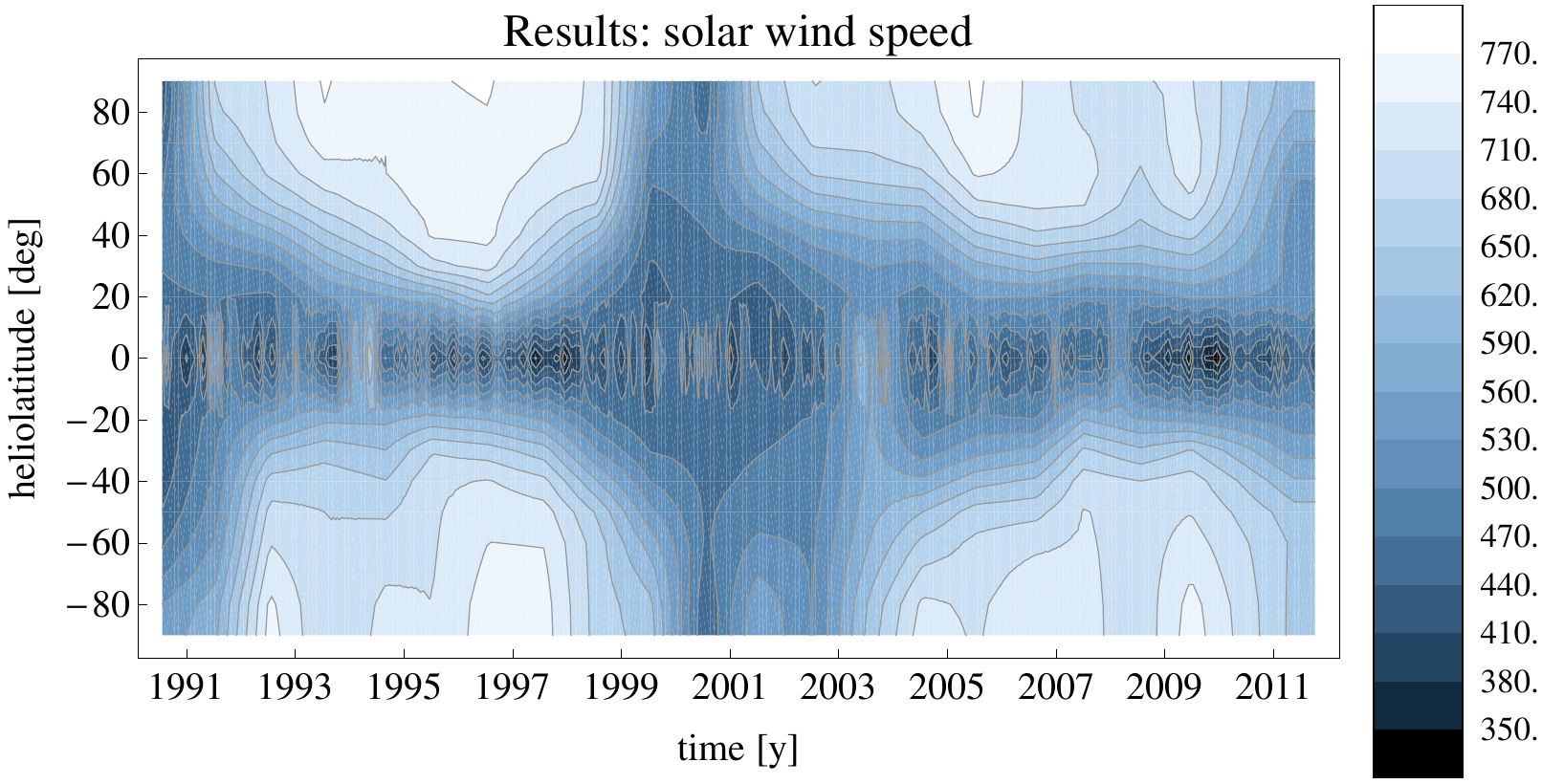}\\
		\includegraphics[scale=0.65]{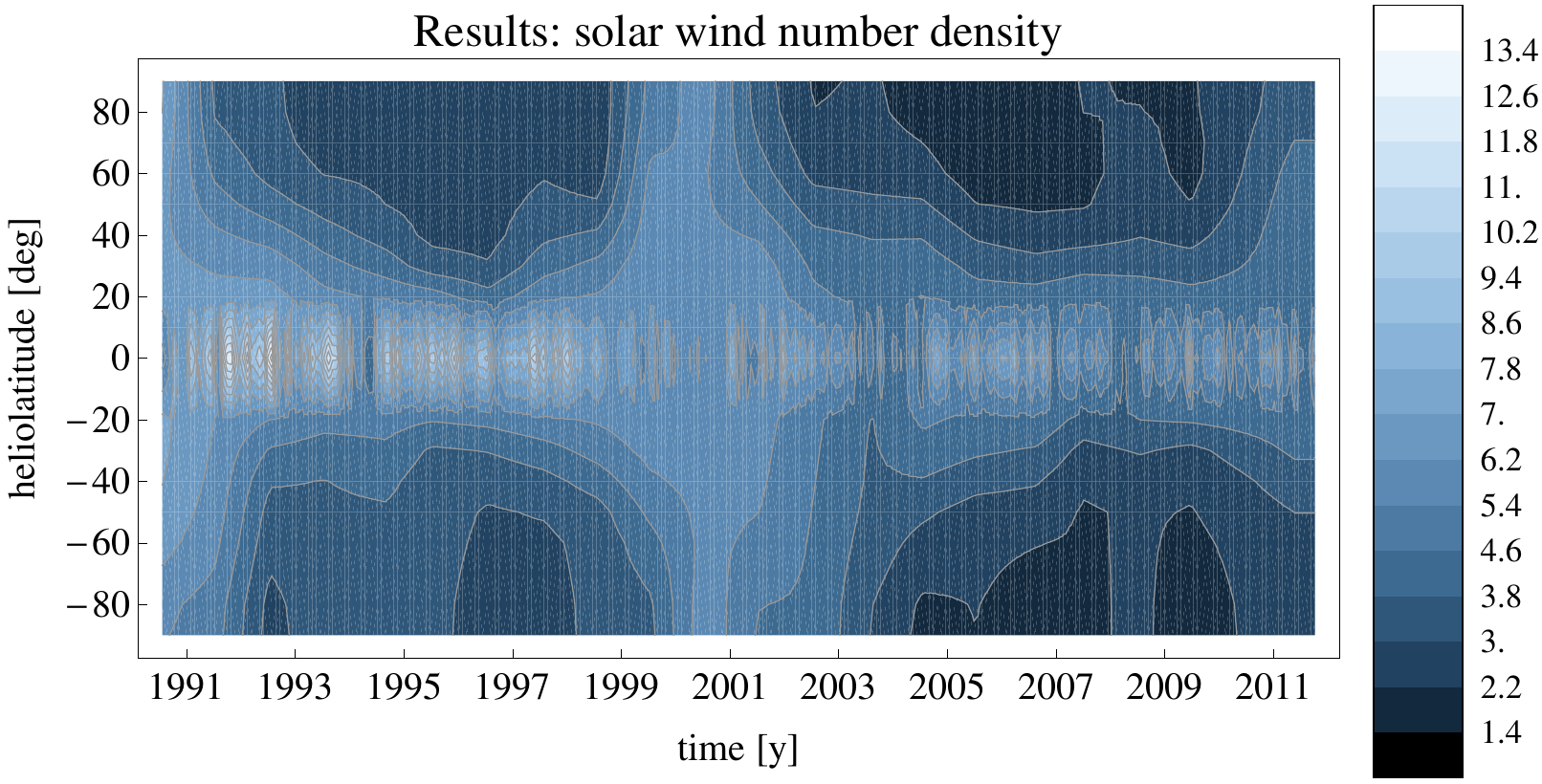}\\
		\caption{Contour maps of the solar wind speed (in km~s$^{-1}$) and number density adjusted to 1~AU (in cm$^{-3}$) shown as a function of time and heliolatitude.}
		\label{figResSpeedDens}
		\end{figure}

The primary results of this study are presented in Figure~\ref{figResSpeedDens}, which shows heliolatitude vs time maps of solar wind speed and number density adjusted to 1~AU. The results confirm that solar wind speed is bimodal during solar minimum, slow at latitudes close to solar equator (and thus to the ecliptic plane) and fast at the poles. 

The heliolatitude structure evolves with the solar activity cycle and becomes flatter when the activity is increasing. The structure is approximately homogeneous in heliographic latitude only during a short time interval during the peak of solar activity, when solar wind at all heliolatitudes is slow (see the panel for 2000 in Figure~\ref{figIPSprofiles}) and highly variable. 

Shortly after the activity maximum the bimodal structure reappears and the fast wind at the poles is observed again, but switchovers from the slow to fast wind close to the poles may still occur during the high activity period: compare the panels for 2001, 2002 and 2003 in the aforementioned figure and see the solar activity level depicted with the blue line in Figure~\ref{figUlyssesComposite}. 

During the descending and ascending phases of solar activity there is a wide band of slow and variable solar wind that is on both sides of the equator and extends to midlatitudes. The fast wind is restricted to polar caps and upper midlatitudes. 

At solar minimum, the structure is sharp and stable during a few years straddling the turn of solar cycles, with high speed at the poles and at midlatitudes and a rapid decrease at the equatorial band. Thus, apart from short time intervals at the maximum of solar activity, the solar wind structure close to the poles is almost flat, with a steady fast speed value typical for wide polar coronal holes, which is in perfect agreement with the measurements from \textit{Ulysses} \citep{phillips_etal:95c, mccomas_etal:00a, mccomas_etal:06a}. 

The bimodal structure of solar wind speed is also reconstructed in solar wind density. The variable dense flows are at low latitudes and rarefied near the poles. The solar wind density changes are anticorrelated with speed. The dense plasma flows are recorded at all heliographic latitudes only during the peak of solar maximum phase (see 2000 in Figure~\ref{figResDens}). For other years the solar wind with a low number density appears at higher latitudes with the minimum values during solar minimum. 

Figure~\ref{figResDens} also shows that during minimum of SC~23 the structure of density was more narrow around equator than it was during the last minimum of SC~24. It seems that the slow and dense plasma flows typical for solar minimum conditions extend to higher latitudes (about 10$^\circ$ farther) than it was during the previous cycle. It means that the secular changes in solar wind density are very well reflected in our results (the profile width at midlatitudes is wider during the most recent years). 

The maps of solar wind speed and density in Figure~\ref{figResSpeedDens} show also a slight hemispheric asymmetry, that seems to reverse from one SC to another.

The evolution of solar wind parameters as a function of time and heliolatitude is needed in the modeling of the heliosphere and in the interpretation on measurements such as ENAs observations by IBEX \citep{mccomas_etal:09a} and heliospheric Lyman-$\alpha$ glow analysis \citep{quemerais_etal:06b, lallement_etal:10b}. Having the density and speed we can easily calculate solar wind flux:
		\begin{equation}
		\label{eqFluxDef}		
		F \left(\phi_j, t_i\right)= v_{\mathrm{p}}\left( \phi_j, t_i \right) n_{\mathrm{p}} \left( \phi_j, t_i \right),
		\end{equation}
and charge-exchange rate between solar wind protons and stationary H atoms:
		\begin{equation}
		\label{eqCXrateDef}		
		\beta_{\mathrm{CX}} \left(\phi_j, t_i\right)= n_{\mathrm{p}} \left( \phi_j, t_i \right) v_{\mathrm{p}} \left( \phi_j, t_i \right) \sigma_{\mathrm{CX}}\left( v_{\mathrm{p}}\left(\phi_j, t_i \right) \right),
		\end{equation}
where $\sigma_{\mathrm{CX}}$ is the cross section for charge-exchange rate 		 \citep{lindsay_stebbings:05a}, and dynamic pressure:
		\begin{equation}
		\label{eqDynPressDef}		
		p_{\mathrm{dyn}} \left(\phi_j, t_i\right)= \frac{1}{2} m_p n_{\mathrm{p}} \left( \phi_j, t_i \right) v^{2}_{\mathrm{p}} \left( \phi_j, t_i \right),
		\end{equation}
where $j$ numbers the 10-degree heliolatitude bins and $i$ the number of Carrington rotation in our time grid.

\begin{figure*}[!h]
		\centering	
		\includegraphics[scale=0.65]{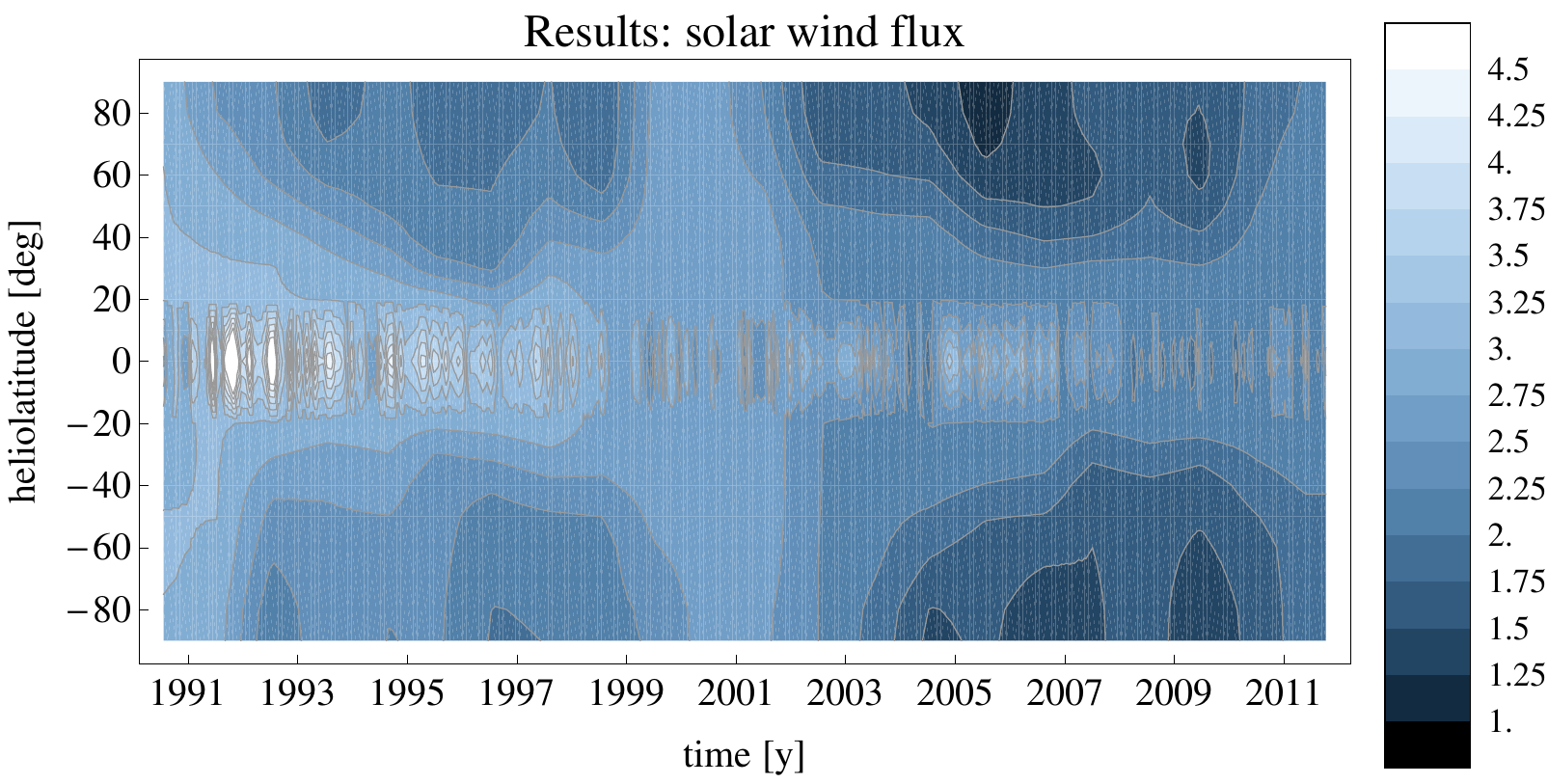}\\
		\includegraphics[scale=0.65]{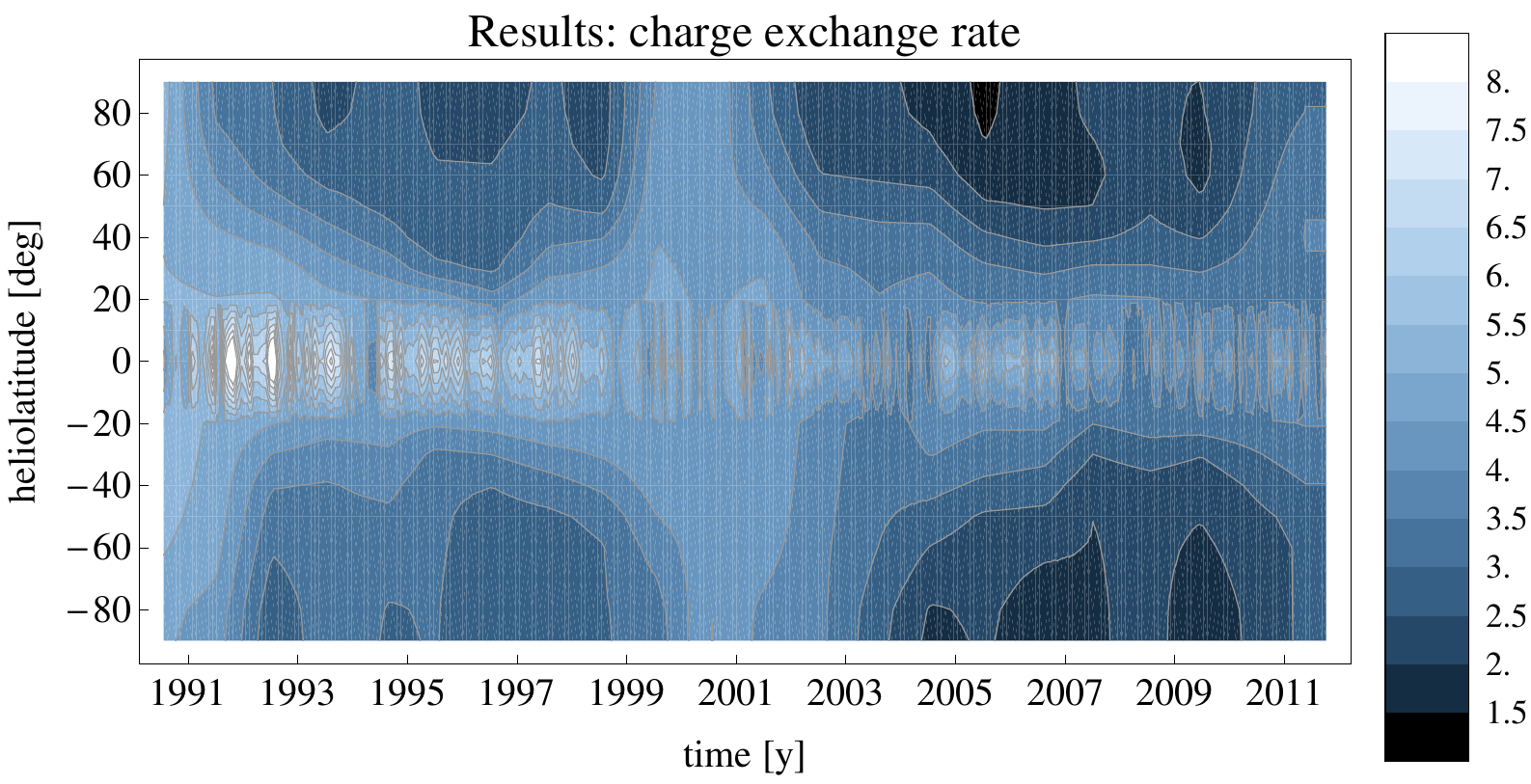}\\
		\includegraphics[scale=0.65]{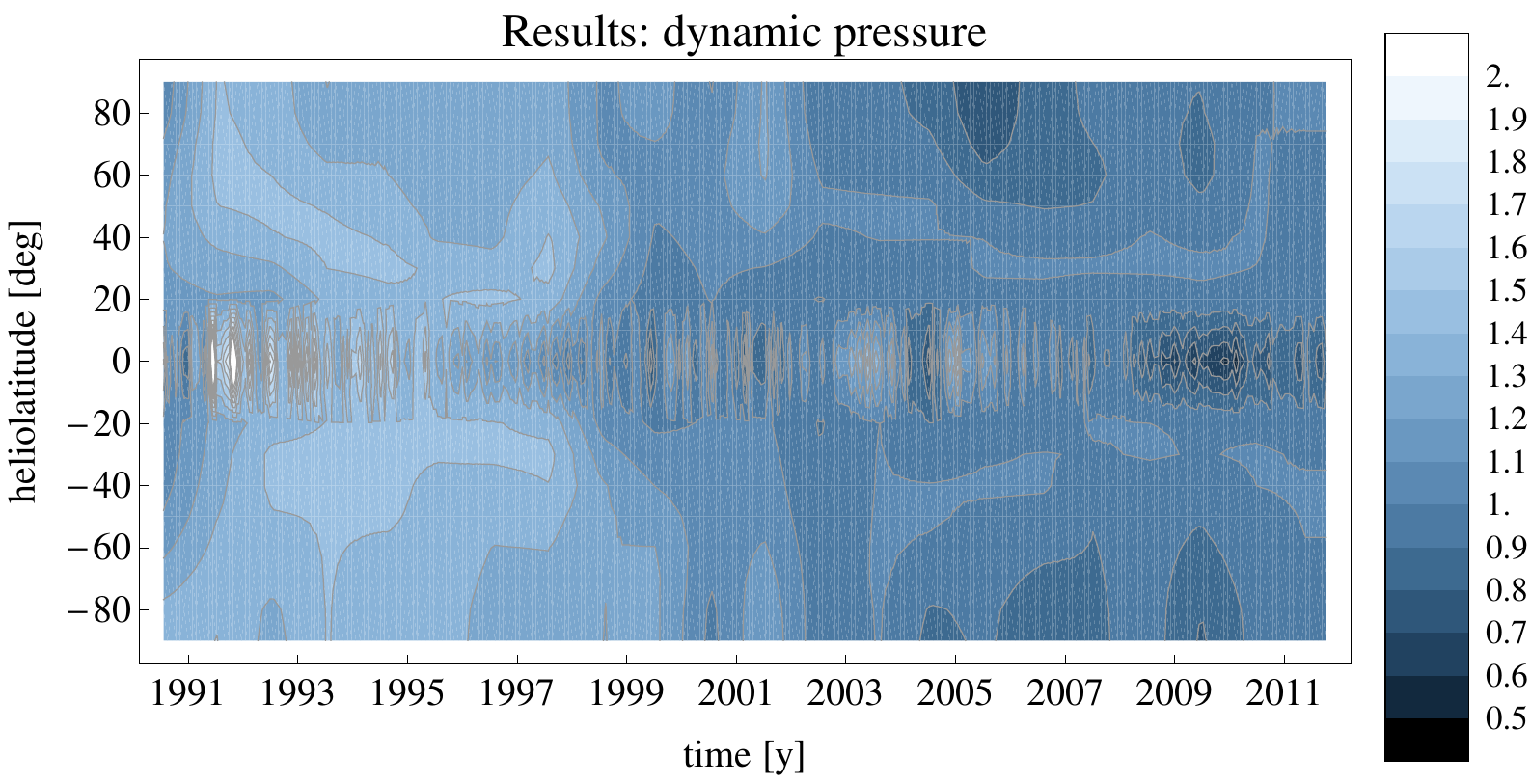}\\
		\caption{Contour maps of the solar wind flux (in $10^8$~cm$^{-2}$~s$^{-1}$), charge-exchange (in $10^{-7}$~s$^{-1}$) and dynamic pressure (in nPa) shown as a function of time and heliolatitude.}
		\label{figResFluxCXDynPress}
		\end{figure*}

Figure~\ref{figResFluxCXDynPress} shows contour maps of solar wind flux, charge-exchange rate and dynamic pressure for the years since 1990 to the end of 2011 and for heliolatitude from $-90^\circ$ to $90^\circ$. The gap in 2010 is filled by the average value calculated from 2009 and 2011. The flux features a clear secular drop after the last solar maximum. The bimodal structure during the current solar minimum seems to be even better defined than during the previous one. The structure at solar maximum is quite flat and seen longer than in the case of speed. 

The charge exchange rate basically follows the behavior of the flux, with a clear latitudinal contrast during low activity period, an almost flat structure at solar maximum, and the secular drop after the most recent solar maximum.

Dynamic pressure behaves different than the flux. While variations in time are clearly visible, the latitudinal structure is much less pronounced and does not vary much during the solar cycle. Basically, dynamic pressure is almost spherically symmetric (with possible exceptions in the polar caps, which, however, cease closer to the poles than in the case of the flux), and the most striking feature is the secular drop in the strength, which begins earlier than in the flux, namely about 1998. Such a behavior has pronounced consequences for the shape of the termination shock, which should not feature a very strong latitudinal variation in size, but which should now be significantly closer to the Sun than during the previous solar minimum. 

\section{Discussion}		
To check the credibility of our results we compare them with the \textit{Ulysses} data from all scans. The \textit{Ulysses} data are prepared by splitting the hourly data into full Carrington rotations and calculating average values. Next we linearly interpolate our model values to the times and heliolatitudes corresponding to the Carrington rotation-averaged \textit{Ulysses} data. 

\begin{figure*}[!h]
		\centering	
		\begin{tabular}{cc}
		\includegraphics[scale=0.35]{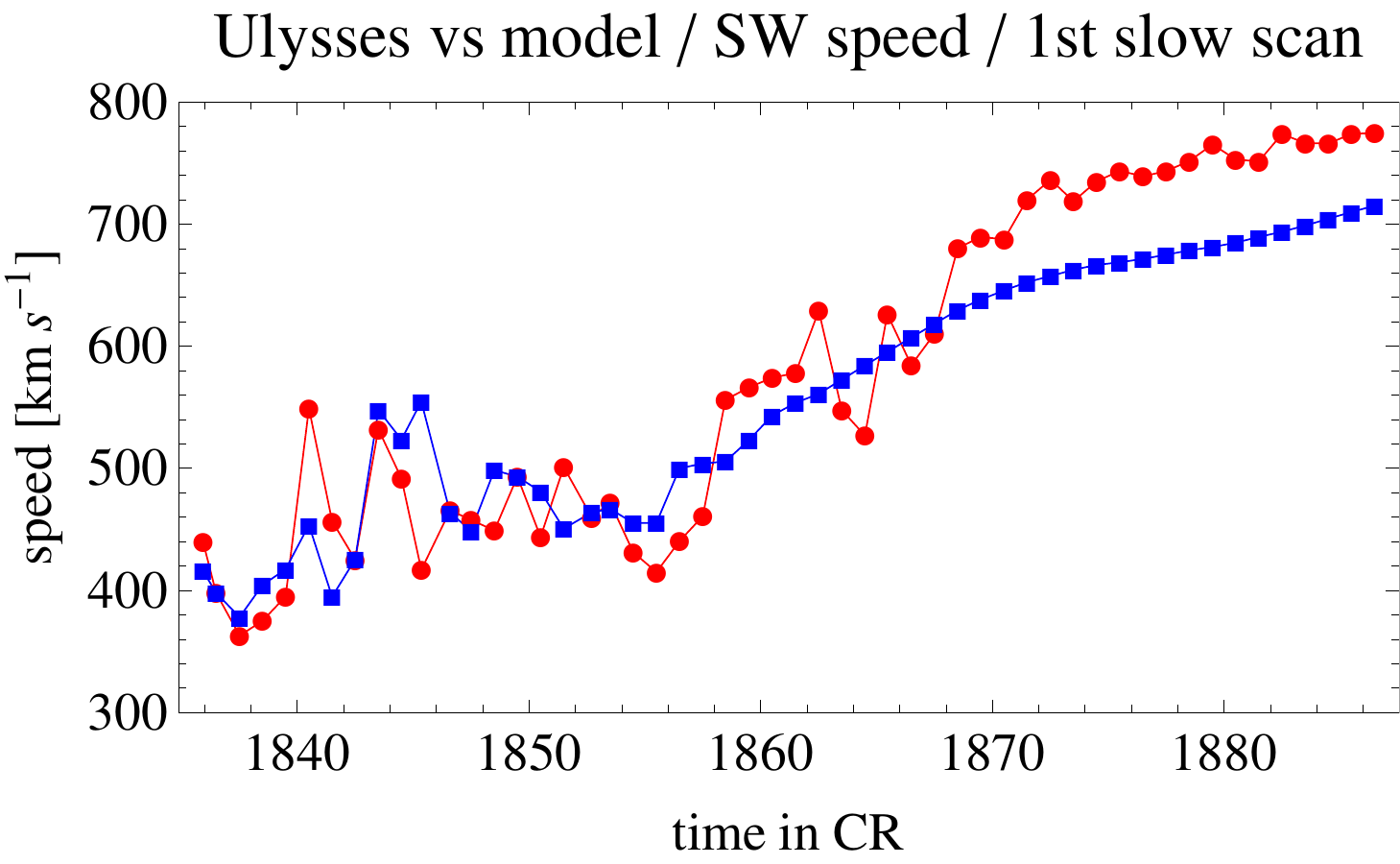} & 	\includegraphics[scale=0.35]{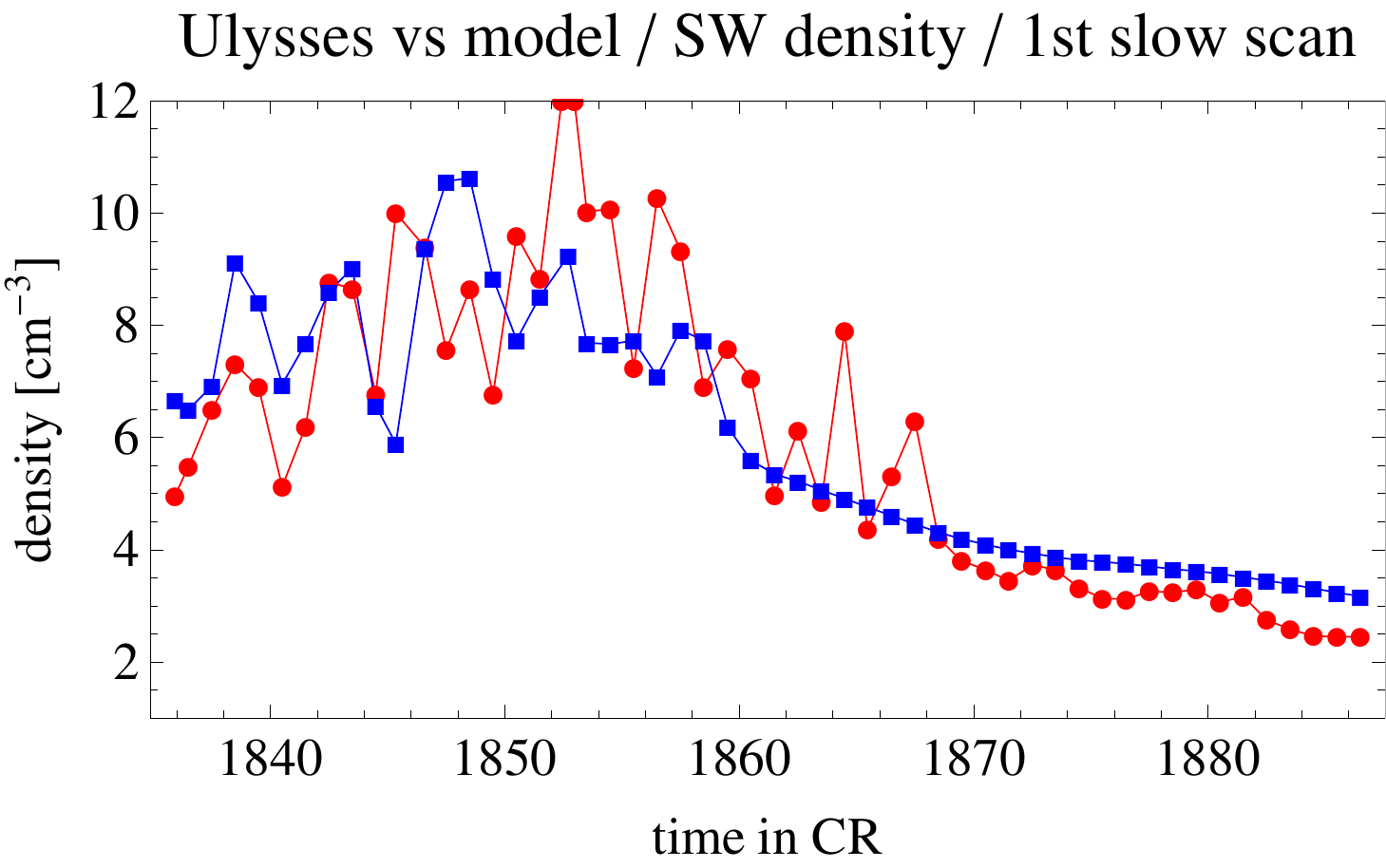}\\
		\includegraphics[scale=0.35]{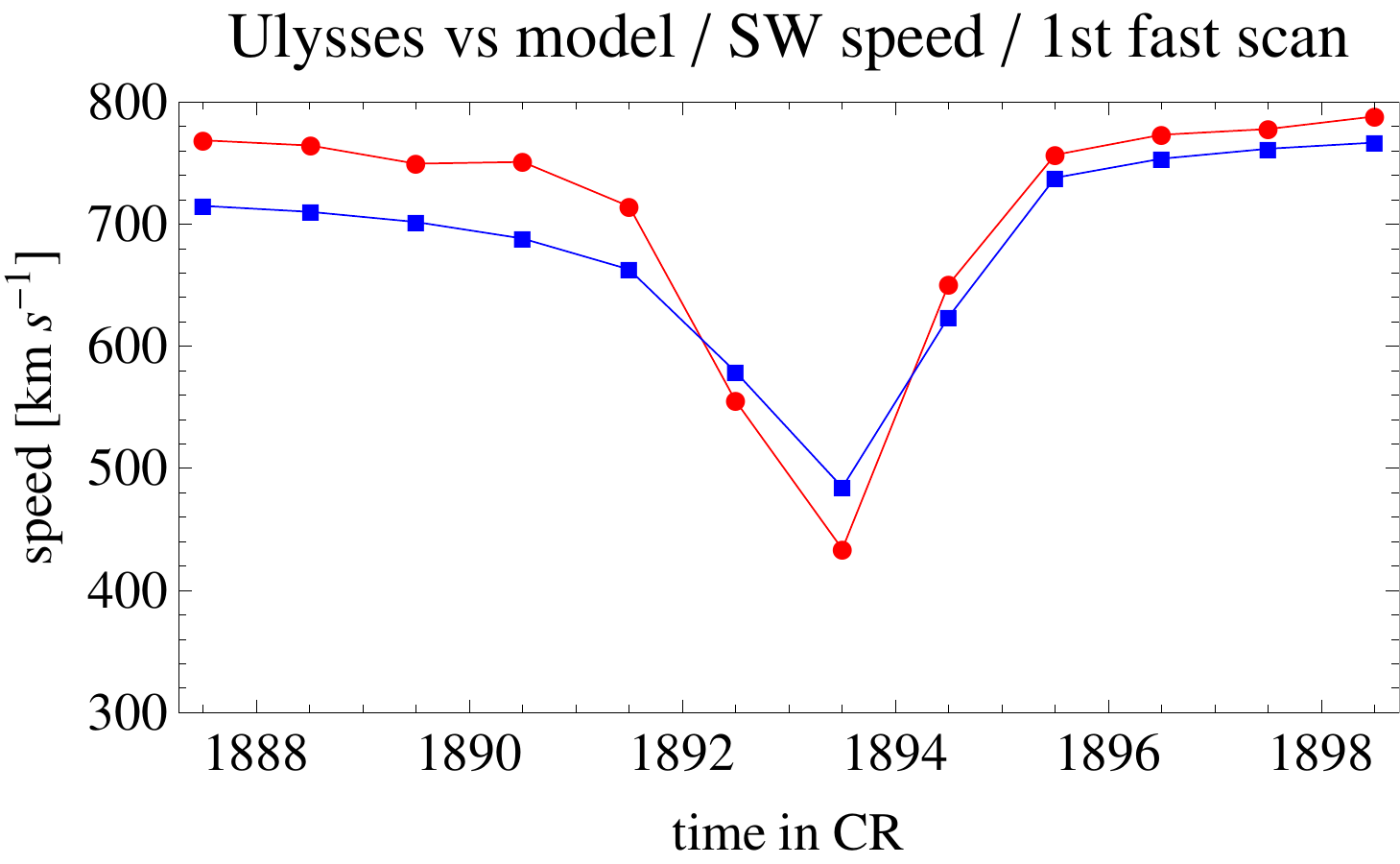} & 	\includegraphics[scale=0.35]{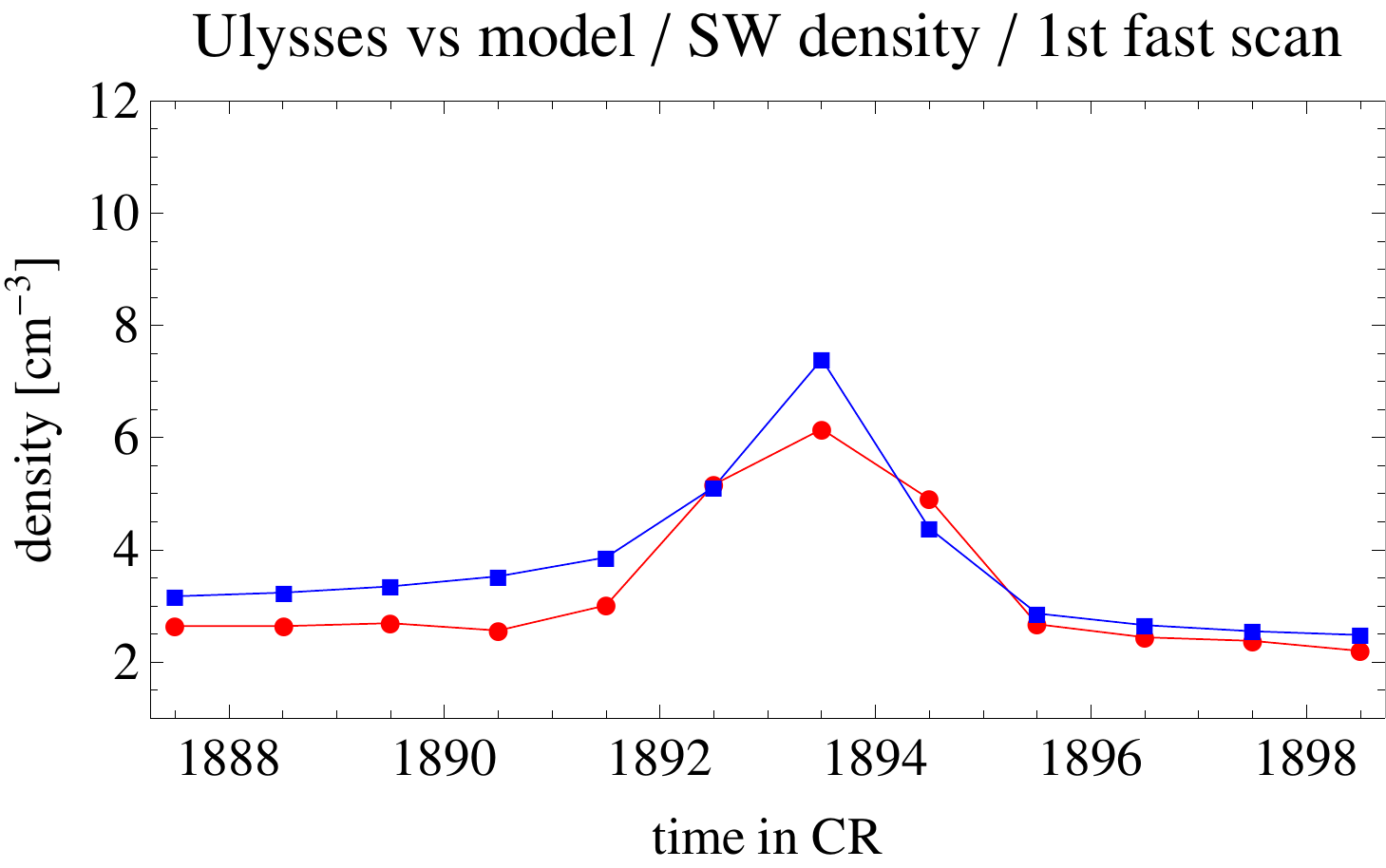}\\
		\includegraphics[scale=0.35]{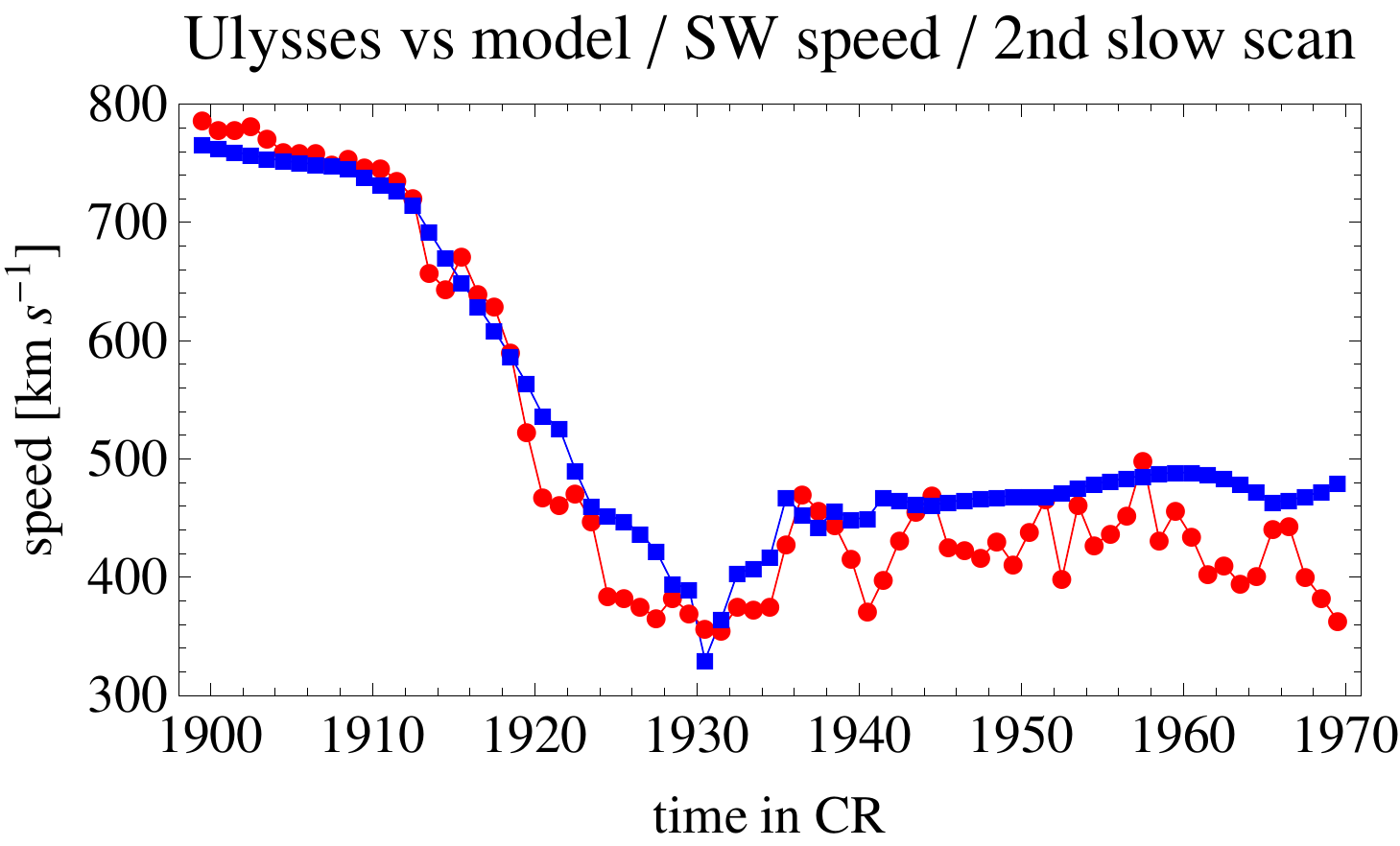} & 	\includegraphics[scale=0.35]{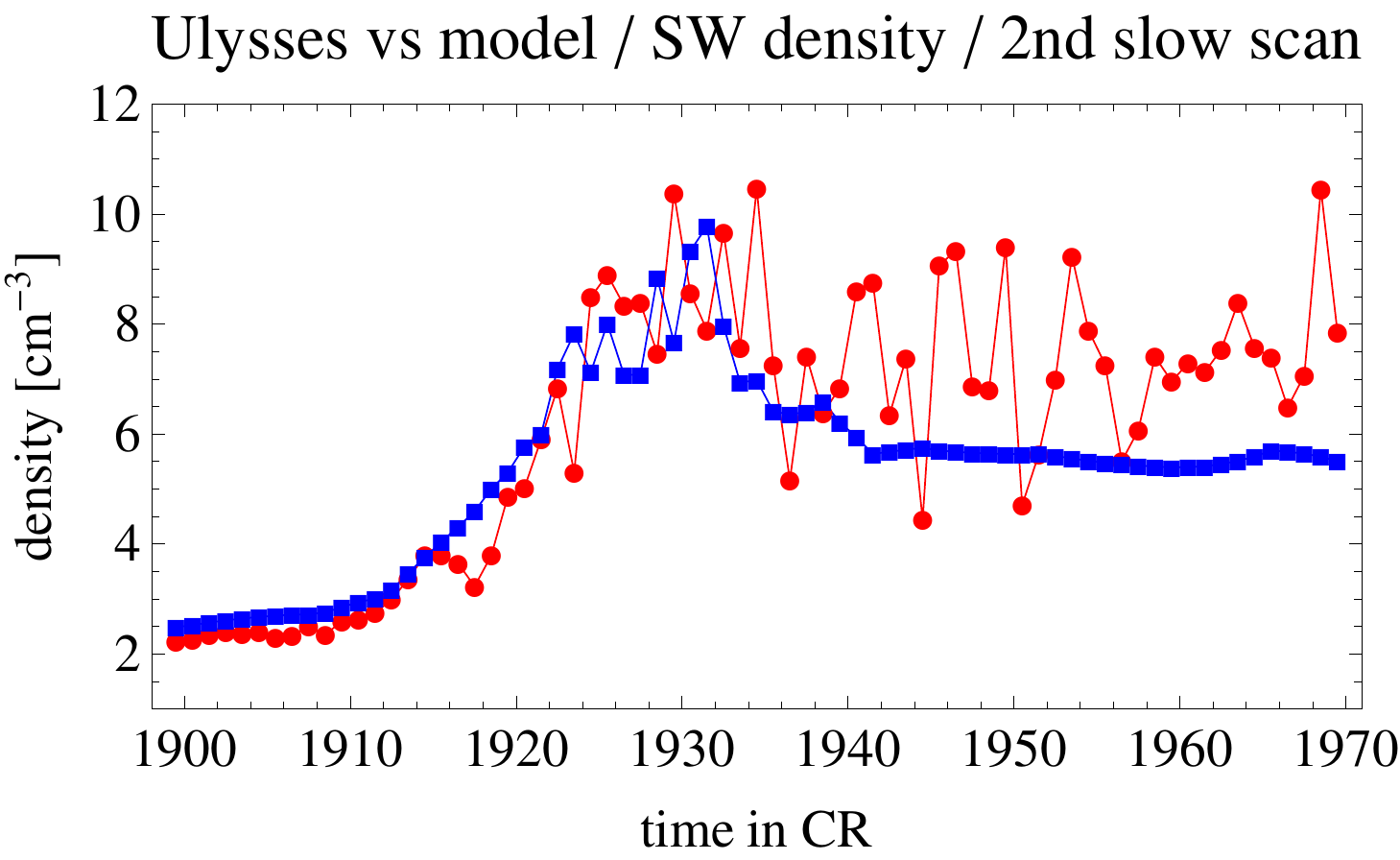}\\
		\includegraphics[scale=0.35]{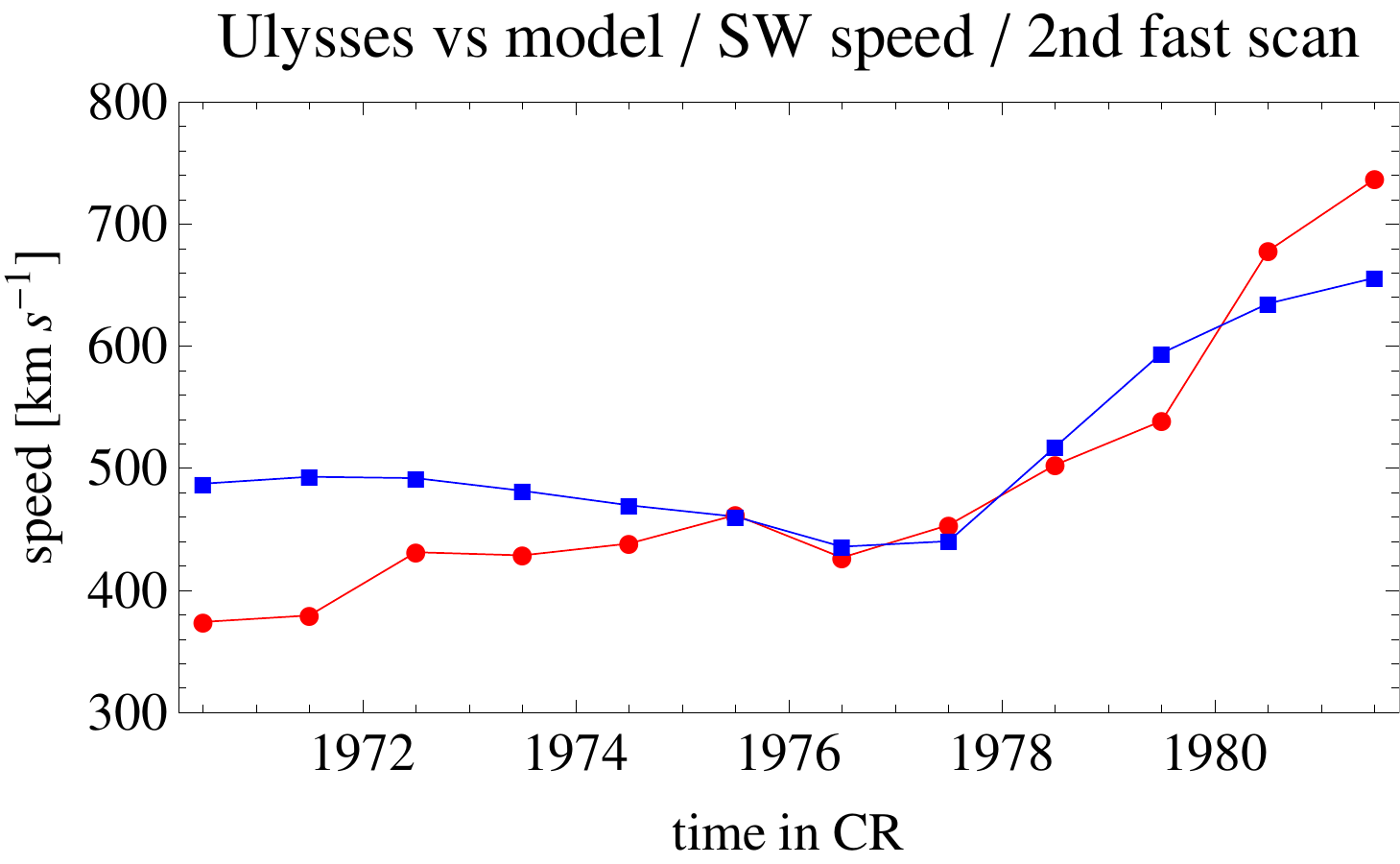} & 	\includegraphics[scale=0.35]{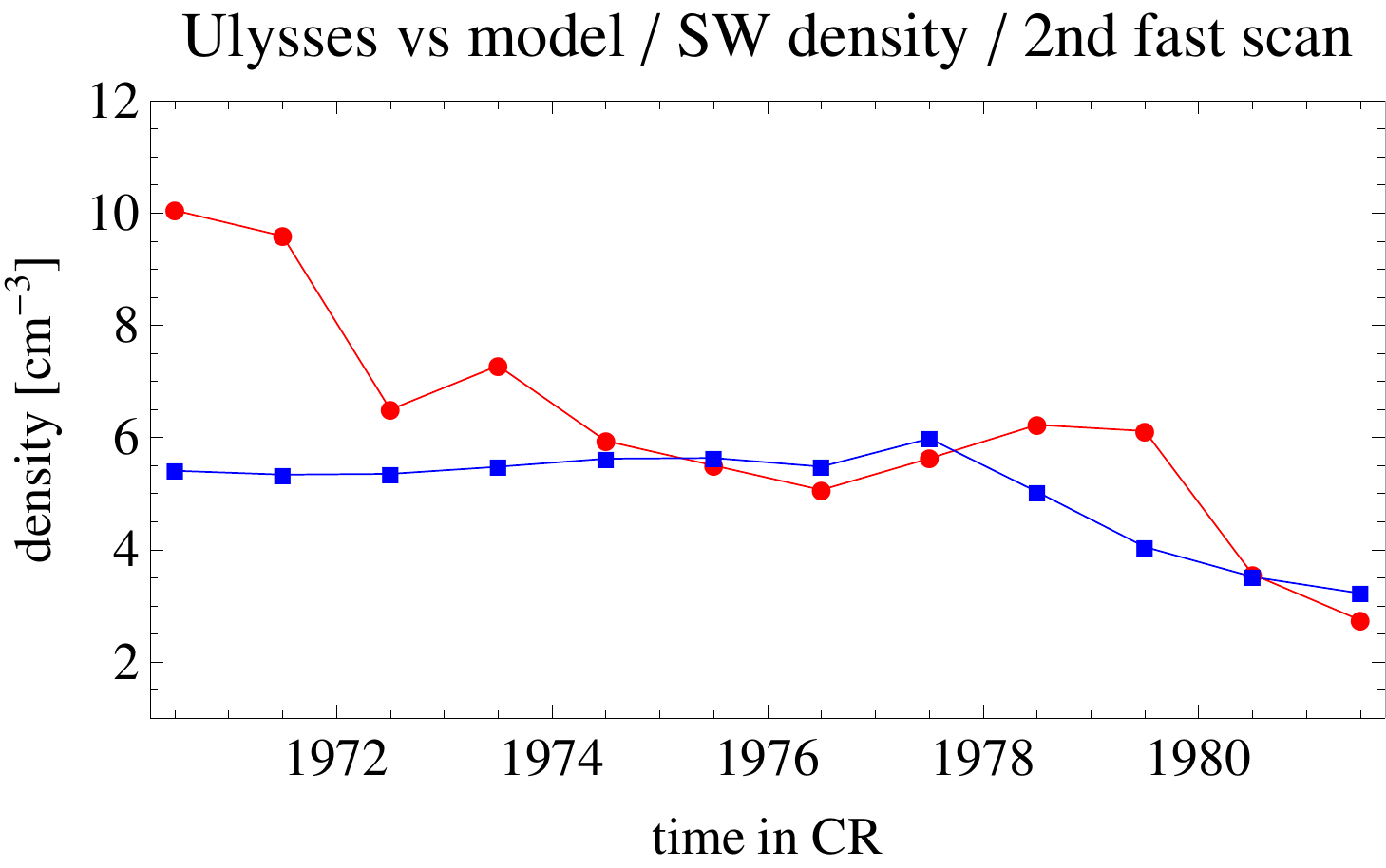}\\
		\includegraphics[scale=0.35]{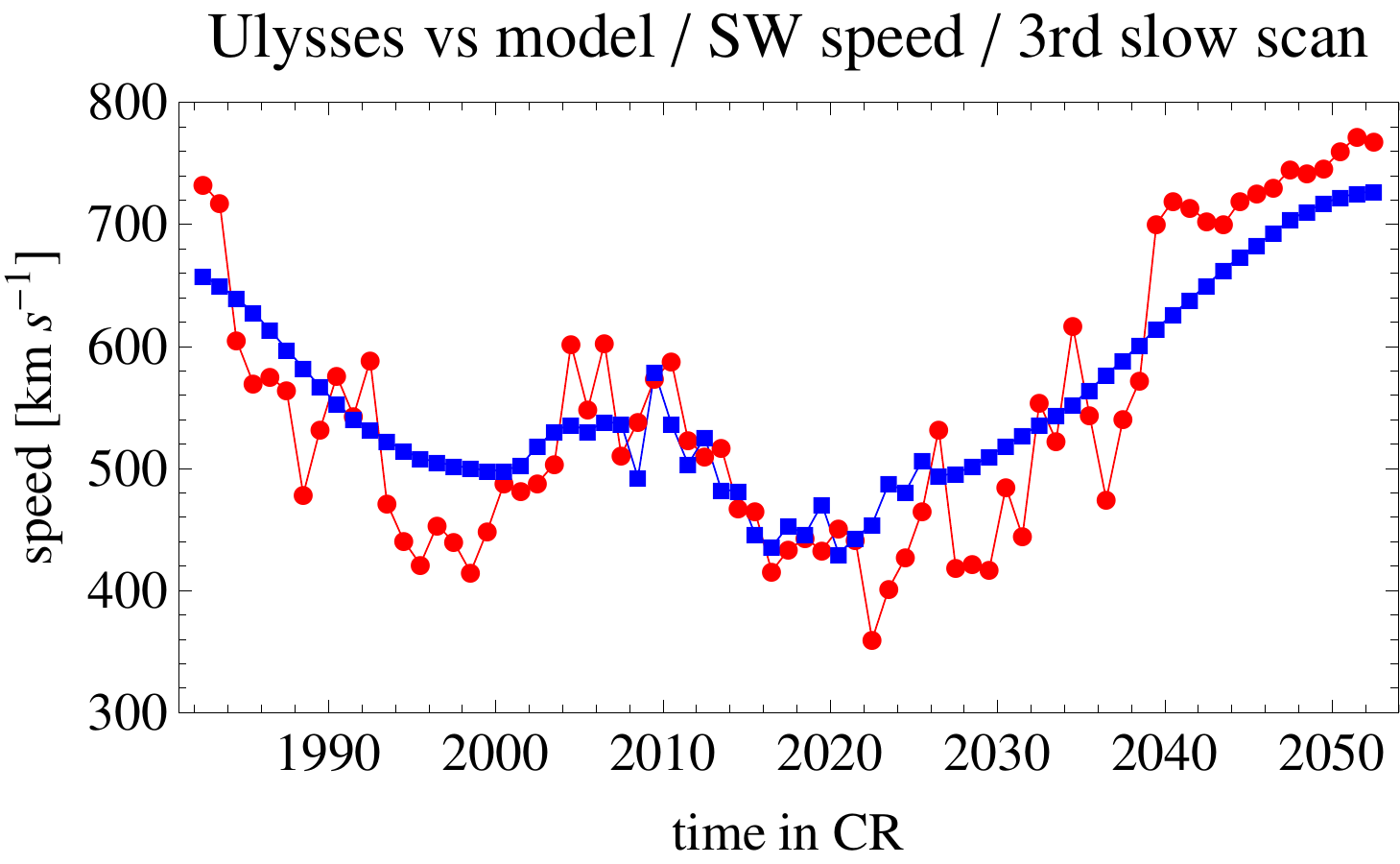} & 	\includegraphics[scale=0.35]{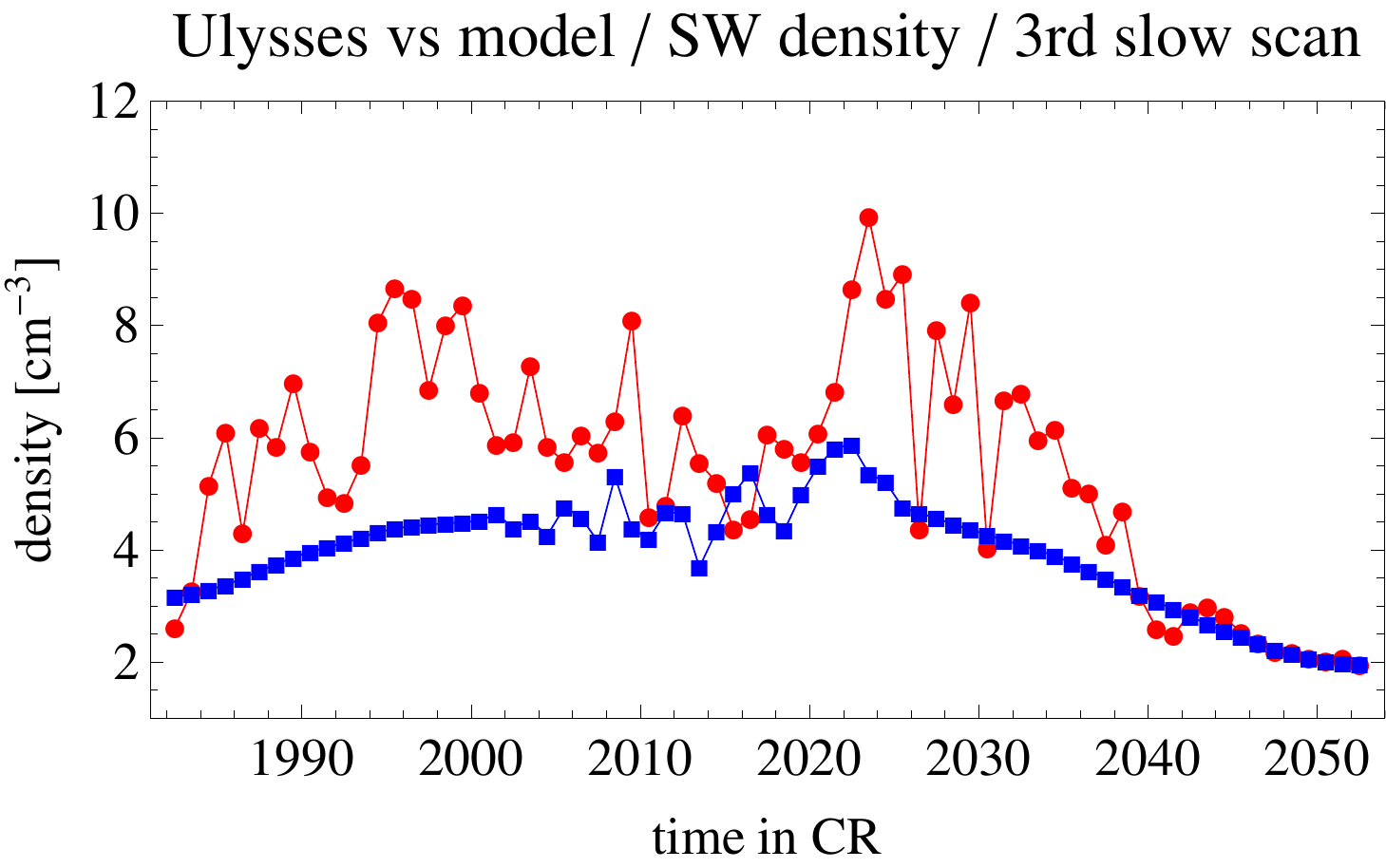}\\
		\includegraphics[scale=0.35]{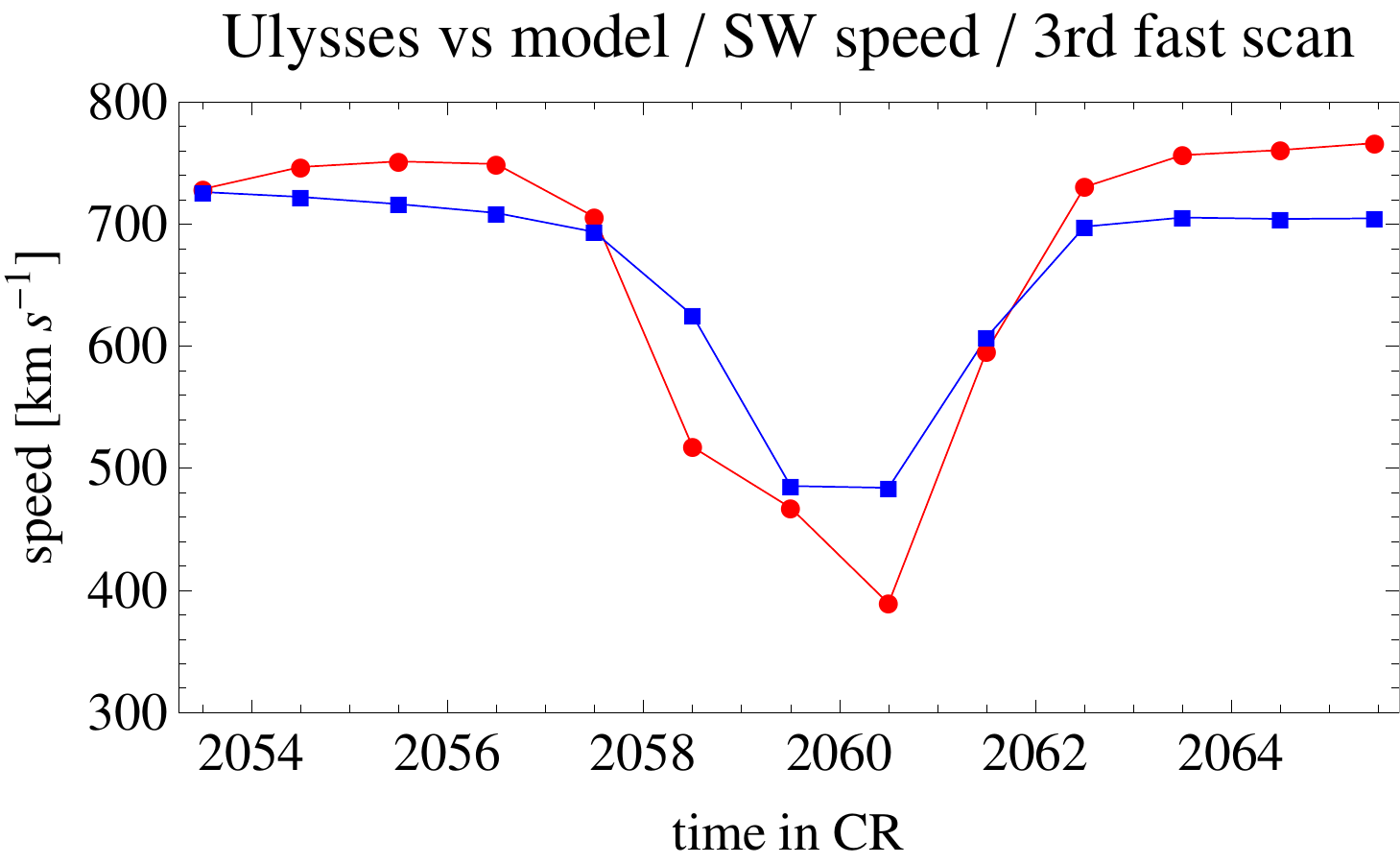} & 	\includegraphics[scale=0.35]{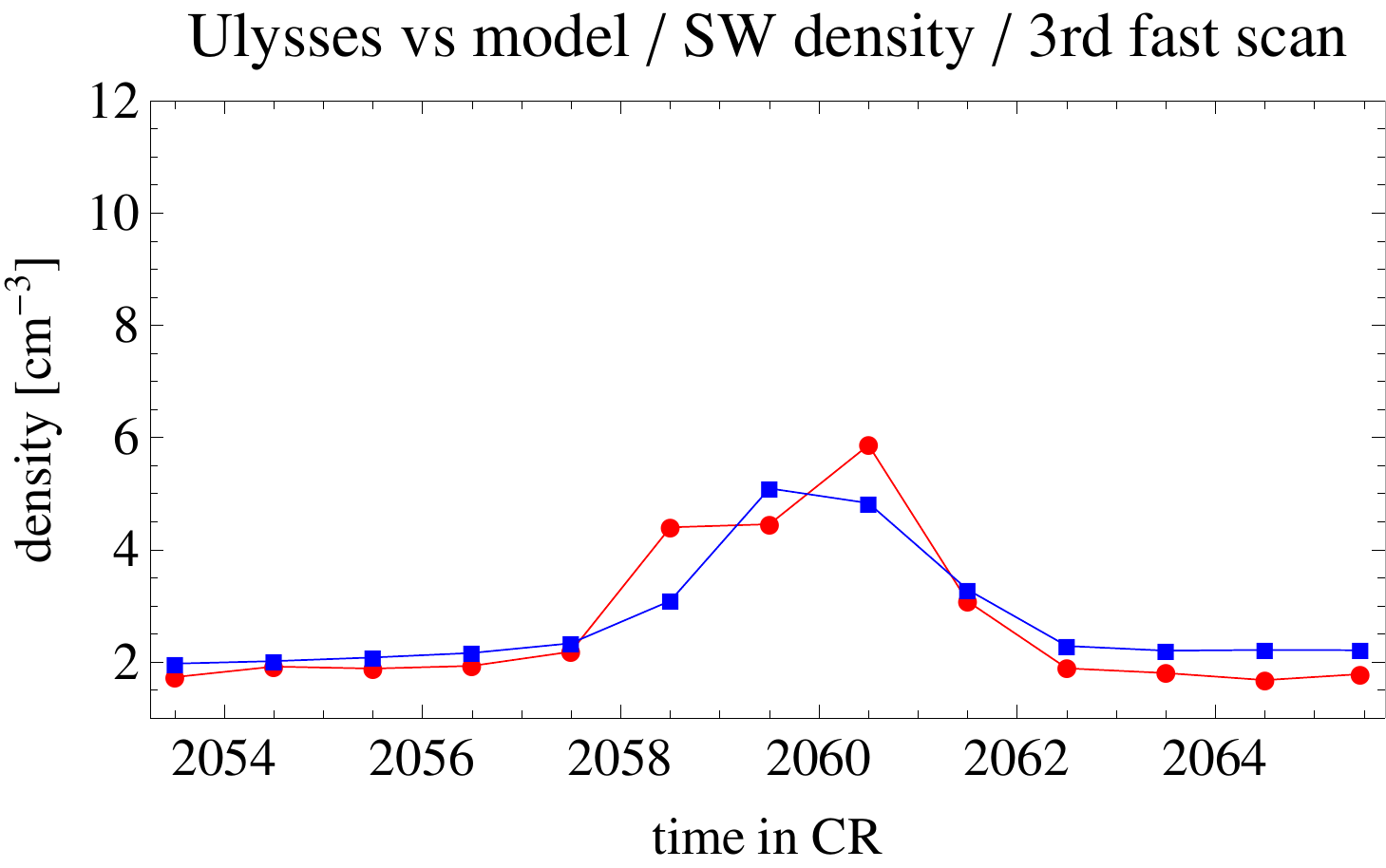}\\
		\includegraphics[scale=0.35]{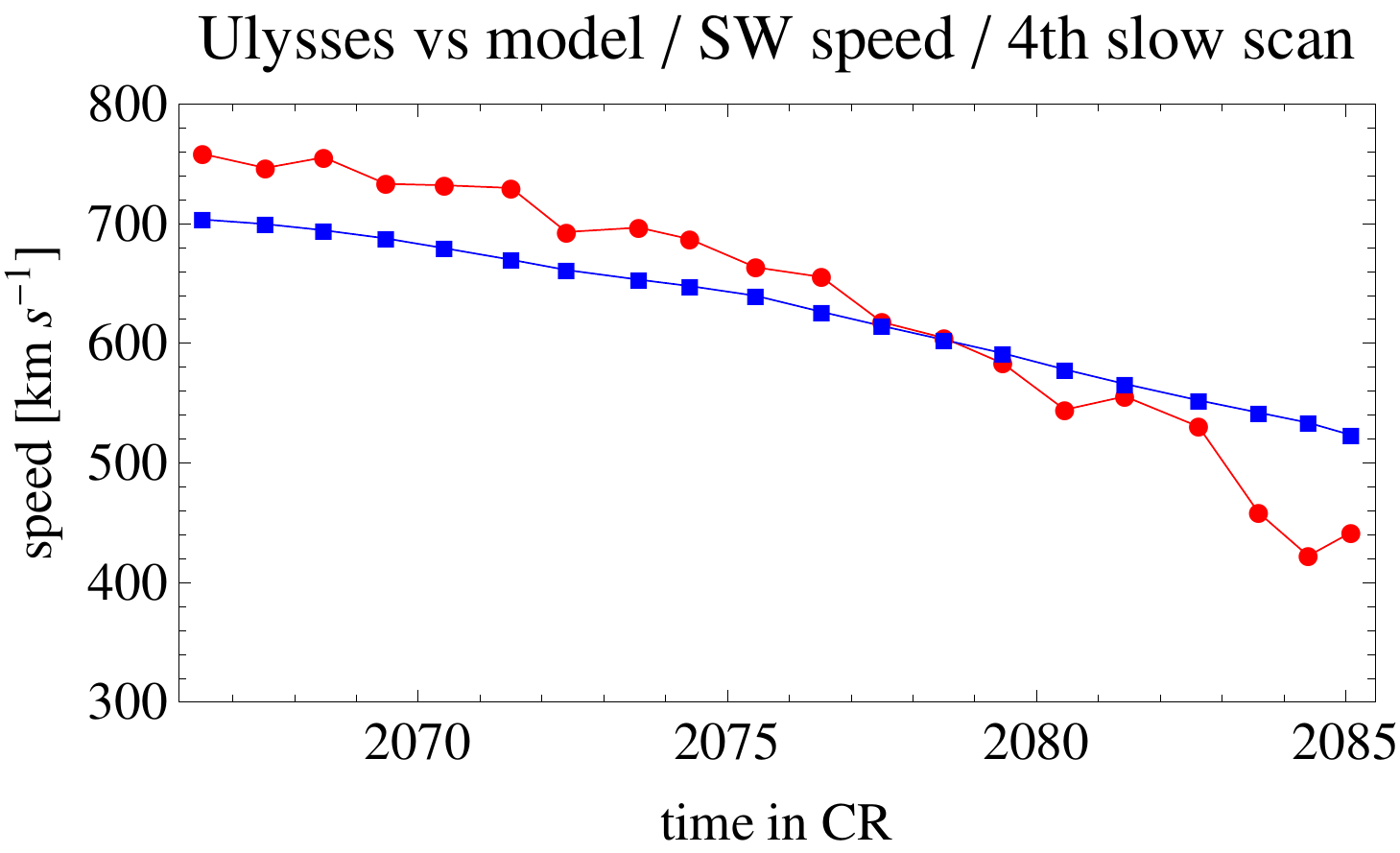} & 	\includegraphics[scale=0.35]{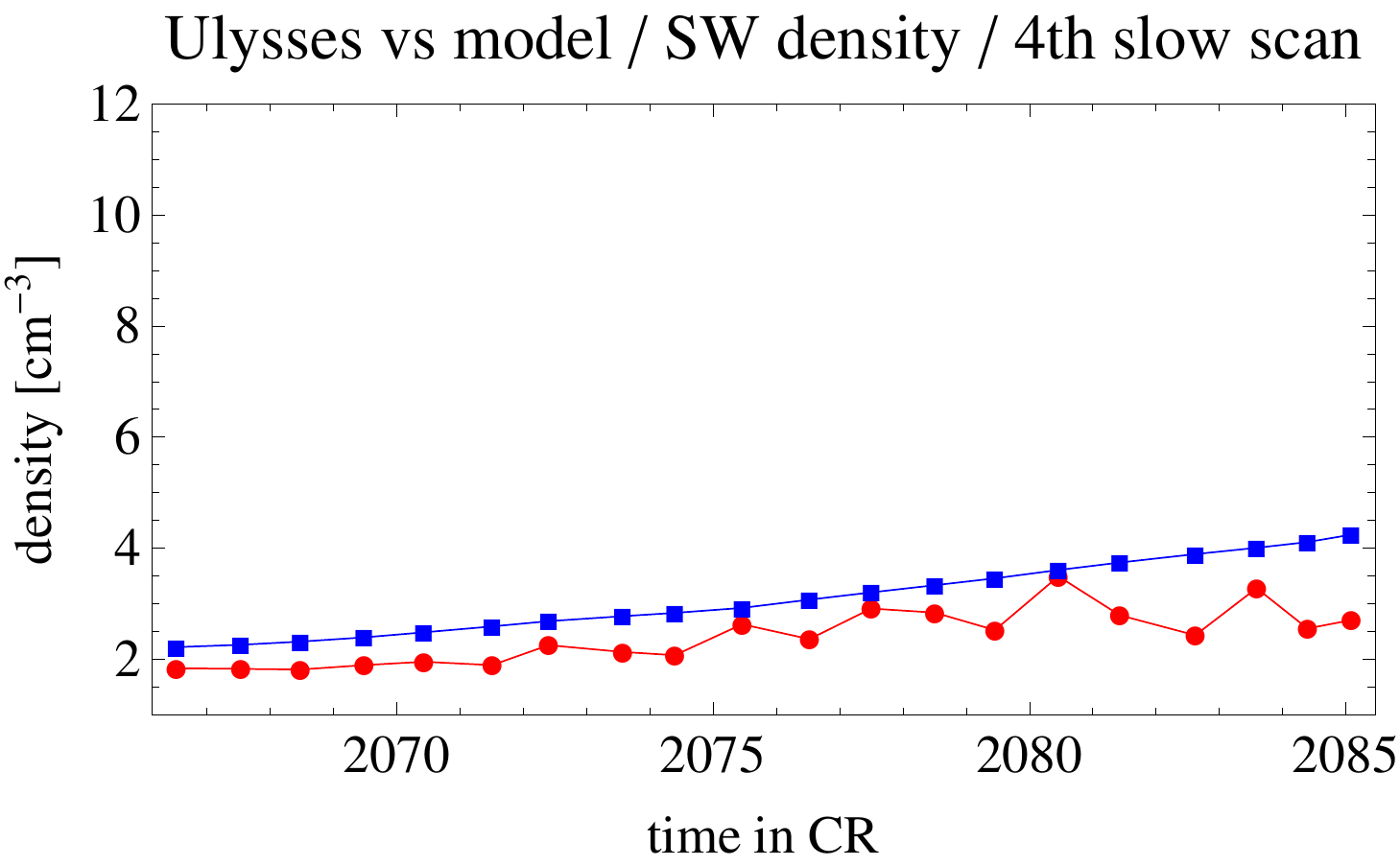}\\		
		\end{tabular}
		\caption{Comparison of \textit{Ulysses} \textit{in~situ} measurements for slow and fast scans with model results for solar wind speed and adjusted density. Presented are Carrington rotation averaged data. Blue - model, red - \textit{Ulysses}.}
		\label{figCompareModelWithUlysses}
		\end{figure*}

A comparison is presented in Figure~\ref{figCompareModelWithUlysses}. The agreement is quite good in the ecliptic parts of the \textit{Ulysses} orbit and satisfactory for higher latitudes. The model retrieves the fast solar wind speed, but some discrepancies exist for the slow and variable solar plasma flows. We have a better agreement in solar wind speed than in density, which is understandable, since the density values were derived from the already approximate speed values using approximate statistically-derived solar wind density-speed relations. 

The worst agreement between \textit{Ulysses} density measurements and our model is during the third slow scan (during descending phase of solar activity), when \textit{Ulysses} \textit{in~situ} measurements for this solar cycle phase give highly variable values, nearly $50\%$ greater than our model predictions. The source of the disagreement might be connected with the density-speed correlation formula we adopted for this time interval: for solar maximum we use an average formula from the two fast scans during solar minimum assuming that the correlation does not change with the solar cycle.

The overall agreement, however, is much better. The residuals of speed are typically $10\%$, not exceeding $30\%$, and typical residuals of density are $20-30\%$, not exceeding $60\%$. The sign of the residuals varies, which suggests there is no systematic global bias in our method. Given all the uncertainties in the absolute calibration of both \textit{in~situ} measurements and IPS observations and the relative simplicity of our approach, we believe such a level of agreement between the model and the measurements is quite good and hard to improve on without an additional data source.

Such a source of additional information might result from an inversion of photometric maps of the Lyman-$\alpha$ helioglow obtained from SWAN/SOHO observations \citep{lallement_etal:10a}, aimed at calculation of the total ionization rate of neutral interstellar hydrogen in the inner heliosphere as a function of heliolatitude and Carrington rotation. With this information, one might be able to follow the idea presented by \citet{bzowski_etal:12b} and independently calculate the profiles of solar wind density. 

In this analysis we assumed that the solar wind parameters obtained from OMNI-2, \textit{Ulysses} and IPS are directly comparable, i.e. that there is no systematic change in solar wind speed with the solar distance between the region from the solar wind acceleration region (a dozen solar radii) to 1~AU, from which the IPS solar wind speeds are retrieved, and \textit{Ulysses}, which measured between $\sim 1.4$ and $\sim 5.5$~AU. Also, we assume there is no distance-related change in the density other than the simple $1/r^2$ scaling. 

The assumption of purely radial expansion of solar wind implies that the latitude structure we inferred in this paper does not change until the termination shock. This may not be exactly true, as pointed out by \citep{fahr_scherer:04b}, who argue that the pickup ions (PUIs) pressure induces nonradial flows at large  heliocentric distances. Such flows would cause changes both in the radial component of solar wind speed and in the local density. A more thorough study of this effect requires, in our opinion, using a multifluid, 3D and time dependent MHD modeling with turbulence and boundary/initial conditions taken from observations. Such a model, to our knowledge, is still in development \citep{usmanov_etal:11a}. Our results seem to be well suited as the boundary conditions for such modeling.

The assumption of a $1/r^2$ drop in density and constancy of speed is increasingly invalid with an increase of solar distance because of the interaction with neutral interstellar gas, which results in the creation of pickup ions and slowdown and heating of the distant solar wind. These phenomena were extensively discussed by \citet{fahr_rucinski:99,fahr_rucinski:01a,fahr_rucinski:02a,fahr:07a,lee_etal:09a,richardson_etal:95a, richardson_etal:08a,richardson_etal:08b}.

However, for the global modeling of the heliosphere and calculation of survival probabilities it is the total flux of solar wind, being a sum of the core solar wind and pickup ions, that counts most. In this respect, the total solar wind flux quite exactly follows the $1/r^2$ scaling due to the continuity conditions, as discussed by \citet{bzowski_etal:12b}. This is because a vast majority of PUIs are created due to charge exchange and thus are not new members of the solar wind population.

		\clearpage

\section{Summary and outlook}
We have combined \textit{in~situ} and IPS remote-sensing observations to retrieve the heliolatitude structure at 1~AU of solar wind speed and density and its evolution in time and produced a time series of heliolatitude profiles of solar wind speed and density for each year between 1990 and 2009, averaged over 10-degree heliolatitude bins. 

We carefully checked the agreement between the solar wind speed measurements available in the OMNI-2 data base and \textit{Ulysses} fast latitude scans and IPS measurements of solar wind speed in the ecliptic plane and found it to be excellent. We verified the agreement between the yearly-averaged heliolatitude profiles of solar wind speed available from Computer Assisted Tomography processed the IPS observations and the speed profiles obtained from the three fast latitude scans from \textit{Ulysses}. Thus, we adopted the IPS-derived yearly speed profiles with some additional smoothing as representative for solar wind starting from the maximum of solar activity in 1990 until the end of available data in 2011 (see Figure~\ref{figModel4and2forUlysses}). 

Based on the measurements from \textit{Ulysses} obtained from the first and third fast latitude scans, performed during the previous and most recent solar minima, we established approximate linear correlations between solar wind density and speed that can be used only to retrieve the heliolatitude structure of solar wind density. We found that the slope was practically unchanged between the two cycles, but the intercept changed because of the global reduction in solar wind density observed since the solar minimum in 2001. Using these correlations, we calculated the yearly solar wind density profiles based on the smoothed velocity profiles (see Figure~\ref{figResDens}).

To facilitate the use of our time and heliolatitude model of solar wind structure in the interpretation of heliospheric measurements, we calculated bilinearly interpolated profiles of solar wind speed and density on a Carrington rotation period grid and replaced the equatorial values obtained from the aforementioned procedure with Carrington rotation averages of the solar wind speed and density available from the OMNI-2 time series (see Figure~\ref{figResSpeedDens}), and $\pm 10^\circ$ and $\pm 90^\circ$ bins by appropriate interpolation values.

Further, we calculated some quantities frequently used in the heliospheric studies, including solar wind flux, charge-exchange rate with neutral H, and solar wind dynamic pressure (see Figure~\ref{figResFluxCXDynPress}). 

The results presented in this paper under the assumption about the radial solar wind flow can be applied in global heliospheric modeling, where on one hand one needs to know the long-term variations in solar wind ram pressure and production rate of pickup ions, but on the other hand short-scale variations in the parameters are less important. They can also be applied in the interpretation of in-ecliptic heliospheric measurements such as observations of H ENA flux by IBEX or observations of the Lyman-$\alpha$ helioglow, where most important are effects of the solar wind interaction with hydrogen atoms within $\sim 10$~AU from the Sun. A homogeneous treatment of long time series of solar wind observations enables a direct application of our results to the interpretation of heliospheric experiments from 1990 until present. 

The model can still be improved by additional sources of data and by a more advanced modeling of solar wind evolution in time and solar distance. Improving the model will most likely have to be an iterative process. For example, the current model could be used as input to a procedure used to fit a ionization rate model to the global helioglow observations by SWAN/SOHO, and from the result of such fitting an approximation of solar wind density profiles and their evolution in time could be retrieved and used to improve our solar wind evolution model. Studies of the time and heliographic latitude behavior of solar wind is thus still a work in progress.

\begin{acks}
The authors acknowledge the use of solar wind speed data from IPS observations carried by STEL Japan, NASA/GSFC's Space Physics Data Facility's ftp service for \textit{Ulysses}/SWOOPS (\url{ftp://nssdcftp.gsfc.nasa.gov/spacecraft_data/ulysses/plasma/swoops/ion/}) and OMNI-2 data collection (\url{ftp://nssdcftp.gsfc.nasa.gov/spacecraft_data/omni/}). The F$_{10.7}$ solar radio flux was provided by the NOAA and Pentincton Solar Radio Monitoring Program operated jointly by the National Research Council and the Canadian Space Agency (\url{ftp://ftp.ngdc.noaa.gov/STP/SOLAR_DATA/SOLAR_RADIO/FLUX/Penticton_Adjusted/} and \url{ftp://ftp.geolab.nrcan.gc.ca/data/solar_flux/daily_flux_values/}). \\
J.S. and M.B. were supported by the Polish Ministry for Science and Higher Education grants NS-1260-11-09 and N-N203-513-038. Contributions from D.M. were supported by NASA's IBEX Explorer mission. 
\end{acks}

\end{article} 

\bibliographystyle{spr-mp-sola}

\end{document}